\documentclass[sigconf]{acmart}

\usepackage{algorithm}
\usepackage{tabularx}

\newif\ifdraft
\draftfalse

\usepackage{xparse}
\usepackage{xspace}
\usepackage{tikz}
\usepackage{float}
\usepackage{multirow}
\usepackage{multicol}
\usepackage{colortbl}
\usepackage{array}
\newcommand{\PreserveBackslash}[1]{\let\temp=\\#1\let\\=\temp}
\newcolumntype{C}[1]{>{\PreserveBackslash\centering}p{#1}}
\newcolumntype{R}[1]{>{\PreserveBackslash\raggedleft}p{#1}}
\newcolumntype{L}[1]{>{\PreserveBackslash\raggedright}p{#1}}

\usepackage{microtype}
\usepackage{graphicx}
\usepackage{mathtools}
\usepackage{float}
\usepackage{subcaption}
\usepackage{booktabs} %
\usepackage[shortlabels]{enumitem}

\usepackage{amssymb}%
\usepackage{pifont}%

\usepackage{bbold}

\usepackage{amssymb,enumitem}

\setlist[itemize]{leftmargin=*}
\setlist[enumerate]{leftmargin=*}

\newcommand*{\rej}{{\ooalign{\lower.3ex\hbox{$\sqcup$}\cr\raise.4ex\hbox{$\sqcap$}}}}

\usepackage{stackengine}

\usepackage{xcolor}

\newcommand{\ie}{\textit{i.e.,}\@\xspace}
\newcommand{\eg}{\textit{e.g.,}\@\xspace}

\DeclareRobustCommand\encircle[1]{\tikz[baseline=(char.base)]{\node[shape=circle,fill,inner sep=1pt] (char) {\textcolor{white}{#1}}}}

\graphicspath{{images/}}

\usepackage{booktabs,arydshln}
\makeatletter
\def\adl@drawiv#1#2#3{%
        \hskip.5\tabcolsep
        \xleaders#3{#2.5\@tempdimb #1{1}#2.5\@tempdimb}%
                #2\z@ plus1fil minus1fil\relax
        \hskip.5\tabcolsep}
\newcommand{\cdashlinelr}[1]{%
  \noalign{\vskip\aboverulesep
           \global\let\@dashdrawstore\adl@draw
           \global\let\adl@draw\adl@drawiv}
  \cdashline{#1}
  \noalign{\global\let\adl@draw\@dashdrawstore
           \vskip\belowrulesep}}
\makeatother

\usepackage{cleveref}

\newcommand{\nlp}[1]{}

\newcolumntype{x}[1]{>{\centering\arraybackslash\hspace{0pt}}p{#1}}

\newcommand{\ours}{\texttt{ADAGE}\xspace}
\ifdraft
\usepackage[textsize=tiny]{todonotes}
\newcommand{\mynote}[1]{\textcolor{red}{[note: #1]}}

\newcommand{\mytodo}[1]{\textcolor{red}{[todo: #1]}}
\newcommand{\mycomment}[1]{\textcolor{red}{[comment: #1]}}
\newcommand{\chris}[1]{\textcolor{red}{Chris: #1}}
\newcommand{\ahmad}[1]{\textcolor{darkpastelgreen}{[Ahmad: #1]}}
\newcommand{\vinith}[1]{\textcolor{blue}{Vinith: #1}}
\newcommand{\yunxiang}[1]{\textcolor{cyan}{Yunxiang: #1}}
\newcommand{\xiao}[1]{\textcolor{blue}{xiao: #1}}
\definecolor{chocolate(traditional)}{rgb}{0.48, 0.25, 0.0}

\definecolor{darkpastelgreen}{rgb}{0.01, 0.75, 0.24}
\newcommand{\natalie}[1]{\textcolor{darkpastelgreen}{natalie: #1}}
\definecolor{pistachio}{rgb}{0.58, 0.77, 0.45}

\newcommand{\jonas}[1]{\textcolor{violet}{[Jonas: #1]}}

\newcommand{\adelin}[1]{\textcolor{red}{[Adelin: #1]}}
\newcommand{\mohammad}[1]{\textcolor{red}{[Mohammad: #1]}}

\definecolor{amber(sae/ece)}{rgb}{1.0, 0.49, 0.0}
\newcommand\adam[1]{{\textcolor{red}{[Adam: #1]}}}
\newcommand{\sierra}[1]{\textcolor{blue}{[Sierra: #1]}}
\newcommand{\armin}[1]{\textcolor{cyan}{[Armin: #1]}}
\newcommand{\nikita}[1]{\textcolor{cyan}{[Nikita: #1]}}
\newcommand{\franzi}[1]{\textcolor{purple}{[Franzi: #1]}}

\else

\usepackage[disable]{todonotes}
\newcommand{\chris}[1]{}
\newcommand{\franzi}[1]{}
\newcommand{\vinith}[1]{}
\newcommand{\adam}[1]{}
\newcommand{\yunxiang}[1]{}
\newcommand{\natalie}[1]{}
\newcommand{\jonas}[1]{}
\newcommand{\adelin}[1]{}
\newcommand{\mynote}[1]{}
\newcommand{\xiao}[1]{}
\newcommand{\mytodo}[1]{}
\newcommand{\mycomment}[1]{}
\newcommand{\ahmad}[1]{}
\newcommand{\mohammad}[1]{}
\newcommand{\sierra}[1]{}
\newcommand{\armin}[1]{}
\newcommand{\nikita}[1]{}

\fi

\usepackage{amsmath,amsfonts,bm}

\def\eqref#1{equation~\ref{#1}}

\def\1{\bm{1}}

\DeclareMathAlphabet{\mathsfit}{\encodingdefault}{\sfdefault}{m}{sl}
\SetMathAlphabet{\mathsfit}{bold}{\encodingdefault}{\sfdefault}{bx}{n}

\usepackage{algpseudocode}
\DeclareMathOperator*{\argmax}{arg\,max}

\usepackage{threeparttable}
\Crefname{algocf}{Algorithm}{Algorithms}
\AtBeginDocument{%
  }

\acmYear{2026}\copyrightyear{2026}
\setcopyright{cc}
\setcctype[4.0]{by}
\acmConference[ASIA CCS '26]{ACM Asia Conference on Computer and Communications Security}{June 1--5, 2026}{Bangalore, India}
\acmBooktitle{ACM Asia Conference on Computer and Communications Security (ASIA CCS '26), June 1--5, 2026, Bangalore, India}
\acmDOI{10.1145/3779208.3806086}
\acmISBN{979-8-4007-2356-8/26/06}

\author{Jing Xu}
\affiliation{%
  \institution{CISPA Helmholtz Center for Information Security}
  \country{}
}
\email{jing.xu@cispa.de}

\author{Franziska Boenisch}
\affiliation{%
  \institution{CISPA Helmholtz Center for Information Security}
  \country{}
}
\email{boenisch@cispa.de}

\author{Adam Dziedzic}
\affiliation{%
  \institution{CISPA Helmholtz Center for Information Security}
  \country{}
}
\email{adam.dziedzic@cispa.de}



\begin{document}

\newcommand{\ourtitle}{
ADAGE: Active Defenses Against GNN Extraction
}
\title{\ourtitle}

\begin{abstract}
Graph Neural Networks (GNNs) achieve high performance in various real-world applications, such as drug discovery, traffic states prediction, and recommendation systems. 
The fact that building powerful GNNs requires a large amount of training data, powerful computing resources, and human expertise turns the models into lucrative targets for model stealing attacks. 
Prior work has revealed that the threat vector of stealing attacks against GNNs is large and diverse, as an attacker can leverage various heterogeneous signals ranging from node labels to high-dimensional node embeddings to create a local copy of the target GNN at a fraction of the original training costs. 
This diversity in the threat vector renders the design of effective and \emph{general} defenses challenging and existing defenses usually focus on one particular stealing setup.
Additionally, they solely provide means to identify stolen model copies rather than preventing the attack. 
To close this gap, we propose the first and general \texttt{Active Defense Against GNN Extraction (ADAGE)}. 
\ours builds on the observation that stealing a model's full functionality requires highly diverse queries to leak its behavior across the input space. Our defense monitors this query diversity and progressively perturbs outputs as the accumulated leakage grows. 
In contrast to prior work, \ours can prevent stealing across \textit{all common attack setups}. 
Our extensive experimental evaluation using six benchmark datasets, four GNN models, and three types of adaptive attackers shows that \ours 
penalizes attackers to the degree of rendering stealing impossible, whilst preserving predictive performance on downstream tasks.  
\ours, thereby, contributes towards securely sharing valuable GNNs in the future.
\end{abstract}

\maketitle

\renewcommand{\shortauthors}{Xu et al.}

\section{Introduction}
Many real-world datasets can be represented as graphs, such as social, transport, or financial networks.
To train on graph data, Graph Neural Networks (GNNs) have been introduced~\citep{kipf2016semi, hamilton2017inductive, xu2018powerful, velivckovic2017graph} which achieve high performance in many applications, \eg node classification~\citep{kipf2016semi}, graph classification~\citep{vishwanathan2010graph, shervashidze2011weisfeiler}, link prediction~\citep{zhang2018link}, and recommendations~\citep{fan2019graph}. 
Since training performant GNNs requires large amounts of training data, powerful computing resources, and human expertise, the models become lucrative targets for model stealing attacks~\citep{tramer2016stealing}. In model stealing attacks, an attacker leverages query-access to a target model and uses the query data and corresponding model outputs to train a local surrogate model (\ie a \emph{"stolen copy"}) with similar task performance, often at a fraction of the original training costs. 

\begin{figure*}[t!]
    \centering
    \includegraphics[width=0.75\textwidth]{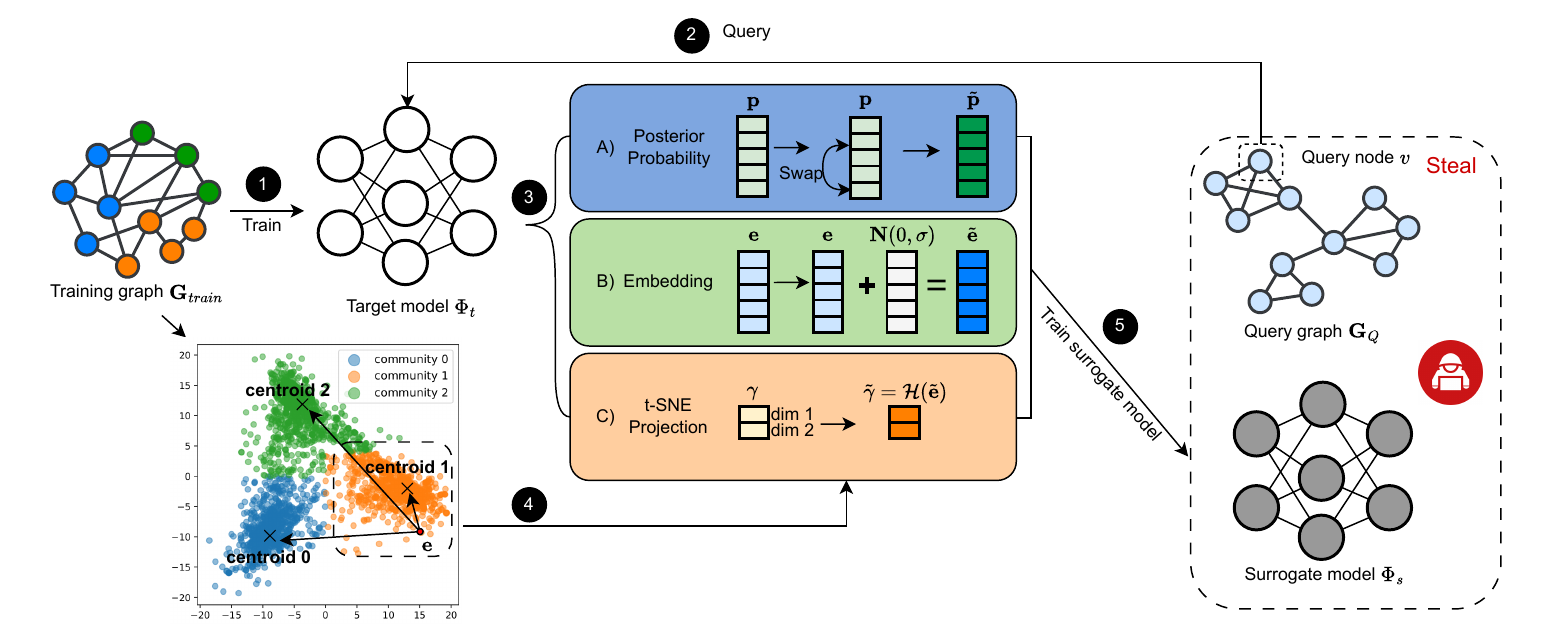}
    \caption{\textbf{Overview of our \ours.} \encircle{1} Target model $\Phi_t$ is trained on training graph $\mathbf{G}_{train}$. Model owner detects communities in $\mathbf{G}_{train}$, and computes each community's centroid in the embedding space. \encircle{2} Attacker queries $\Phi_t$  with node $v$ from a query graph  $\mathbf{G}_Q$. \encircle{3} Based on the setup, $\Phi_t$ yields either A)~a predicted posterior probability, B) a high-dimensional node embedding, or C) a low-dimensional projection. \encircle{4} Based on the internal representations of $v$, the model owner identifies its nearest community in $\mathbf{G}_{train}$. Based on the fraction of total communities already covered by the attacker's queries, the output of $\Phi_t$ is perturbed with an adequate strength, where the type of perturbation depends on the output. %
    The more communities covered, the higher the perturbation. \encircle{5} The perturbed output is returned to the attacker whose trained surrogate model decreases in performance as the defense strength increases over time.}
    \label{fig:defense_overview}
\end{figure*}

Model stealing attacks have shown significant effectiveness aga-\\inst GNNs in prior work~\citep{defazio2019adversarial, shen2022model, wu2022model}. \citet{shen2022model} underscores the \textit{wide range of threat vectors} exploited by such attacks, demonstrating that adversaries can target GNNs through various model outputs including class probabilities,  node embeddings, or even low dimensional projections of these node embeddings. 
Given this variety of threat vectors and the attacks' success, we require defenses that are  \textit{general} and \textit{prevent}  GNN stealing. However, prior defenses are typically limited to one specific scenario, focusing, for example, on transductive or inductive GNNs and targeting only a single type of output. Moreover, these defenses aim to detect stolen models post-attack through techniques like watermarking or fingerprinting~\citep{zhao2021watermarking, xu2023watermarking, waheed2023grove}, rather than preventing the stealing while it is taking place.

To address these limitations, we propose \texttt{Active Defense Aga-\\inst GNN Extraction} as a novel defense mechanism. Unlike prior approaches, \ours is the first method that is both \textit{general}, \ie capable of defending against a wide range of stealing scenarios involving diverse model outputs, and \textit{active}, meaning it can \textit{proactively prevent} GNN model stealing while it is happening. \ours leverages the observation that, to steal a GNN with its full functionality, the attacker has to query the target model with diverse data covering diverse regions of the model's input space. 
This is because the more diverse the queries, the more information about the model's functionality can be exposed. 
We further identify that, in GNNs, this query diversity can be well approximated through the lens of different \textit{communities} in the underlying graph.

Leveraging the model owner's access to the underlying training graph of the GNN and the GNN-internal query representations, we design \ours to monitor the fraction of communities in the underlying graph that a user’s queries to the GNN have already covered. 
Based on this information, \ours dynamically calibrates the defense strength, introducing increasing perturbation to the model output (\eg node labels, embeddings, or projections) as more communities are queried. 
We present an overview of our \ours framework in \Cref{fig:defense_overview}. 

Our thorough experimental evaluation on six benchmark datasets and four GNN architectures highlights that with \ours, the model outputs returned to attackers degrade the performance of stolen model copies, while maintaining downstream task performance.

In summary, we make the following contributions:
\begin{itemize}
    \item We propose \ours, the first \textit{general} and \textit{active} defense to prevent GNN model stealing.
    \item We thoroughly evaluate \ours on six datasets and four different GNN models to show that \ours prevents model stealing in all common stealing setups while maintaining high predictive performance on downstream tasks.
    \item We highlight that \ours is effective to prevent model stealing in both node classification and link prediction tasks. 
    \item We assess \ours against three different types of adaptive attackers and show that our defense remains effective.
\end{itemize}

\section{Background}
We first introduce the notations and fundamental concepts used in this paper. 
\subsection{Notations}
\label{sec:notations}

We define $\mathbf{G}=(\mathcal{V},\mathcal{E},\mathbf{X})$ as a undirected, unweighted, attributed graph, where $\mathcal{V}=\left \{v_1, v_2, \dots, v_n \right \}$ denotes the set of nodes, $\mathcal{E}\subseteq \left \{ (v, u)|v, u \in \mathcal{V} \right \}$ denotes the set of edges, $\mathbf{X}$ denotes the node attribute matrix. We denote $\mathbf{A} \in \left \{ 0, 1 \right \} ^ {n \times n}$ as the adjacency matrix, where $A_{vu}=1, \forall(v, u)\in \mathcal{E}$. \Cref{tab:notation} provides an overview on the notation used in this paper for the readers' convenience. 
We use lowercase letters to denote scalars, calligraphic letters to denote sets, and bold uppercase letters to denote matrices.
\begin{table}[ht]
\small
 \centering
\caption{\textbf{Notation} used throughout the work.}
\label{tab:notation}
 \begin{tabular}{c|l} 
 \hline
 \textbf{Notations} & \textbf{Descriptions} \\ \hline
 $\mathbf{G}=(\mathcal{V},\mathcal{E},\mathbf{X})$ & graph \\ \hline
 $v, u \in \mathcal{V}$ & node  \\ \hline
 $c_i \in \mathbb{C}$ & node class \\ \hline
 $n =|\mathcal{V}|$ & number of nodes \\ \hline
 $d$ & dimension of a node embedding vector \\ \hline
 $m$ & dimension of a node feature vector \\ \hline
 $\mathbf{A} \in \left \{ 0, 1 \right \} ^ {n \times n}$ & adjacency matrix \\ \hline
 $\mathbf{X} \in \mathbb{R}^{n \times m}$ & node feature matrix \\ \hline
 $\mathbf{R}$ & query response \\ \hline
  $\bm{\Theta}\in \mathbb{R}^{n \times |\mathbb{C}|}$ & predicted posterior probability matrix \\ \hline
 $\mathbf{E} \in \mathbb{R}^{n\times d}$ & node embedding matrix \\ \hline
 $\bm{\Upsilon} \in \mathbb{R}^{n \times 2}$ & 2-dimensional t-SNE projection matrix \\ \hline
 $\mathbf{G}_{train} / \mathbf{G}_{test}$ & training/test graph \\ \hline
 $\mathbf{G}_Q$ & query graph \\ \hline
 $\delta$ & query rate \\ \hline
 $K$ & number of communities \\ \hline
 $\mathcal{N}_v$ & neighborhood of $v$ \\ \hline
 $\Phi_t/\Phi_s$ & target/surrogate GNN model \\ \hline
 $\hat{\Phi}_t/\hat{\Phi}_s$ & target/surrogate encoder \\ \hline
 $\mathcal{C}_t/\mathcal{C}_s$ & target/surrogate classification head \\ \hline
\end{tabular}
\end{table}

\subsection{Preliminaries}
\label{sec:preliminaries}
\paragraph{Graph Neural Networks.} 
GNNs have achieved significant success in processing graph data. 
GNNs take a graph $\mathbf{G}=(\mathcal{V},\mathcal{E},\mathbf{X})$ as an input, and learn a representation vector (embedding) $\mathbf{z}_v$ for each node ${v} \in \mathbf{G}$, or the representation for the entire graph, $\mathbf{z}_\mathbf{G}$. 
Modern GNNs (\eg GCN~\citep{kipf2017semi}, GraphSAGE~\citep{hamilton2017inductive}, and GAT~\citep{velickovic2018graph}) follow a neighborhood aggregation strategy, where one iteratively updates the representation of a node by aggregating representations of its neighbors. After $l$ iterations of aggregation, a node's representation captures both structure and feature information within its $l$-hop network neighborhood~\citep{xu2018powerful}. 
GNNs then either output node or graph representations. The node representations can be used for various downstream tasks, such as node classifications, recommendation engines, and visualizations. %
GNN models for node classification tasks can be trained through two learning settings, \ie \textit{transductive} learning and \textit{inductive} learning, where in transductive learning, we input the entire graph for training and mask the labels of the "test" data nodes, while in inductive learning, the test graph is disjoint from the input training graph.

Formally, the $l$-th layer of a GNN is:
\begin{equation}
    \mathbf{z}_v^{(l)} = \sigma(\mathbf{z}_v^{(l-1)}, AGG(\{ \mathbf{z}_u^{(l-1)}; u \in \mathcal{N}_v \})), \forall l \in [L],
    \label{eqn:2.2-1}
\end{equation}
where $\mathbf{z}_v^{(l)}$ is the representation of node $v$ computed in the $l$-th iteration. $\mathcal{N}_{v}$ are neighbors of node $v$, and the $AGG(\cdot)$ is an aggregation function that can vary for different GNN models. $\mathbf{z}_v^{(0)}$ is initialized as node feature, while $\sigma$ is an activation function.
For the graph classification task, the READOUT function pools the node representations for a graph-level representation $\mathbf{z}_\mathbf{G}$:
\begin{equation}
    \mathbf{z}_\mathbf{G} = READOUT({\mathbf{z}_v;v \in \mathcal{V}}).
    \label{eqn:2.2-2}
\end{equation}
READOUT can be a simple permutation invariant function,\eg summation, or a sophisticated graph-level pooling function~\citep{ying2018hierarchical, zhang2018end}. 

\paragraph{Stealing GNNs.} The two different setups for training of GNNs also reflect in the setup for their stealing attacks: 
Attackers who aim at stealing transductive GNNs are assumed to have access to the training graph of the target model---often an unrealistic assumption. 
In contrast, attackers who steal inductive GNNs %
are assumed to query the target model with a separate query graph.
In the \textit{transductive setup}, \citet{defazio2019adversarial} proposed GNN stealing that relies on training a surrogate model on perturbated subgraphs and their labels output by the target model, similar to model stealing in non-graph settings, \eg~\citep{tramer2016stealing,papernot2017practical}. \citet{wu2022model} %
extended the attack to more diverse attackers with different degrees of background knowledge. 
In the more realistic \textit{inductive setup} for GNN stealing, there exist currently three state-of-the-art attacks proposed by \citet{shen2022model}.
All attacks assume that the attacker has access to a query dataset $\mathbf{G}_Q$ and obtains the query response $\mathbf{R}_Q$ from the target model $\Phi_t$ which they use to train a surrogate model $\Phi_s$ that mimics the behavior of $\Phi_t$. The query response $\mathbf{R}_Q$ can be A) a \textbf{predicted posterior probability} matrix, B) a \textbf{node embedding} matrix,  or C) a \textbf{t-SNE projection} matrix of the node embedding matrix:

\indent \textbf{A) Predicted Posterior Probabilities.}  The query graph $\mathbf{G}_Q$ contains the adjacency matrix $\mathbf{A}_Q$ and the node feature matrix $\mathbf{X}_Q$. 
As discussed in the threat model, in this stealing setup, the target model consists of a backbone encoder $\hat{\Phi}_t$ which outputs a high-dimensional representation of the query node and a classification head $\mathcal{C}_t$ which outputs a predicted posterior probability. Also, the surrogate model consists of a backbone encoder $\hat{\Phi}_s$ and a classification head $\mathcal{C}_s$. 
Specifically, given a query graph $\mathbf{G}_Q$, $\hat{\Phi}_t$ and $\hat{\Phi}_s$ takes all nodes' $\mathit{l}$-hop subgraphs from $\mathbf{G}_Q$ and outputs high-dimensional representation for each query node, as 
\begin{equation}
\begin{split}
    \mathbf{E} = \hat{\Phi}_t(\mathbf{X}_Q, \mathbf{A}_Q), \mathbf{\hat{E}} = \hat{\Phi}_s(\mathbf{X}_Q, \mathbf{A}_Q)\text{.}
\end{split}
    \label{equ:node_embedding_steal}
\end{equation}
Then, with classification head $\mathcal{C}_t$, the target model can output predicted posterior probabilities for each query node, and the surrogate model (including the encoder $\hat{\Phi}_s$ and classification head $\mathcal{C}_s$) is trained by minimizing the Cross-Entropy loss between the posterior probabilities from the surrogate and target models as
\begin{equation}
\begin{split}
    \boldsymbol{\Theta} &= \mathcal{C}_t(\mathbf{E}), \boldsymbol{\hat{\Theta}} = \mathcal{C}_s(\mathbf{\hat{E}}) \\
    \mathcal{L}_C &= \mathtt{Cross\_Entropy}(\boldsymbol{\Theta}, \boldsymbol{\hat{\Theta}})\text{.}
\end{split}
    \label{equ:posterior_prob}
\end{equation}
\indent \textbf{B) High-dimensional Node Embeddings.} In addition to predicted posterior probabilities, the target model may also directly output the high-dimensional node embeddings, i.e., $\mathbf{E}$. 
With the model output of high-dimensional node embeddings, the goal of the surrogate model is to mimic the behavior of the target model by minimizing the RMSE loss between the output of the surrogate model (\ie $\mathbf{\hat{E}}$) and $\mathbf{E}$ as%
\begin{equation}
\begin{split}
    \mathcal{L}_R &= \mathtt{RMSE}(\mathbf{\hat{E}}, \mathbf{E})\text{.}
\end{split}
    \label{equ:loss}
\end{equation}
where $n_Q$ represents the number of nodes in the query graph $\mathbf{G}_Q$. 

\indent \textbf{C) Low-dimensional t-SNE Projections.} The output of the target model can also concist of low-dimensional t-SNE projections, where each row is a 2-dimensional vector. t-SNE projections are widely returned in the scenarios of graph visualizaiton~\citep{huang2020gnnvis}, transfer learning~\citep{zhu2021transfer}, federated learning~\citep{he2021fedgraphnn}, fine-tuning pretrained GNNs~\citep{hu2019strategies}, and model partitioning where the target model is split into local and cloud parts bridged by embeddings information~\citep{song2019auditing}. The training procedure of the surrogate model is similar to that with model outputs of high-dimensional node embeddings, \ie RMSE loss is used to optimize the surrogate model, as
\begin{equation}
\begin{split}
    \boldsymbol{\Upsilon} &= \mathcal{H}(\mathbf{E}), \boldsymbol{\hat{\Upsilon}} = \mathcal{H}(\mathbf{\hat{E}}) \\
    \mathcal{L}_R &= \mathtt{RMSE}(\boldsymbol{\hat{\Upsilon}}, \boldsymbol{\Upsilon}) \text{,}
\end{split}
\label{equ:t-sne_steal}
\end{equation}
where $\mathcal{H}$ denotes the t-SNE projecting transformation.
To provide a holistic and general defense against GNN stealing, \ours provides protection for all of the three stealing setups.

\section{Our Active Defense}%
\label{sec:method}
In this section, we first introduce the threat model, outlining the adversary’s objectives, capabilities, and knowledge within the context of the state-of-the-art GNN stealing attack framework. Then, we describe the defender’s capabilities and goals. Finally, we present our proposed active defense framework.

\begin{figure*}[!htpb]
    \centering
    \includegraphics[width=0.6\textwidth]{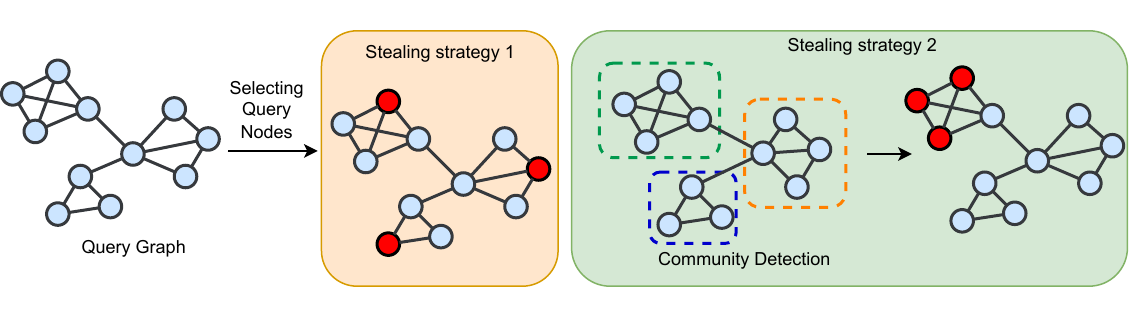}
    \caption{\textbf{Query node selecting strategies.} 
In stealing strategy 1, the query nodes are sampled from the query graph uniformly at random while in stealing strategy 2, the query nodes which are similar to each other are sampled. Here, we utilize a community detection algorithm to sample similar nodes.}
    \label{fig:query_node_selecting}
\end{figure*}

\begin{figure*}[!hbtp]
\centering

    \begin{subfigure}[t]{0.5\textwidth}
         \centering
         \includegraphics[width=\textwidth]{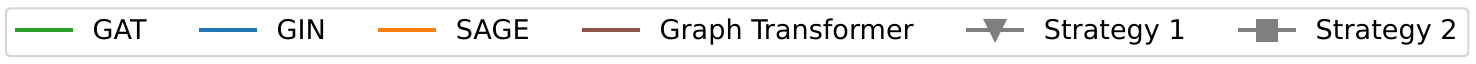}
     \end{subfigure}
     \vfill
    \begin{subfigure}[t]{0.32\textwidth}
         \centering
         \includegraphics[width=\textwidth]{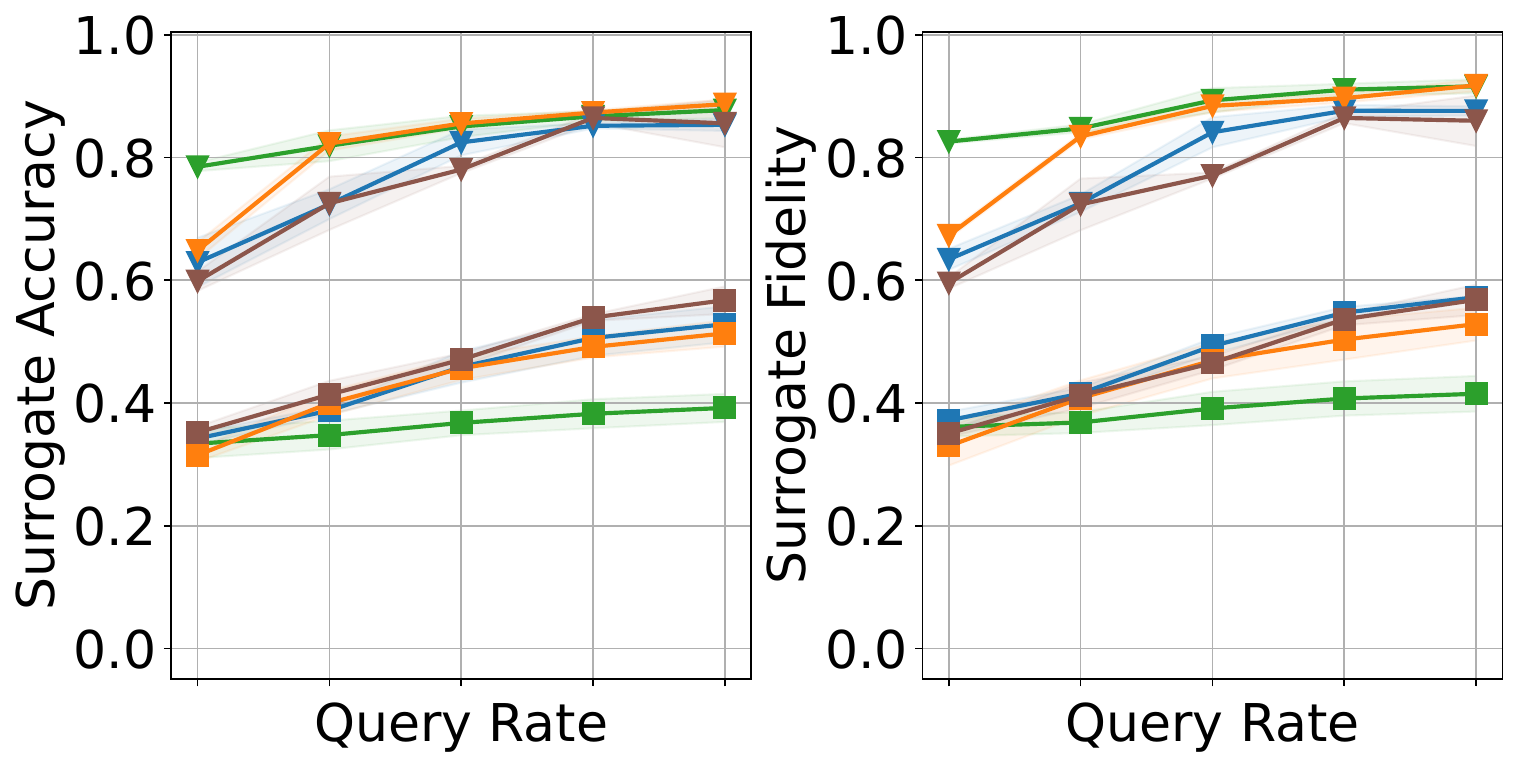}
         \caption{Probabilities.}
     \end{subfigure}
     \begin{subfigure}[t]{0.32\textwidth}
         \centering
         \includegraphics[width=\textwidth]{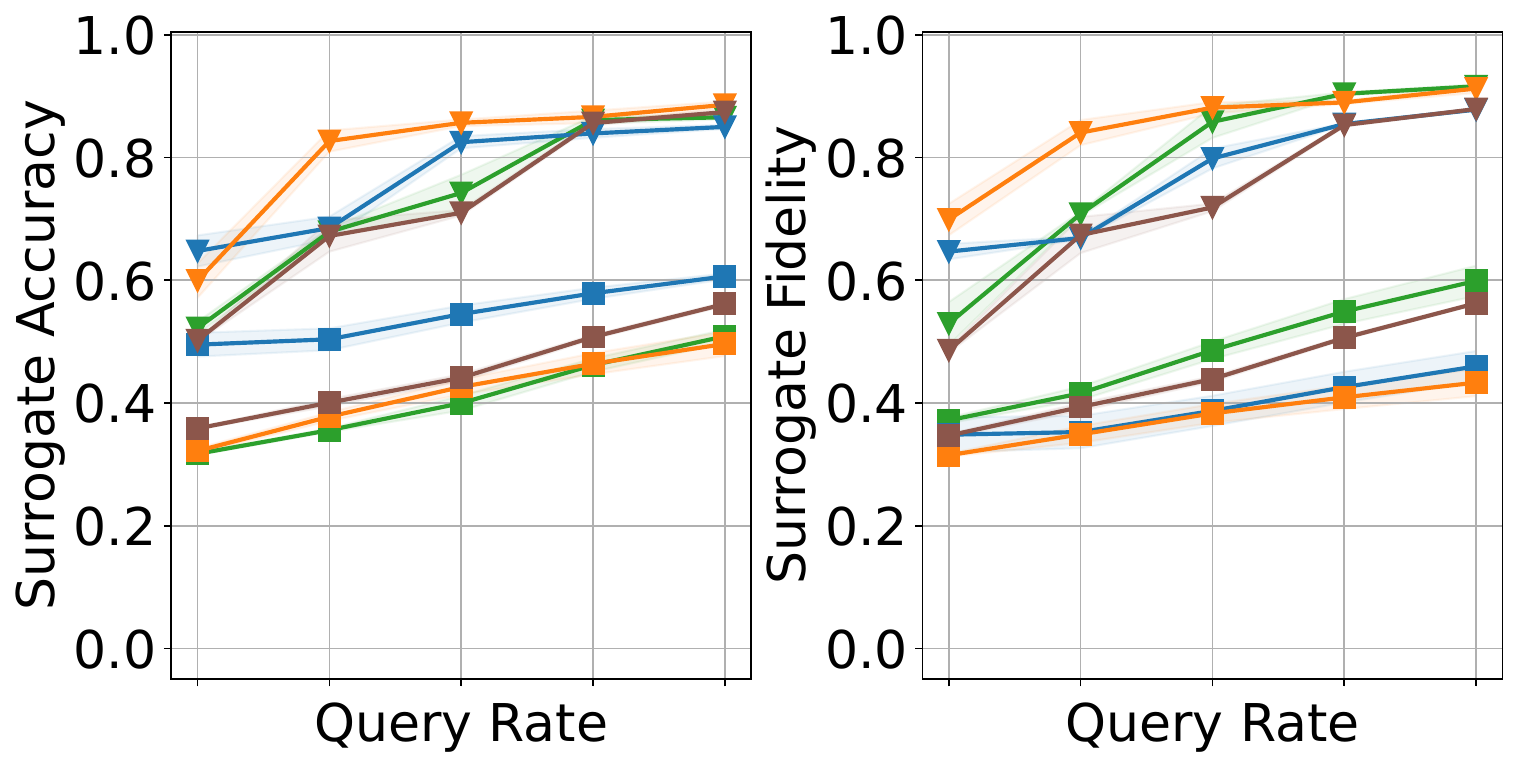}
         \caption{Embeddings.}
     \end{subfigure}
     \begin{subfigure}[t]{0.32\textwidth}
         \centering
         \includegraphics[width=\textwidth]{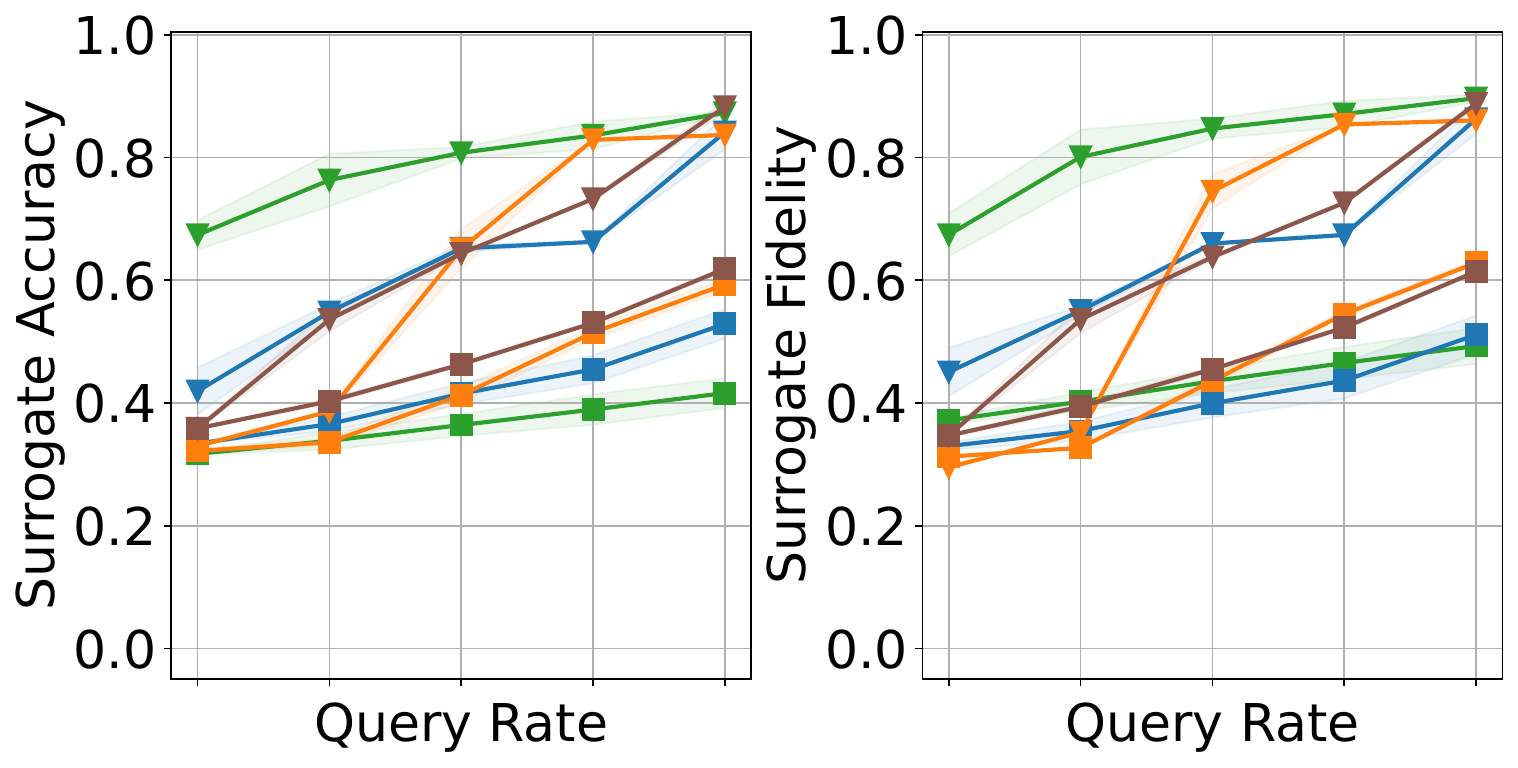}
         \caption{Projections.}
     \end{subfigure}

\caption{\textbf{Performance of the surrogate model based on different stealing strategies (ACM dataset).}
Over all stealing attack setups, we observe that stealing with more diverse nodes \textbf{(strategy~1)}, yields higher accuracy and fidelity (\ie surrogate model's similarity to the target model) than querying with less diverse nodes \textbf{(strategy 2)}. 
Overall, stealing with random queries from diverse communities improves the performance of the surrogate model.
}
\label{fig:attack_performance}
\end{figure*}

{\subsection{Threat Model} %
We consider the three GNN stealing attack setups from~\citet{shen2022model} where the target model either outputs A) predicted posterior probabilities, B) high-dimensional node embeddings,  or C) low-dimensional t-SNE projections of node embeddings.  
Note that following the common GNN architectural standards, in all three cases, the target model consists of a backbone encoder $\hat{\Phi}_t$ which outputs a high-dimensional representation of the query node. 
Additionally, depending on the setup, a classification head for node classification or a projection layer to project the high-dimensional outputs to a low-dimensional space is added to the encoder. 
We assume that each user queries the model through a user account, which enables the model owner to track the accumulated query diversity. 

\noindent \textbf{Adversary's Goal:} 
Our adversary can pursue two different goals, namely stealing the target model's \textit{functionality} or matching its \textit{behavior} as closely as possible~\citep{jagielski2020highfidelity}. 
The success of functionality stealing is quantified by \textit{surrogate accuracy} on the given task and the attacker's goal is to achieve a high task accuracy with their surrogate model.
Matching the stolen model's behavior, in contrast, means that the surrogate model should yield the same predictions as the target model, including the target model's mistakes.
This can be beneficial since a surrogate model with a similar behavior as the target model can be used to launch further attacks~\citep{papernot2017practical}.
The success metric for the adversary here is \textit{surrogate fidelity}, indicating the fraction of queries to which the surrogate model yields the same outputs as the target model.
We assess \ours' success against both stealing goals.

\noindent \textbf{Adversary's Capability:} We assume the adversary has access to a query graph $\mathbf{G}_Q$ and can make queries to the target model. 
We use $\delta$ to denote the percentage of nodes from the query graph that are actually queried by the attacker and denote the resulting selected query graph as $\mathbf{G}_q$. 
For each node of $\mathbf{G}_q$, the attacker observes the corresponding outputs and uses the node and the output to train a local surrogate model. 
The query response, depending on the target models's specification, can be in the form of a predicted posterior probability matrix $\bm{\Theta}$, a node embedding matrix $\mathbf{E}$, or a t-SNE projection matrix of the node embedding matrix $\bm{\Upsilon}$.

\noindent \textbf{Adversary's Knowledge:} Following~\citet{shen2022model}, we assume that the attacker has no knowledge on the target GNN model's parameters and cannot influence its training. 
To model the strongest possible attack (and show that \ours is still effective), we assume that the attacker has knowledge of the target model's architecture and can initialize the surrogate model with the same architecture. 
Additionally, in line with \citet{shen2022model}, we also evaluate the effectiveness of our defense in scenarios where the surrogate model has a different architecture from the target model (\Cref{sec:performance_evaluation}). 
Finally, we assume that the attacker holds a query graph $\mathbf{G}_Q$ from the same distribution as, but non-overlapping with the training graph $\mathbf{G}_{train}$.
This assumption aligns with recent attacks on neural networks ~\citep{he2021stealing, jagielski2020highfidelity, shen2022model}.
A prominent realistic example where such public graphs are available are social networks. 

\noindent \textbf{Defender's Goal \& Capability:} We assume that the defender is the owner of the target GNN model.
Their goal is to prevent the adversary from extracting the functionality and behavior of the target model, \ie they want to lower both surrogate accuracy and surrogate fidelity.
As the owner of the target model, the defender has full access to the target model and the underlying training graph. In addition, they can modify the query responses before returning them to the users to implement the defense.%

\subsection{Intuition of our \ours Defense}
We base our defense on the intuition that the responses of a GNN leak more information, the more diverse the corresponding queries are.
Hence, an attacker who is interested in stealing the \textit{full functionality} of a GNN has to query it with diverse data to obtain the surrogate model with the highest performance (\textit{surrogate accuracy}) and highest similarity to the target model (\textit{surrogate fidelity}).
To illustrate this intuition, we run experiments for stealing a GNN with two different strategies:
In \textbf{stealing strategy 1}, the attacker queries nodes with high diversity, whereas in \textbf{stealing strategy 2}, they query nodes with low diversity (both visualized in \Cref{fig:query_node_selecting}). We detail in \Cref{sub:query_diversity} how query node diversity can be quantified. 
As expected, our results in \Cref{fig:attack_performance} highlight that strategy~1, \ie stealing through diverse queries, is significantly more successful than strategy~2.
This suggests that a defense to prevent GNN model stealing needs to penalize diverse queries that would otherwise benefit an attacker.
At the same time, it should not harm the GNN's predictive performance on particular tasks, such as providing a group of similar users in a social network with targeted advertisements~\citep{facebook_ad_relevance}, or detecting fraud based on localized behavior in transaction graphs~\citep{dou2020enhancing}.%

\subsection{Design  of our \ours}
To implement the above intuition, our \ours consists of two building blocks: (1) the quantification of query diversity to calibrate the penalty strength (see \Cref{sub:query_diversity}), and (2) the design of the penalty itself, depending on the model output type (see \Cref{sub:penalty_design}). %

\subsubsection{Estimation of Information Leakage through Query Diversity}%
\label{sub:query_diversity}
In GNNs, the model owner has access to the underlying training graph $G_{train}$, which serves as a foundation for quantifying query diversity. Specifically, the model owner leverages the graph's \textit{communities} as a key signal. At a high level, the defense operates as follows:
1) Identify the communities within the graph.
2) For each incoming query $q_j$, determine the closest community it belongs to.
3) Track the communities covered by a user's queries over time.
4) Gradually adjust the defense mechanism to impose stronger penalties as the number of queried communities increases.
We provide a detailed explanation of these steps in the following sections.

\noindent \textbf{Communities.}
Formally, a community inside a graph refers to a subset of nodes whose connections among each other are more dense than their connections to other nodes.
Nodes within the same community, \eg users in a social network who grew up in the same state and graduated from the same high school, usually share the same properties and are more similar to each other than to other nodes, \eg users with different backgrounds.
Communities naturally occur in all real-world graphs, such as social networks, citation networks, and biological networks~\citep{steinhaeuser2008community,chintalapudi2015survey,ciglan2013community}. 
We acknowledge that some graphs (\eg random or weakly modular graphs) may not yield strong community structures. However, GNNs are typically applied to structured real-world domains, \eg recommendation systems, and fraud detection of financial transaction networks, where the structural information of the graphs is present and meaningful.  
We further calculated the modularity ($\mathcal{M}$) of each graph dataset used in our experiments, which ranges from 0.1531 to 0.8895 (as shown in \Cref{table:datasets}), demonstrating that ADAGE remains robust across diverse community structures. 
Therefore, communities provide a reliable and inherently available signal for diversity in graphs.

\noindent \textbf{Community Detection.} 
To detect communities in the underlying graph $\mathbf{G}_{train}$, we rely on a community detection algorithms that yields $K$ communities given $\mathbf{G}_{train}$. 
Concretely, we use the Louvain Community Detection Algorithm as it is one of the most stable community detection algorithms in the top rankings, and it outperforms other known community detection methods in terms of computation time~\citep{rustamaji2024community, blondel2008fast}. 
We also perform an ablation study using a different community detection algorithm in \Cref{sec:performance_evaluation}. Our results indicate that while also other community detection algorithms are effective for our defense, the Louvain Community Detection Algorithm outperforms them in terms within the defense while requiring lower computational complexity. 
After detecting communities in the underlying graph, we calculate the centroid of each community to obtain the set of community centroids $\Omega=\left \{ \omega_1, \dots, \omega_K \right \}$.

\noindent \textbf{Tracking Query Diversity.} 
Tracking and quantifying a user’s query diversity involves recording the number of distinct communities their queries fall into. For each new query to the GNN, this requires identifying the closest community, logging it, and calculating query diversity as the \textit{fraction of total communities covered} up to that point  (which we denote by $\tau$). 
The closest community is identified by calculating the Euclidean distance between the query embedding $\mathbf{e}$ and each community centroid $w_i$, and selecting the community with the smallest distance. 
Then, we insert the closest community into the set of occupied community indices $\mathcal{I}$ up to the previous query. 
Finally, the fraction $\tau$ of currently occupied communities is calculated by the number of currently occupied community divided by the number of communities $K$. 
We detail the calculation of the fraction $\tau$ in \Cref{alg:adage} (line 2-12). This $\tau$ serves us to estimate the incurred information leakage from the GNN to the user.
It is important to note that \ours does not classify individual queries as benign or malicious. Instead, it calculates the accumulated diversity of the given queries and applies corresponding penalties to the model outputs.  
Our approach is stateful, meaning that costs are incurred per user rather than per query, as the penalties are applied based on the user’s overall query diversity. 

\noindent \textbf{Calibrating the Penalty.} 
We map increasing fractions of covered communities $\tau$ to higher penalties. 
Therefore, we pass $\tau$ as an argument to the perturbation functions that are applied to the model outputs within the defense as specified in the next section.

\begin{algorithm}[htpb]
\small
\caption{ADAGE}
\label{alg:adage}
\textbf{Input}: Current query $q_j$, backbone encoder $\hat{\Phi}_t$, classification head $\mathcal{C}_t$, t-SNE transformation head $\mathcal{H}$, set of community centroids $\Omega=\left \{ \omega_1, \dots, \omega_K \right \}$, number of communities $K$ \\
\textbf{Output}: Perturbed model output of query $q_j$ \\
\textbf{Current state:} set of occupied community indices $\mathcal{I}$ up to the previous query $q_{j-1}$ (computed on queries $\{q_1,q_2,\dots,q_{j-1}\}$)
\begin{algorithmic}[1] %

\State
\State \textbf{// Calculate Occupied Communities (\Cref{sub:query_diversity})}
\State $\mathbf{e} = \hat{\Phi}_t(q_j)$ \Comment{embedding of query $q_j$}
\State min\_dist$\leftarrow\infty$, min\_index$\leftarrow-1$
\For{$i \gets 1$ to $K$}                
    \State {$d=$ EuclideanDistance$(\mathbf{e}, \omega_i)$}
    \If{$d<$min\_dist}
    \State min\_index$=i$; min\_dis$=d$
    \EndIf
\EndFor
\State $\mathcal{I} = \mathcal{I} \cup i$ 
\State $\tau = \frac{|\mathcal{I}|}{K}$ \Comment{fraction of occupied communities}
\State
\State \textbf{// Add Penalty to Model Output (\Cref{sub:penalty_design})}
\If {Stealing setup A)}
\State $\tilde{\mathbf{p}}$ = \texttt{LABEL\_FLIPPING}($\mathbf{e}$, $\mathcal{C}_t$, $\tau$)
\ElsIf{Stealing setup B)}
\State $\tilde{\mathbf{e}} = \mathbf{e} + \mathcal{N}(0, \sigma_\tau \mathbf{I})$ %
\ElsIf{Stealing setup C)}
\State $\tilde{\mathbf{\gamma}} = \mathcal{H}(\mathbf{\tilde{e}})$ %
\EndIf

\State
\State // \texttt{LABEL\_FLIPPING}
\State $\mathbf{p}=\mathcal{C}_t(\mathbf{e})$ \Comment{prediction probabilities}
\State $\rho = h_\eta(\tau)$ \Comment{label flipping probability}
\State $i = \argmax(\mathbf{p})$
\State $j \leftarrow \text{random index}\in \{1, \cdots, |\mathbf{p}| \} $
\State $\tilde{\mathbf{p}}: p_i \leftrightarrow p_j$ \Comment{swap with probability $\rho$} 

\State \textbf{return} $\tilde{\mathbf{p}}$ or $\tilde{\mathbf{e}}$ or $\tilde{\mathbf{\gamma}}$
\end{algorithmic}
\end{algorithm}

\subsubsection{Penalty Design}
\label{sub:penalty_design}
Depending on the type of model output, we need to design different forms of penalties, all calibrated according to the fraction of currently occupied communities $\tau$, as shown in \Cref{alg:adage} (line 14-21). 

\noindent \textbf{A) Predicted Posterior Probabilities.} A naive application of the defense could simply add Gaussian noise to the output probabilities to perturb their values.
However, prior work has shown that supervised models can be effectively stolen using just the top-1 predicted label instead of prediction probabilities~\citep{orekondy2019prediction}.
Since under decent amounts of noise, the noisy top-1 predicted label would remain the same as the original one (only the distance between the highest probability and other labels' probabilities would be reduced), we found this approach to be ineffective in preventing model stealing. 
Instead, we incur label flips directly with a probability $\rho$ calibrated through the function:
\begin{equation}
    \rho=h_{\eta}(\tau) = \frac{1}{1+\exp^{\eta \times (1-2\times\tau)}}\text{,}
\label{equ:cost_label_stealing}
\end{equation}
where $\eta$ compresses the curve to obtain low penalties for a small fraction of occupied communities $\tau$ and very high penalties for large fractions $\tau$. %
Given the prediction probabilities $\mathbf{p}$ from the target model, we perturb the model output by swapping the probability of the predicted class $i$ with that of a randomly selected class $j$, \ie $p_i \leftrightarrow p_j$, with the probability $\rho$ as defined in \Cref{equ:cost_label_stealing}. 
We provide a detailed motivation for the design of \Cref{equ:cost_label_stealing} in \Cref{appendix:cost_label_flip}. 
The full label flip mechanism is outlined in \Cref{alg:adage} (line 23-28), which returns the perturbated predictions to the user.

\noindent \textbf{B) Node Embeddings.}
For models that output high-dimensional output representations, we can indeed add Gaussian noise as a penalty. 
The standard deviation of the added noise is calibrated according to the fraction of occupied communities $\tau$.  
In order to maintain utility on downstream tasks, we follow the idea of~\citet{dubinski2023bucks} and instantiate an exponential function to derive the standard deviation as
\begin{equation}
    \sigma_\tau = f_{\lambda, \alpha, \beta}(\tau) = \lambda \times (\exp^{ln\frac{\alpha}{\lambda} \times \tau \times \beta^{-1}}-1),
    \label{equ:calibration}
\end{equation}
where $\tau$ is the fraction of communities queried.  
The $\lambda < 1$ compresses the curve of $f$ to obtain low $\sigma_\tau$ for a small number of queried communities.
The $\alpha$ specifies the desired penalty strength (ideally configured such that embeddings returned with this penalty are so noisy that they cannot be used for stealing) and $\beta$ specifies at what fraction of communities queried, we want to reach this level of penalty.
For instance, if we want to enforce a $\sigma$ of $1$ at $90\%$ of occupied communities (\ie for $\tau=0.9$), we would need to set $\alpha=1$ and $\beta=0.9$. 
Finally, after perturbation, instead of the original embedding $\mathbf{e}$, we return
\begin{equation}
    \tilde{\mathbf{e}} = \mathbf{e} + \mathcal{N}(0, \sigma_\tau \mathbf{I})\text{,}
    \label{equ:pertur_embedding}
\end{equation}
where Gaussian noise is applied independently to each component of $\mathbf{e}$.

\noindent \textbf{C) t-SNE Pmbeddings.}
For attack setup based on low-dimensional t-SNE projections, we perturb the internal embeddings B) with \Cref{equ:pertur_embedding} and then project those to the lower dimensional space
\begin{equation}
    \tilde{\mathbf{\gamma}} = \mathcal{H}(\mathbf{\tilde{e}}).
\end{equation}
where $\mathcal{H}$
denotes the transformation from high-dimensional embeddings to low-dimensional t-SNE projections.

\section{Empirical Evaluation}
\label{sec:experical_evaluation}
In this section, we conduct a comprehensive analysis of the proposed active defense against state-of-the-art GNN model stealing attacks. We begin by introducing the experimental setup and presenting the evaluation results of \ours from both the attacker's perspective and that of downstream task utility. Next, we explore the impact of the surrogate architecture and the community detection algorithm on the defense performance. Finally, we compare \ours with the current state-of-the-art baseline defense.

\subsection{Experimental Setup}
\label{sec:experimental_setup}
\noindent \textbf{Datasets.} %
To evaluate our defense, we use six public standard benchmarks for GNNs~\citep{hamilton2017inductive, kipf2016semi, xu2018powerful}, including ACM~\citep{wang2019heterogeneous}, DBLP~\citep{pan2016tri}, Pubmed~\citep{sen2008collective}, Citeseer Full (abbreviated as Citeseer)~\citep{giles1998citeseer}, Amazon Co-purchase Network for Photos (abbreviated as Amazon)~\citep{mcauley2015image}, and Coauthor Physics (abbreviated as Coauthor)~\citep{shchur2018pitfalls}.
Specifically, ACM and Amazon are networks where nodes represent the papers/items, with edges indicating connections between two nodes if they have the same author or are purchased together. DBLP, Pubmed, and Citeseer are citation networks where nodes represent publications and edges denote citations among these publications. Coauthor is a user interaction network, with nodes representing the users and edges indicating interactions between them. Statistics of these datasets are summarized in \Cref{table:datasets}. 
For each dataset, we randomly sample $20\%$ of nodes as the training data $\mathbf{G}_{train}$ for $\Phi_t$ and $30\%$ nodes as the query graph $G_Q$. From $G_Q$, we select a fraction $\delta$ of nodes for our attack. The remaining nodes of the graph are used as test data $\mathbf{G}_{test}$ to evaluate the target model $\Phi_t$, surrogate model $\Phi_s$, and also the performance of the surrogate model after applying our defense. This setting matches the inductive learning on evolving graphs as laid out in~\citet{hamilton2017inductive, shen2022model}. 

\begin{table}[!h]
\small
 \centering
 \caption{\textbf{Statistics of datasets.} $|\mathcal{V}|, |\mathcal{E}|, m, |\mathbb{C}|$, $\mathcal{M}$ denote the number of nodes, number of edges, dimension of a node feature vector, number of classes, and graph modularity, respectively. %
 }
\begin{tabular}{ccccccc} 
 \hline
 Dataset & $|\mathcal{V}|$ & $|\mathcal{E}|$ & $m$ & $|\mathbb{C}|$ & $\mathcal{M}$\\
 \hline
 ACM & $3,025$ & $26,256$ & $1,870$ & $3$ & $0.1531$\\
 DBLP & $17,716$ & $105,734$ & $1,639$ & $4$ & $0.4366$ \\
 Pubmed & $19,717$ & $88,648$ & $500$ & $3$ & $0.7705$ \\
 Citeseer & $4,230$ & $5,358$ & $602$ & $6$ & $0.8895$\\
 Amazon & $7,650$ & $143,663$ & $745$ & $8$ & $0.7381$\\
 Coauthor & $34,493$ & $495,924$ & $8,415$ & $5$ & $0.6681$ \\
 \hline
\end{tabular}
\label{table:datasets}
\end{table}

\noindent \textbf{Models and Hyperparameters.} We use four widely-used GNNs architectures, \ie GIN~\citep{xu2018powerful}, GAT~\citep{velivckovic2017graph}, GraphSAGE~\citep{hamilton2017inductive} and Graph Transformer~\citep{ijcai2021p214} for the target and surrogate model in our evaluation. %
For the attack setup B, where the target model outputs high-dimensional node embeddings, the surrogate model trains a backbone encoder $\hat{\Phi}_s$ and a classification head $\mathcal{C}_s$ with label information of the query graph. Finally, $\hat{\Phi}_s$ and $\mathcal{C}_s$ are combined to calculate the accuracy and fidelity of the test data. 
Hyperparameters used in training target and surrogate models are shown in \Cref{Table:parameter_setting_target} and \Cref{Table:parameter_setting_surrogate} (\Cref{app:hyperparameter_models}), respectively. 
We set $\alpha=1$, $\lambda=10^{-6}$, $\eta=10$ and specifically per dataset the number of communities $K$ and the percentage of occupied communities $\beta$ (as shown in \Cref{table:hyperparameter}, \Cref{app:hyperparamters}). %
More details on hyperparameter selection are provided in \Cref{app:hyperparamters}. The results are averaged over five independent trials. 

\noindent \textbf{Evaluation Metrics.} We evaluate \textit{accuracy} and \textit{fidelity} of the surrogate model, following the two adversaries defined by \citet{jagielski2020highfidelity}. 
Formally, \textit{surrogate accuracy} is defined as the number of correct predictions made divided by the total number of predictions made, while \textit{surrogate fidelity} is defined as the number of predictions agreed by both the surrogate and target models divided by the total number of predictions made. Both metrics are normalized between 0 and 1, with higher scores implying better performance. 

\begin{figure*}[!h]
\centering
    \begin{subfigure}[b]{0.5\textwidth}
         \centering
         \includegraphics[width=\textwidth]{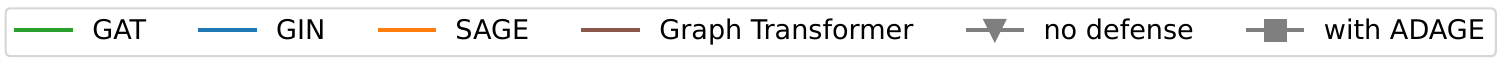}
     \end{subfigure}
     \vfill
     \begin{subfigure}[t]{0.32\textwidth}
         \centering
         \includegraphics[width=\textwidth]{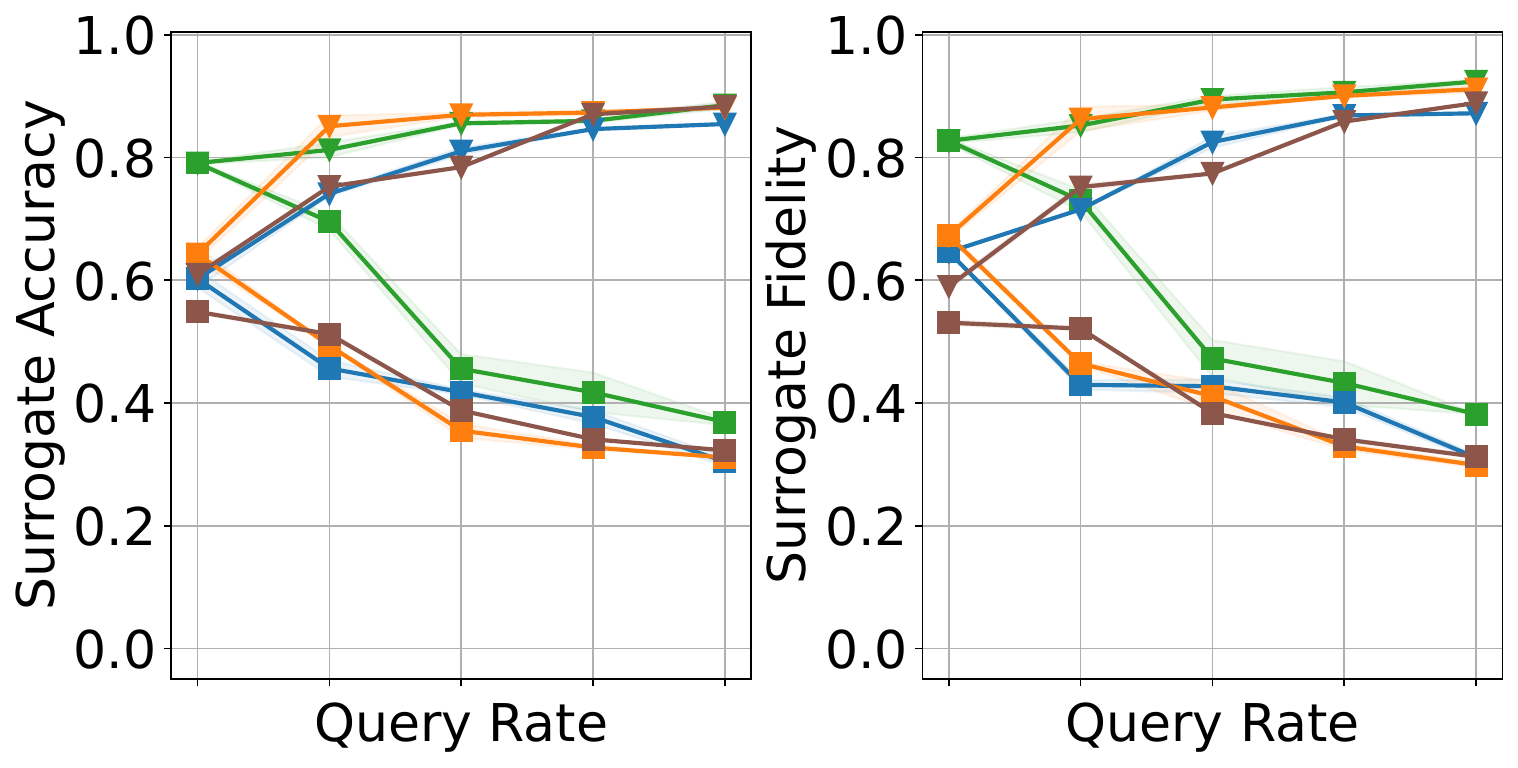}
         \caption{Probabilities.}
     \end{subfigure}
     \begin{subfigure}[t]{0.32\textwidth}
         \centering
         \includegraphics[width=\textwidth]{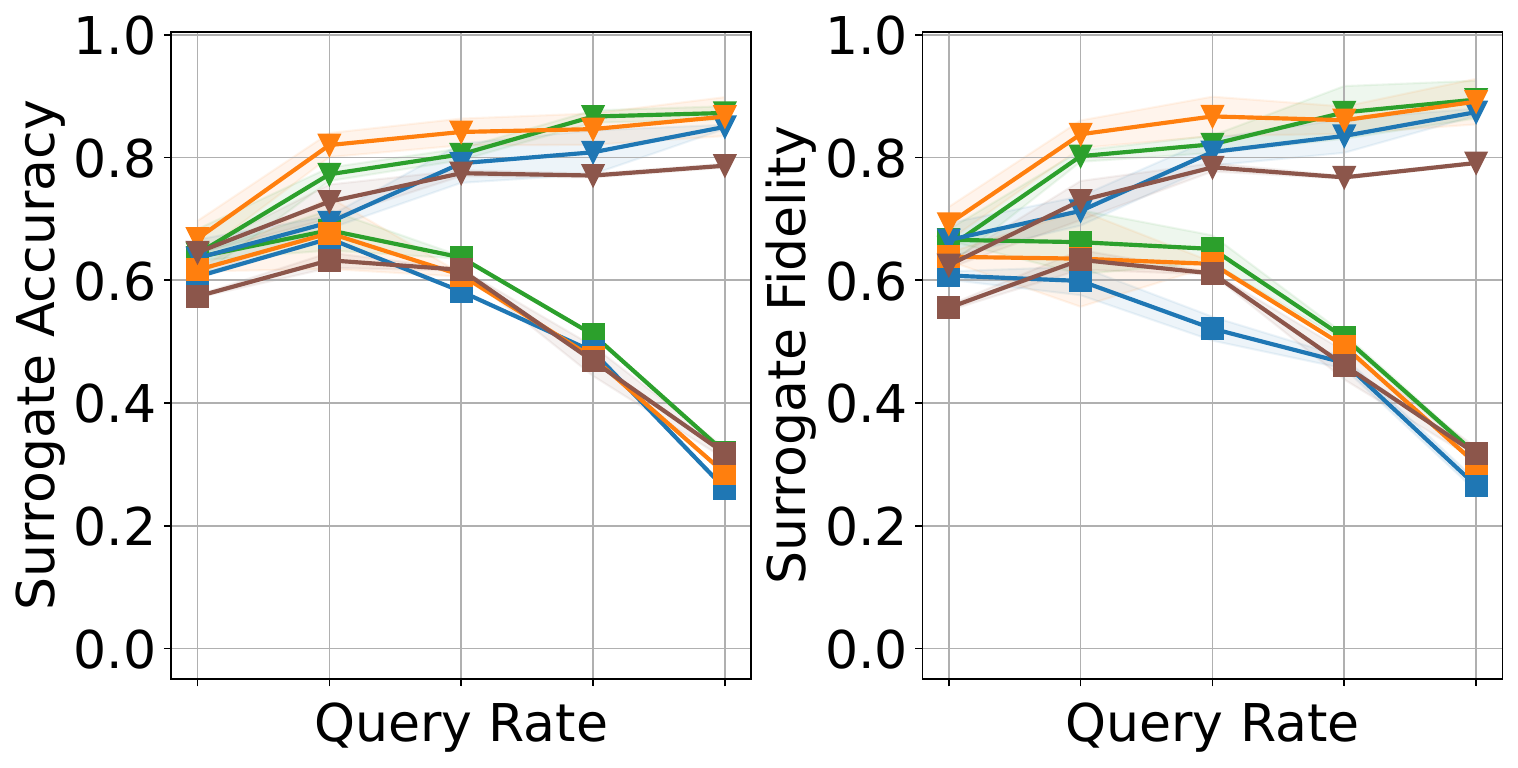}
         \caption{Embeddings.}
     \end{subfigure}
     \begin{subfigure}[t]{0.32\textwidth}
         \centering
         \includegraphics[width=\textwidth]{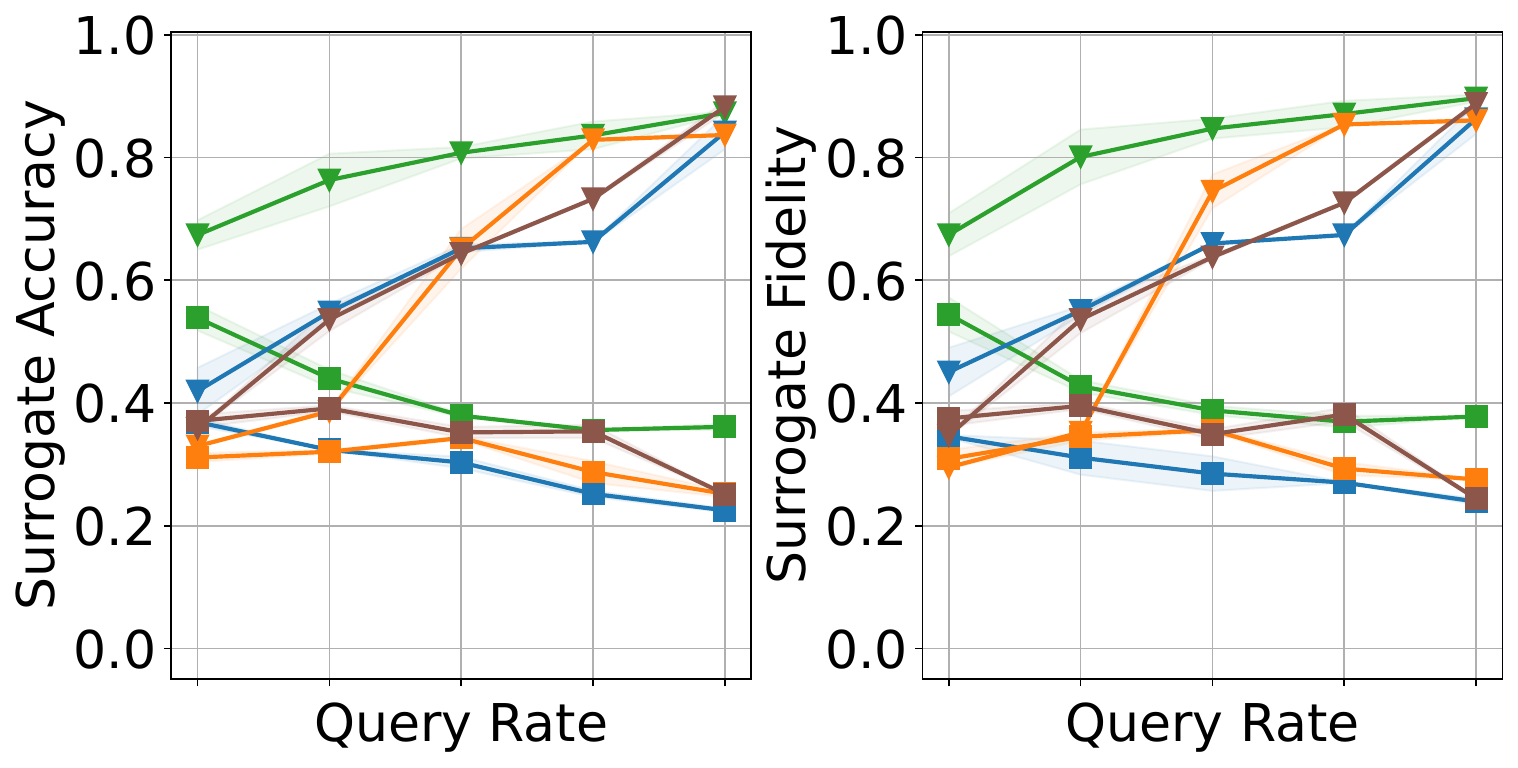}
         \caption{Projections.}
     \end{subfigure}
\caption{\textbf{Performance of the surrogate model with and without our defense (ACM dataset).} Overall, our defense degrades the stealing performance of the surrogate model, especially when the query rate is high.}
\label{fig:adversary_performance_acm}
\end{figure*}

\subsection{Performance Evaluation}
\label{sec:performance_evaluation}

\noindent \textbf{Attackers.} \Cref{fig:adversary_performance_acm} illustrates the stealing performance under the three attack setups from \citep{shen2022model}, with and without applying \ours. We take the results of ACM as an example and provide the results for other datasets in \Cref{app:adversary_performance}. 
The figure reveals that increasing the query rate results in a corresponding increase in surrogate accuracy and fidelity for all three attack setups before applying \ours. 
At the highest query rate (\ie $\delta=0.25$) which still refers to a small number of query nodes, \eg $226$ nodes for ACM dataset, %
the surrogate model can achieve significant performance, more than $83\%$ accuracy in all cases. 
Upon applying \ours, the stealing performance experiences a significant reduction, particularly with increasing query rates. 
To further illustrate the degradation in stealing performance for the attacker, we present the testing accuracy of the attacker both before and after applying \ours in \Cref{table:results_with_without_defenses_acm}. 
These results are obtained using the largest query rate (\ie $\delta=0.25$), as it represents the most challenging scenario for our defense. 
Notably, when \ours is employed, the testing accuracy for the attacker decreases significantly. 
For instance, in attack setup A, without \ours, the surrogate models (GAT, GIN, GraphSAGE, Graph Transformer) achieve accuracies close to those of the target models (which are $88.53\%$, $85.46\%$, $88.14\%$, $88.30\%$, respectively). %
However, with \ours, these accuracies decrease to $36.90\%$, $30.45\%$, and $31.12\%$, $32.27\%$ respectively, representing at least a $50\%$ accuracy drop in all cases. 
Therefore, our results show \ours can dramatically degrade the performance of surrogate models.

\begin{table*}[!h]
\small
 \centering
 \caption{\textbf{Performance for attacker and target downstream tasks with and without \ours in three attack setups (ACM, $\delta=0.25$, $c_i$ represents a community, GT - Graph Transformer).} Overall, with our defense, the performance for downstream tasks remains high while the performance of the surrogate model is significantly degraded. }
 \label{table:results_with_without_defenses_acm}
\begin{tabular}{cccccccc} 
 \hline
  & User & Dataset & Defense & GAT & GIN & GraphSAGE & GT \\
 \hline
Baseline & N/A & $\mathbf{G}_{test}$ & N/A & $90.04\pm 0.67$ & $88.30\pm0.47$ & $90.75\pm0.92$ & $96.72\pm0.30$ \\
 \hline
 \hline
 \multirow{5}{22mm}{\begin{center}Attack setup A (Probabilities)\end{center}} 
  & Attacker & $\mathbf{G}_{test}$ & NONE & {$88.53\pm0.62$} & {$85.46\pm0.16$} & {$88.14\pm0.12$} & {$88.30\pm0.49$} \\
  & Attacker & $\mathbf{G}_{test}$ & \ours & {$\textbf{36.90}\pm0.51$} & {$\textbf{30.45}\pm0.73$} & {$\textbf{31.12}\pm0.42$} & {$\textbf{32.27}\pm0.12$} \\
  \cline{2-8}
  & Downstream Task 1 & $c_1$ & \ours & {$88.69\pm0.67$} & {$84.24\pm2.19$} & {$89.74\pm1.47$} & {$94.42\pm1.24$} \\
  & Downstream Task 2 & $c_2$ & \ours & {$87.73\pm2.03$} & {$88.10\pm0.72$} & {$90.16\pm1.53$} & {$95.62\pm0.34$} \\
  & Downstream Task 3 & $c_3$ & \ours & {$87.34\pm0.48$} & {$85.68\pm1.07$} & {$88.56\pm0.79$}  & {$95.60\pm0.67$} \\
 \hline
 \hline

 \multirow{5}{22mm}{\begin{center}Attack setup B (Embeddings)\end{center}} 
 & Attacker & $\mathbf{G}_{test}$ & NONE & {$87.26\pm1.09$} & {$85.00\pm0.34$} & {$86.67\pm3.16$} & {$78.67\pm0.32$} \\
 & Attacker & $\mathbf{G}_{test}$ & \ours & {$\textbf{31.96}\pm0.03$} & {$\textbf{26.07}\pm0.17$} & {$\textbf{28.55}\pm0.13$} & {$\textbf{31.78}\pm0.89$} \\
 \cline{2-8} 
 & Downstream Task 1 & $c_1$ & \ours & {$87.94\pm0.07$} & {$89.61\pm0.20$} & {$88.37\pm0.10$}  & {$95.34\pm0.93$} \\
 & Downstream Task 2 & $c_2$ & \ours & {$88.87\pm0.03$} & {$86.22\pm0.41$} & {$89.05\pm0.51$} & {$95.49\pm0.27$} \\
 & Downstream Task 3 & $c_3$ & \ours & {$87.32\pm0.51$} & {$87.19\pm0.33$} & {$89.18\pm0.10$} & {$95.69\pm0.62$} \\
 \hline
 \hline

 \multirow{5}{22mm}{\begin{center}Attack setup C (Projections)\end{center}} 
 & Attacker & $\mathbf{G}_{test}$ & NONE & {$87.28\pm0.19$} & {$84.14\pm2.81$} & {$83.67\pm0.11$} & {$88.27\pm0.94$} \\
 & Attacker & $\mathbf{G}_{test}$ & \ours & {$\textbf{36.12}\pm0.51$} & {$\textbf{22.51}\pm0.32$} & {$\textbf{25.19}\pm0.51$} & {$\textbf{25.16}\pm0.05$} \\
 \cline{2-8}
 & Downstream Task 1 & $c_1$ & \ours & {$86.59\pm0.39$} & {$86.83\pm1.59$} & {$89.10\pm1.78$} & {$95.35\pm0.93$} \\
 & Downstream Task 2 & $c_2$ & \ours & {$89.83\pm1.10$} & {$86.97\pm1.23$} & {$88.94\pm1.87$} & {$95.44\pm0.40$} \\
 & Downstream Task 3 & $c_3$ & \ours & {$89.19\pm1.73$} & {$84.90\pm0.36$} & {$88.55\pm2.18$} & {$95.11\pm0.94$} \\

 \hline
 \hline
\end{tabular}
\end{table*}

\noindent \textbf{Task Performance.} %
In addition to degrading the stealing performance of attacks, our defense ensures a minimal impact on the performance for \textit{targeted} downstream tasks. 
Therefore, we assess the downstream performance (measured in testing accuracy) before and after applying \ours on three randomly selected communities (denoted as $c_1$, $c_2$, and $c_3$) from the testing dataset. 
Evaluating at the community level is important, as many practical applications, such as providing a tailored advertisement to a target group in a social network, naturally operate on specific communities.  
The results are presented in \Cref{table:results_with_without_defenses_acm} for the three attack setups. 
Our results indicate that, in general, \ours has a negligible impact on the testing accuracies of communities compared to the target models. 
For instance, in attack setup A on the GAT model, the testing accuracies achieve approximately $88.69\%$, $87.73\%$, and $87.34\%$ on the three communities, respectively. This represents a less than $3\%$ accuracy drop compared to the target model's accuracy of $90.04\%$. Overall, our results illustrate that our defense maintains the downstream task performance close to that of the original target models.

\begin{figure*}[htpb]
\centering
    \begin{subfigure}[t]{0.35\textwidth}
         \centering
         \includegraphics[width=\textwidth]{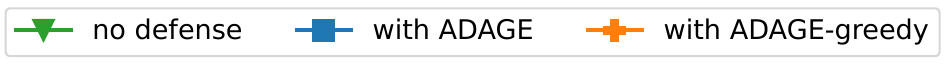}
     \end{subfigure}
     \vfill
     \begin{subfigure}[b]{0.32\textwidth}
         \centering
         \includegraphics[width=\textwidth]{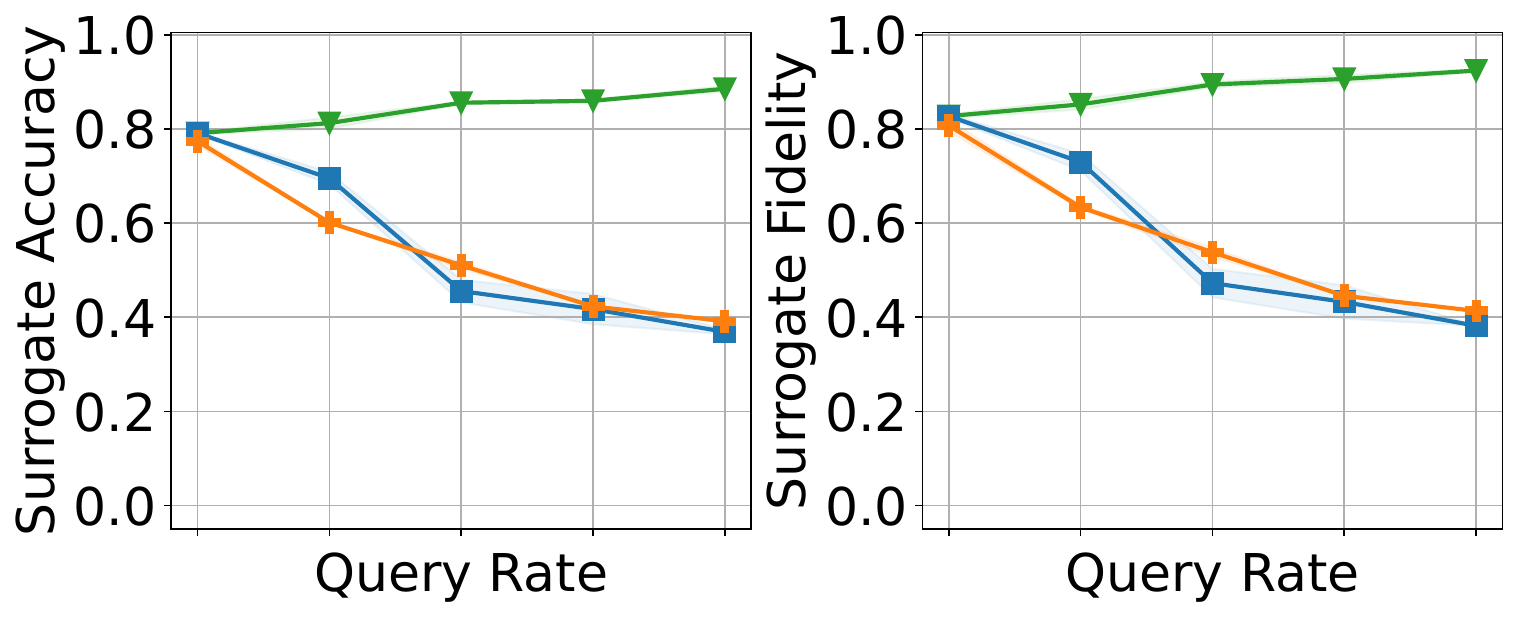}
         \caption{Probabilities.}
     \end{subfigure}
     \begin{subfigure}[b]{0.32\textwidth}
         \centering
         \includegraphics[width=\textwidth]{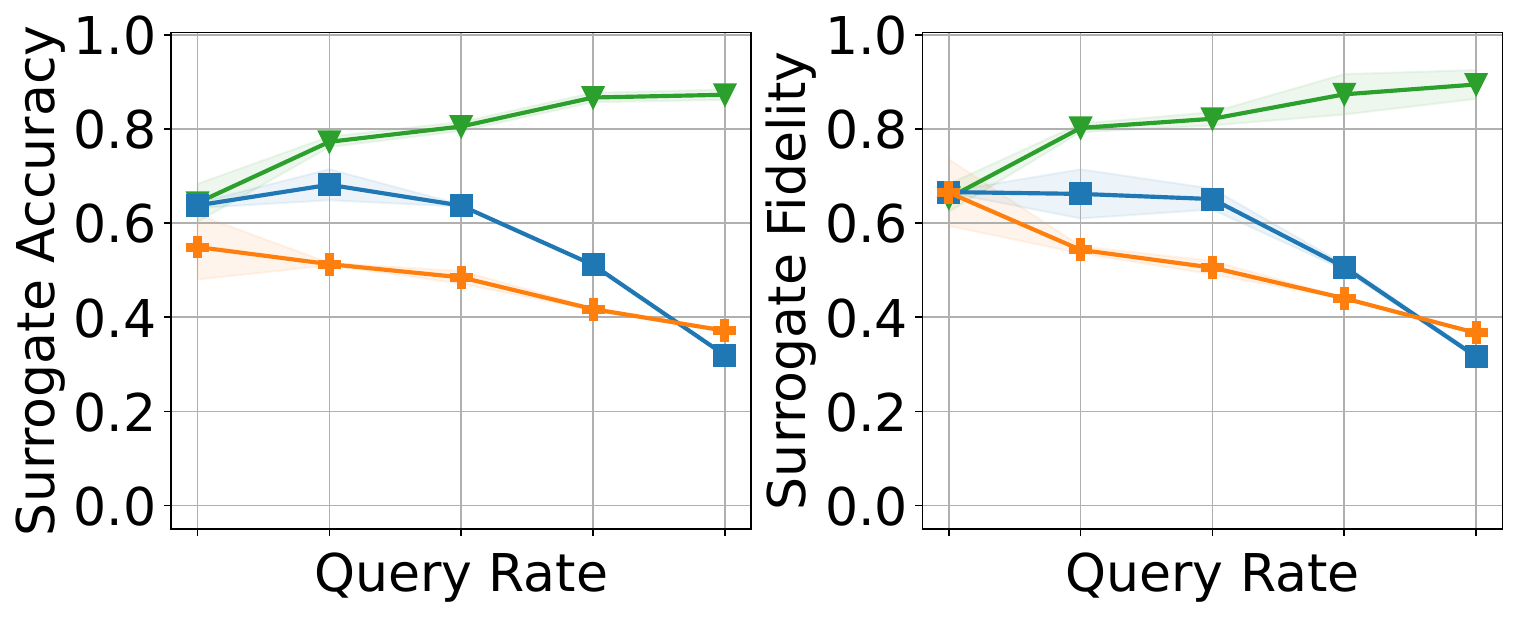}
         \caption{Embeddings.}
     \end{subfigure}
     \begin{subfigure}[b]{0.32\textwidth}
         \centering
         \includegraphics[width=\textwidth]{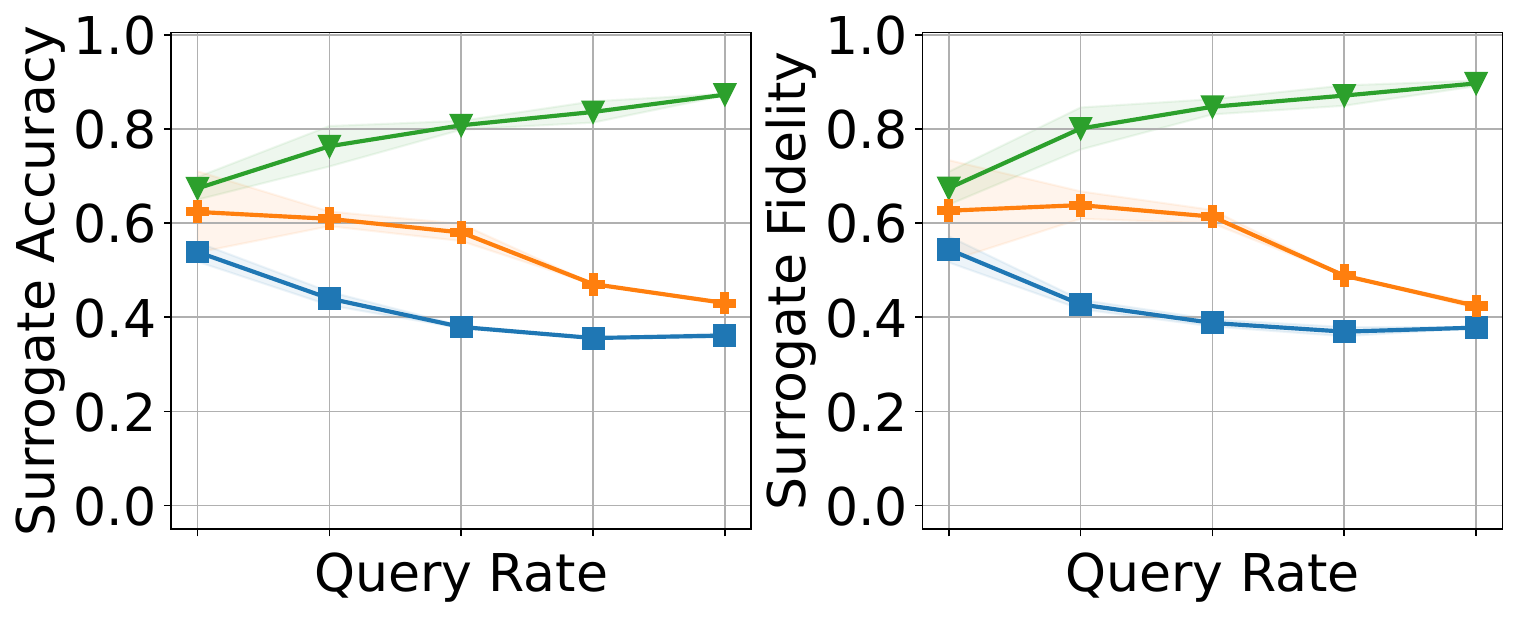}
         \caption{Projections.}
     \end{subfigure}
\caption{\textbf{Ablation on the Choice of the Community Detection Algorithm (ACM dataset, GAT model).} We assess \ours using the Clauset-Newman-Moore greedy modularity maximization method as the community detection method. Overall, the stealing performance with \ours-greedy is similar with \ours. %
}
\label{fig:adversary_performance_greedy_acm}
\end{figure*}

We note that in real-world GNN applications (\eg recommendation or fraud detection), benign users typically interact within a specific region (community) of the graph. While we acknowledge that in some rare cases, a legitimate user may query diverse communities, this represents an exception rather than the norm. To further quantify potential effects in such rare cases, we evaluated the downstream task performance when users cover multiple communities (\eg 1-5 communities), as presented in \Cref{app:multiple_communities}. The results show that the performance for target downstream tasks remains high for legitimate users querying several communities, while attackers, who typically span many communities, are penalized much more strongly due to the exponential scaling of our penalty function design (see \Cref{equ:cost_label_stealing} and \Cref{equ:calibration}).

\noindent \textbf{Ablation Study on Surrogate Architecture.} %
It has been shown that the adversary does not require knowledge about the architecture of target models to conduct model stealing attacks on GNNs~\citep{shen2022model}. 
Therefore, although our threat model assumes that the attacker has knowledge of the target model's architecture, we also evaluate \ours in scenarios where the surrogate model’s architecture differs from that of the target model. 
Given our experimental setup (\ie four GNN architectures), there are 16 different combinations for each stealing setup. 
The stealing performance of these 16 combinations under three attack setups is illustrated in \Cref{fig:adversary_diff_model_acm}, using the ACM dataset as an example. 
The results demonstrate that \ours remains effective in these scenarios. 

\noindent \textbf{Ablation Study on Community Detection Algorithm.} In \ours, we apply the Louvain Community Detection Algorithm to detect communities in the underlying graph. Here, we experiment with a different community detection algorithm, namely the Clauset-Newman-Moore greedy modularity maximization method~\citep{clauset2004finding} which has complexity $O(n \cdot log^2(n))$ ($n$ is the number of nodes in the graph). 
The comparison of stealing performance between applying \ours and \ours-greedy\footnote{Here, \ours-greedy defines \ours which utilizes the Clauset-Newman-Moore greedy modularity maximization method to detect the communities in the underlying graph.} is illustrated in \Cref{fig:adversary_performance_greedy_acm},%
taking the ACM dataset and GAT models as an example. The exact results are shown in \Cref{table:results_with_without_defenses_greedy_acm}. %
As we can observe, the degradation of the stealing performance applying \ours-greedy is comparable to that of \ours. 
This indicates that the Clauset-Newman-Moore greedy modularity maximization method remains effective in our defense, demonstrating the flexibility of \ours. 
However, the Clauset-Newman-Moore greedy modularity maximization method has complexity $O(n \cdot log^2(n))$ which is higher than the Louvain Community Detection Algorithm, \ie $O(n \cdot log(n))$. Consequently, we adopt the Louvain Community Detection Algorithm as the community detection algorithm in \ours to achieve high defense performance and also computational efficiency.

\noindent \textbf{Computational Complexity.} %
In the first building block, \ie the quantification of query diversity (\Cref{sub:query_diversity}), for each incoming user query, the complexity is $O(K)$ where $K$ is the number of communities since we need to compare the embedding of the incoming query to all community centroids. 
The calculation of community centroids is performed once before deployment and is $O(n \cdot log (n))$ since the Louvain Community Detection algorithm identifies communities in $O(n \cdot log (n))$ where $n$ is the number of nodes in the graph. 
Also, it is noted that we do not need to save the user's query history. Instead, it maintains a simple dictionary that saves, for each user identifier, a single floating-point value representing community coverage, which is updated whenever a new query comes in. This design results in low memory usage.

We also report the wall-clock (elapsed) times with and without applying our defense. We take the ACM with our B attack scenario (\ie embedding-based) and 0.25 query rate as the example. We show in \Cref{tab:wall-clock_time} that our defense has an insignificant computational cost. 
Furthermore, we evaluate the complexity of comparing embeddings to all community centroids, as discussed in \Cref{app:complexity_comparing_emgeddings}, confirming the scalability of \ours.

\noindent \textbf{Baseline Comparison.} We compare \ours with the current state-of-the-art baseline defense which adds static noise to perturb GNN outputs (as proposed in \citep{SSLextraction, liu2022stolenencoder, ContSteal}). We experiment with two different amounts of static noise: (1) $\sigma=0.05$ aims at protecting the model while not harming the performance while (2) $\sigma=5$ prioritizes the defense against model stealing, potentially sacrificing the model performance. For this experiment, we use the ACM with our B) attack scenario (\ie embedding-based stealing) as the example, as shown in \Cref{table:results_static_noise_baseline}.

\begin{table*}[ht]
\small
\centering
\caption{\textbf{Performance for attacker and a target downstream task with Static Noise Addition Defenses vs. Our \ours.} Adding a small amount of noise ($\sigma_1=0.05$) results in a negligible drop in performance for both the downstream task (row 6) and attacker (row 2). Adding a large amount of noise prevents stealing (row 3), but also dramatically harms the downstream task performance (row 7). Our \ours overcomes these shortcomings and provides high performance for the downstream task (row 8) while effectively defending the GNNs against stealing attacks (row 4). %
}
\begin{tabular}{c|c|c|c|c|c}
\hline
User & Defense & GAT & GIN & GraphSAGE & Graph Transformer \\ \hline \hline
Attacker & NONE & 88.53$\pm$0.62 & 85.46$\pm$0.16 & 88.14$\pm$0.12 & 88.30$\pm$0.49 \\
Attacker & NOISE $\sigma=0.05$ & 87.45$\pm$0.41 & 83.51$\pm$0.46 & 87.29$\pm$0.53 & 87.21$\pm$0.44 \\
Attacker & NOISE $\sigma=5$ & 36.28$\pm$0.74 & 34.00$\pm$0.73 & 34.59$\pm$0.78 & 35.85$\pm$0.67 \\
Attacker & \ours & 36.90$\pm$0.51 & 30.45$\pm$0.73 & 31.12$\pm$0.42 & 32.27$\pm$0.12 \\ \hline \hline
Downstream Task & NONE & 89.92$\pm$0.21 & 87.32$\pm$0.13 & 90.15$\pm$0.53 & 96.23$\pm$0.29 \\
Downstream Task & NOISE $\sigma=0.05$ & 89.07$\pm$0.59 & 86.82$\pm$0.45 & 89.42$\pm$0.55 & 95.37$\pm$0.49 \\
Downstream Task & NOISE $\sigma=5$ & 37.01$\pm$0.45 & 35.92$\pm$0.39 & 35.15$\pm$0.59 & 42.74$\pm$0.27 \\
Downstream Task & \ours & 88.69$\pm$0.67 & 84.24$\pm$2.19 & 89.74$\pm$1.47 & 94.42$\pm$1.24 \\ \hline \hline
\end{tabular}
\label{table:results_static_noise_baseline}
\end{table*}

Our results show that when we add noise that is small enough to preserve utility for the downstream task ($\sigma=0.05$), it is not strong enough to prevent stealing. On the contrary, when we choose noise that is large enough to prevent stealing ($\sigma=5$), it harms the downstream task performance and makes the GNN unusable. Our \ours method overcomes these shortcomings by adding dynamic amounts of noise based on the users’ queries.

\noindent \textbf{Link Prediction Tasks.} While our evaluation, so far focused on node classification tasks, we show that ADAGE can also defend models exposed for link prediction tasks. We describe the concrete setup and adaptation in \Cref{app:other_tasks}, and show the results in \Cref{table:results_with_without_defenses_link_cora} and \Cref{table:results_with_without_defenses_link_citeseer} for Cora and CiteSeer, respectively. We observe that, similar as for node classification, ADAGE degrades the performance of the stolen model significantly, while maintaining it for the downstream tasks. %

\begin{figure}[!h]
\centering
     \begin{subfigure}[t]{0.15\textwidth}
         \centering
         \includegraphics[width=\textwidth]{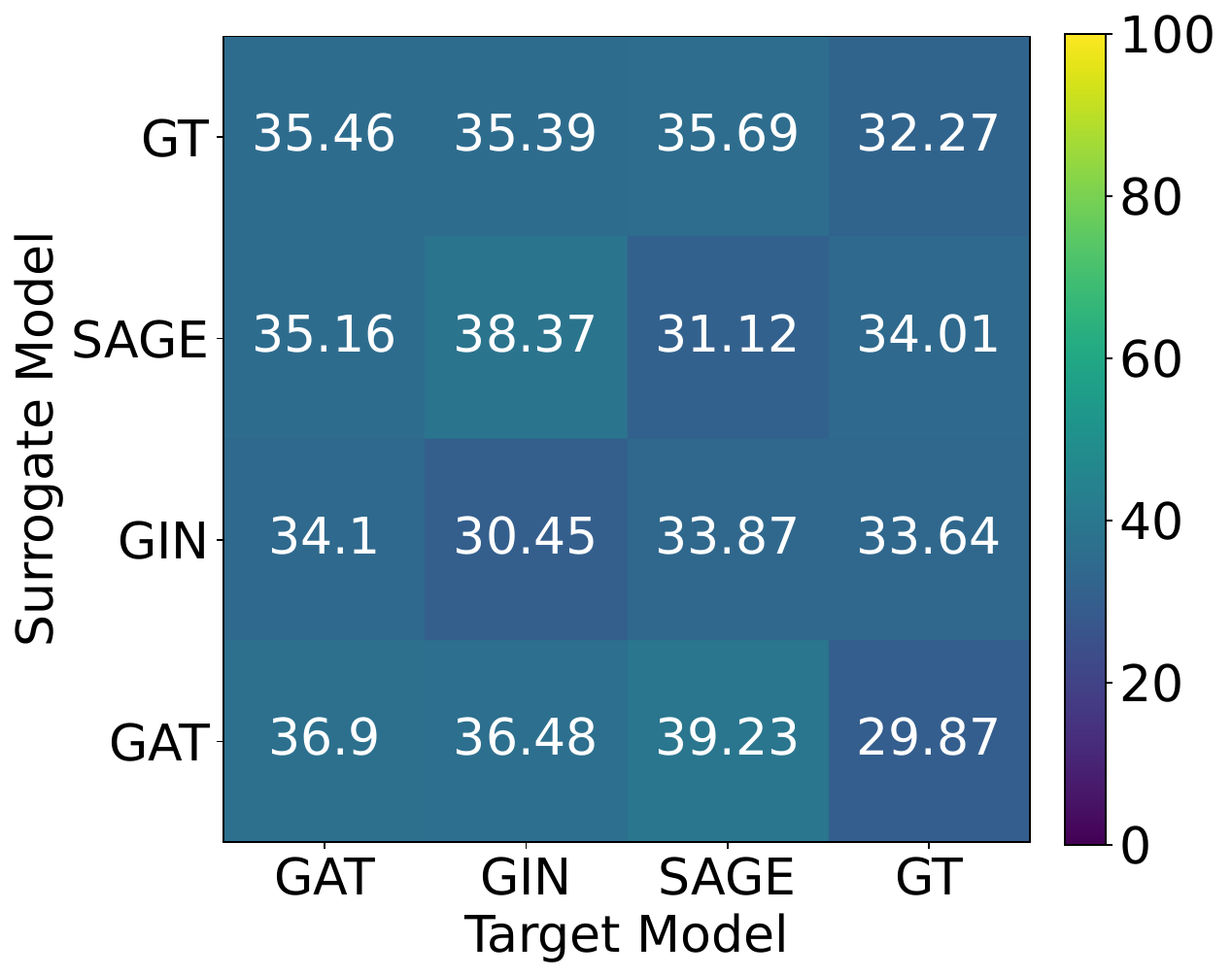}
         \caption{Probabilities.}
     \end{subfigure}
     \begin{subfigure}[t]{0.15\textwidth}
         \centering
         \includegraphics[width=\textwidth]{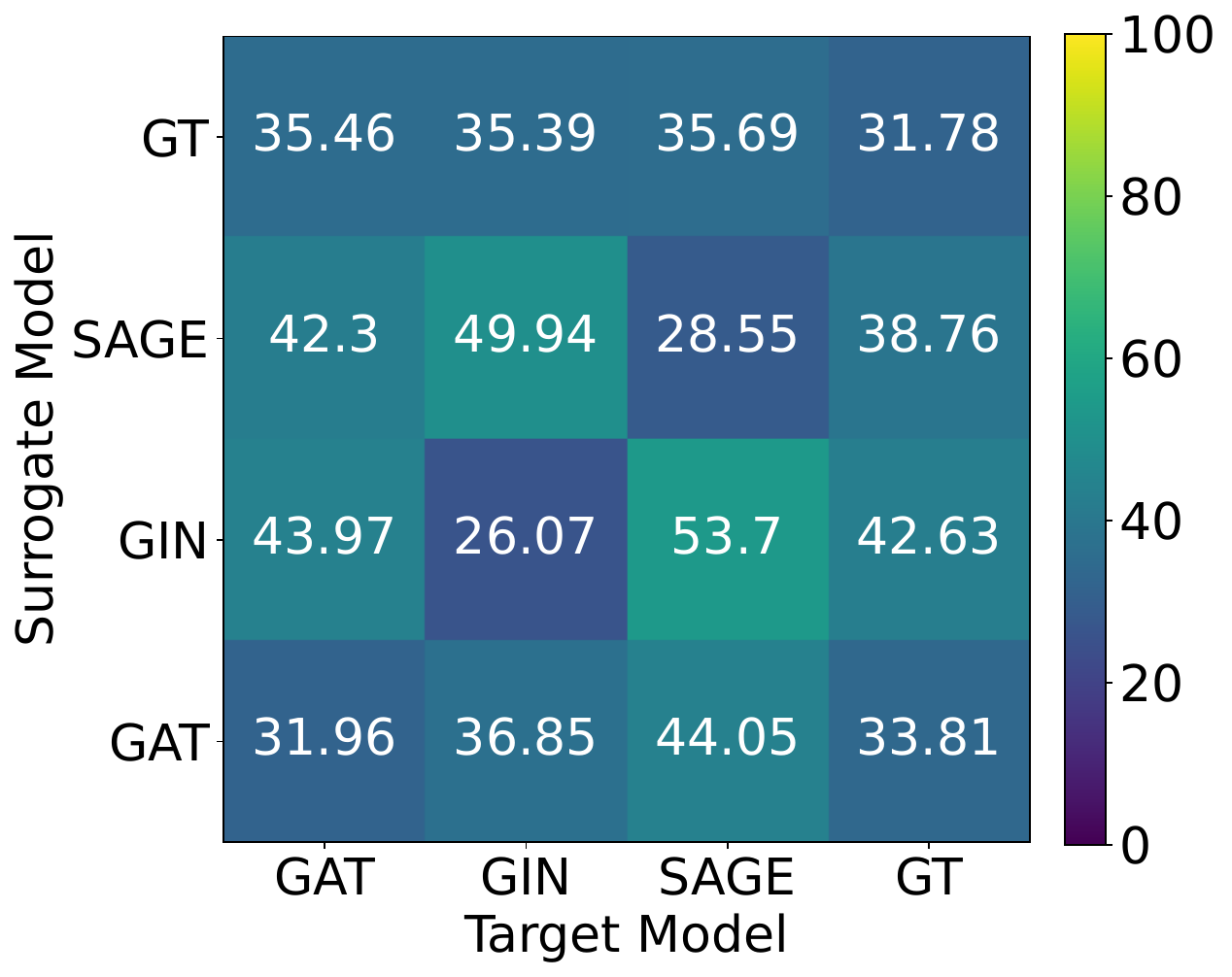}
         \caption{Embeddings.}
     \end{subfigure}
     \begin{subfigure}[t]{0.15\textwidth}
         \centering
         \includegraphics[width=\textwidth]{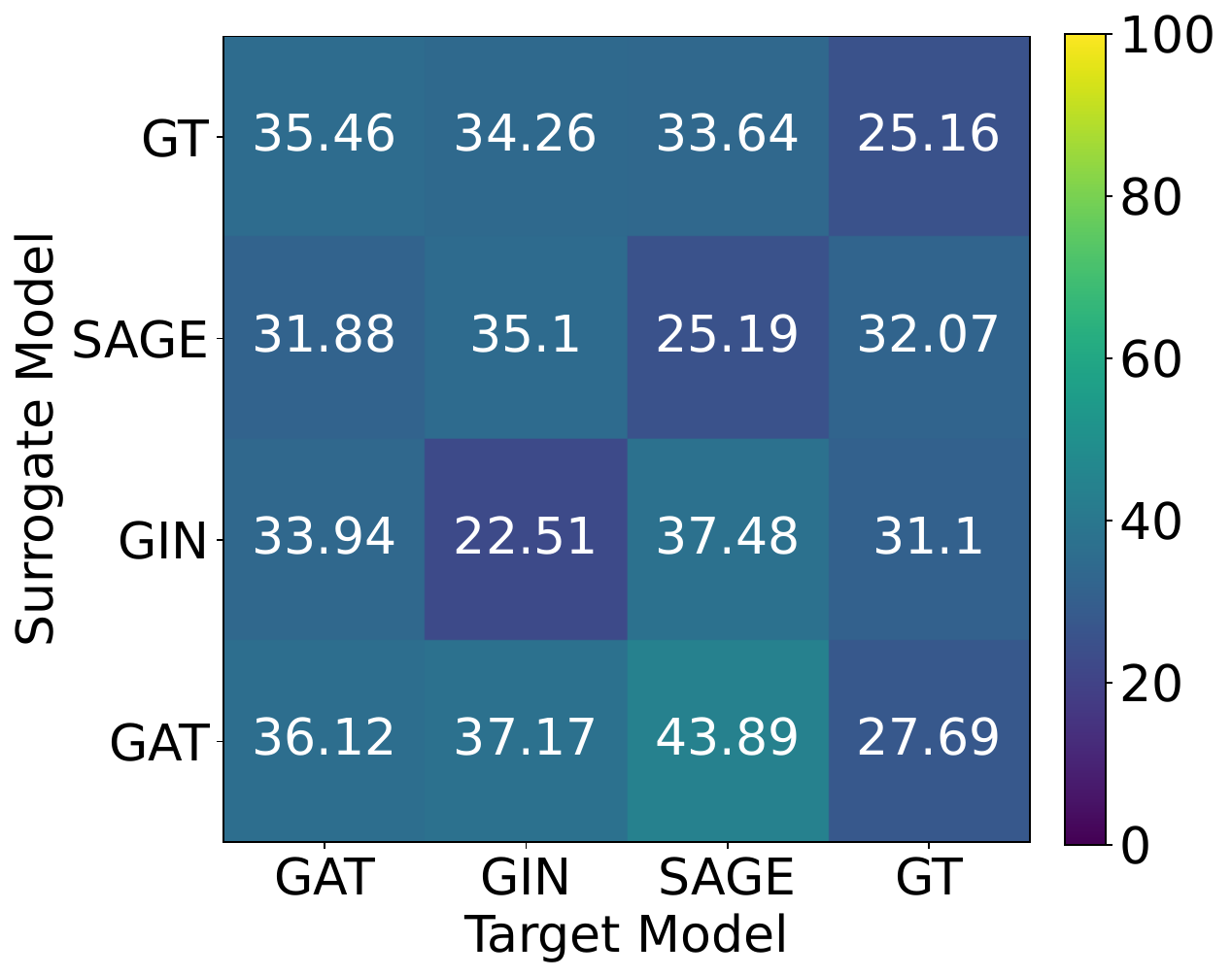}
         \caption{Projections.}
     \end{subfigure}
\caption{\textbf{Performance of the surrogate model with different architectures from the target model (ACM dataset, $\delta=0.25$, GT: Graph Transformer).} Overall, independent of the surrogate model architecture, our defense degrades the stealing performance (surrogate accuracy) dramatically. %
}
\label{fig:adversary_diff_model_acm}
\end{figure}

\begin{table}[ht]
\small
\centering
\caption{\textbf{Wall-clock (elapsed) times with and without applying our \ours (GT - Graph Transformer).} Our defense has an insignificant computational cost, resulting low latency.}
\begin{tabular}{c|c|c|c}
\hline
\parbox{1cm}{Model} & \parbox{2cm}{Time without ADAGE (sec)} & \parbox{2cm}{Time with ADAGE (sec)} & \parbox{1cm}{Time increase} \\ \hline \hline
GAT & 81.367845 & 82.746824 & 1.69\% \\ \hline
GIN & 73.473628 & 75.362746 & 2.57\% \\ \hline
GraphSAGE & 74.316633 & 75.038274 & 0.97\% \\ \hline
GT & 92.483173 & 95.581372 & 3.35\% \\ \hline
\end{tabular}
\label{tab:wall-clock_time}
\end{table}

\section{Adaptive Attackers}
\label{sec:discussions}
Thus far, we have evaluated our proposed \ours defense against state-of-the-art GNN model stealing attacks. In this section, we further evaluate the effectiveness of our defense by investigating three types of potential adaptive attacks, where we relax the constraints on the attacker's knowledge and access to the target model. 

\subsection{Average out Noise-Attacker}
In the first adaptive attack, we assume that the attacker is aware of our defense mechanism, \ie preventing model stealing by adding noise (either by flipping labels or adding Gaussian noise to the representations). %
The attacker attempts to overcome the defense by querying the target model multiple times for the same query node and averaging the model responses to mitigate the effects of the added noise. 
The surrogate performance after applying our defense with various query repeat times (REP) is presented in \Cref{fig:average_noise} for the ACM dataset with the GAT model. The baseline represents the result with REP of $1$.  
As we can observe, the stealing performance remains similar to the baseline when the number of REP is less than $200$. This indicates that our defense can still prevent the stealing even when the attacker repeats each query node up to $200$ times. 
However, as REP rises to $1000$, the stealing performance increases compared with the baseline. 
Nevertheless, even with REP of $2000$, the stealing performance remains lower than that without \ours. For instance, in the node embedding attack setup, there is a degradation around $14\%$ in surrogate accuracy with REP set to $2000$ (with the highest query rate). 
Furthermore, it is important to note that achieving such stealing performance improvement requires substantial query effort from the attacker's side (in case of paid API access also significantly higher monetary access costs), which is opposite to the main goal of a model stealing attack---training a surrogate model with a minimal cost---de-incentivizing stealing. 
Finally, to defend further against this adaptive attack, \ours can be extended to assign the same value of noise to the same query sample so that after averaging the model responses, the noise persists.

\subsection{Knowledge on Communities-Attacker}
In the second adaptive attack, we relax the constraints on the attacker's knowledge on the communities in the underlying graph. Based on their knowledge about the communities, the attacker is assumed to select query nodes predominantly from the same community to minimize the impact of our defense. 
We consider two strengths of attackers, \ie 1) a \textbf{\textit{perfect attacker} (PA)}. This attacker has perfect knowledge of the communities within the underlying training graph of the target model, and 2) a \textbf{\textit{knowledgeable attacker} (KA)}. This attacker only has access to their query graph and additionally knows a) the number of communities ($K$) used to defend the model, and/or b) the community detection algorithm ($Alg.$). Then, based on these two dimensions, we denote $4$ subcategories of knowledgeable attackers, \ie \textbf{KA\_aa, KA\_ab, KA\_ba}, and \textbf{KA\_bb}. We summarize them in \Cref{tab:adaptive_attacker}. 
The stealing performance under this second type of attacker is presented in \Cref{fig:adaptive_attacker_community_knowledge}, with results for the ACM dataset on the GAT model. 
We can observe that with knowledge about communities in the underlying graph, the stealing performance increases as the query rate rises.
This means that this adaptive attacker can indeed mitigate the penalty imposed by our defense compared to the normal attacker. 
However, even with the highest query rate, the stealing performance is notably lower than without our defense. 
For example, the surrogate accuracy of PA in embedding attack setup is $55.17\%$ with a query rate of $0.25$, while without \ours, it reaches $87.26\%$. This demonstrates that the diversity of the query nodes significantly impacts the stealing performance (as in~\Cref{fig:attack_performance}), and the second adaptive attack cannot achieve high attack performance. %

 \begin{figure}[!h]
\centering
\begin{subfigure}[t]{0.35\textwidth}
     \centering
     \includegraphics[width=\textwidth]{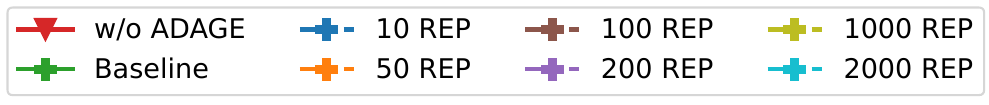}
 \end{subfigure}
 \vfill     
 \begin{subfigure}[t]{0.15\textwidth}
     \centering
     \includegraphics[width=\textwidth]{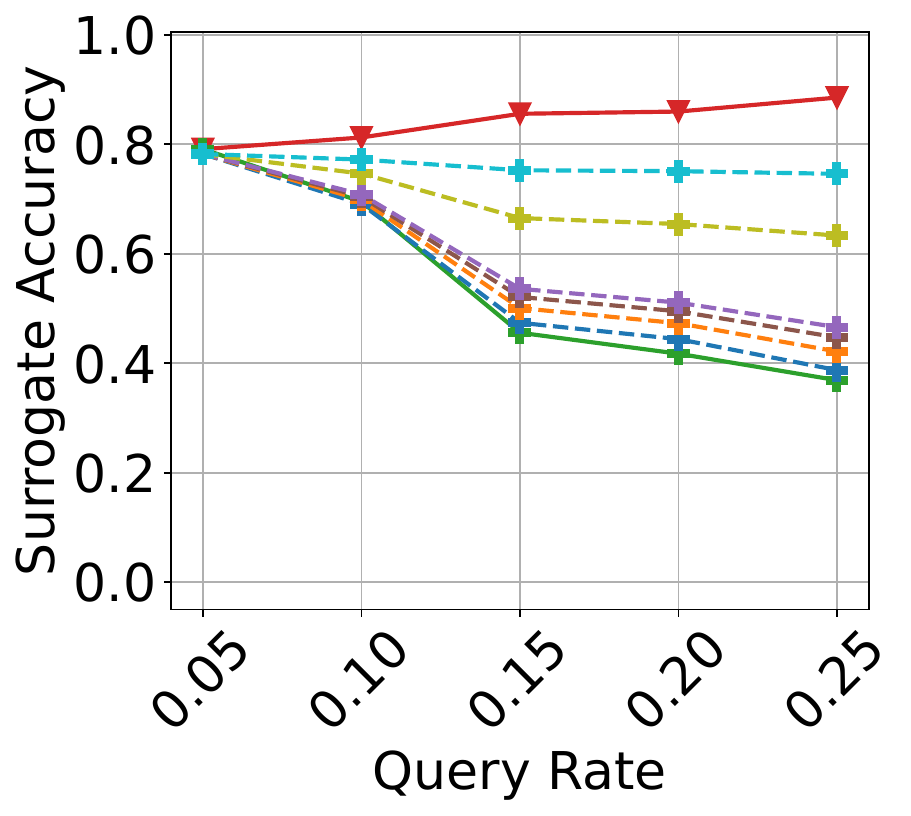}
     \caption{Probabilities.}
 \end{subfigure}
 \begin{subfigure}[t]{0.15\textwidth}
     \centering
     \includegraphics[width=\textwidth]{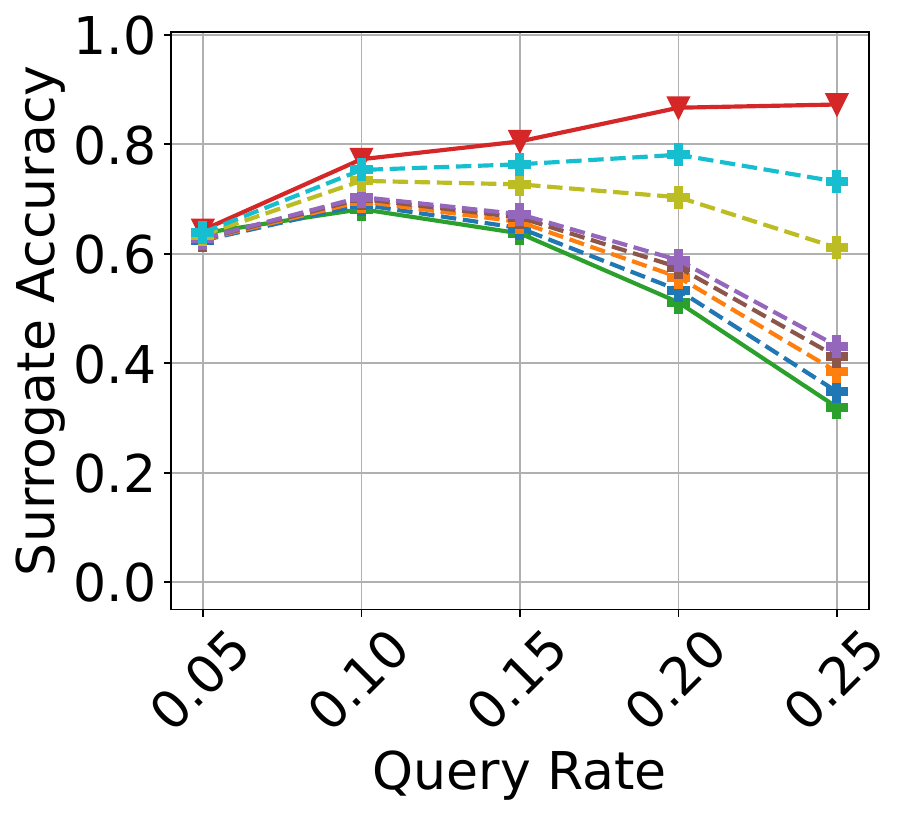}
     \caption{Embeddings.}
 \end{subfigure}
 \begin{subfigure}[t]{0.15\textwidth}
     \centering
     \includegraphics[width=\textwidth]{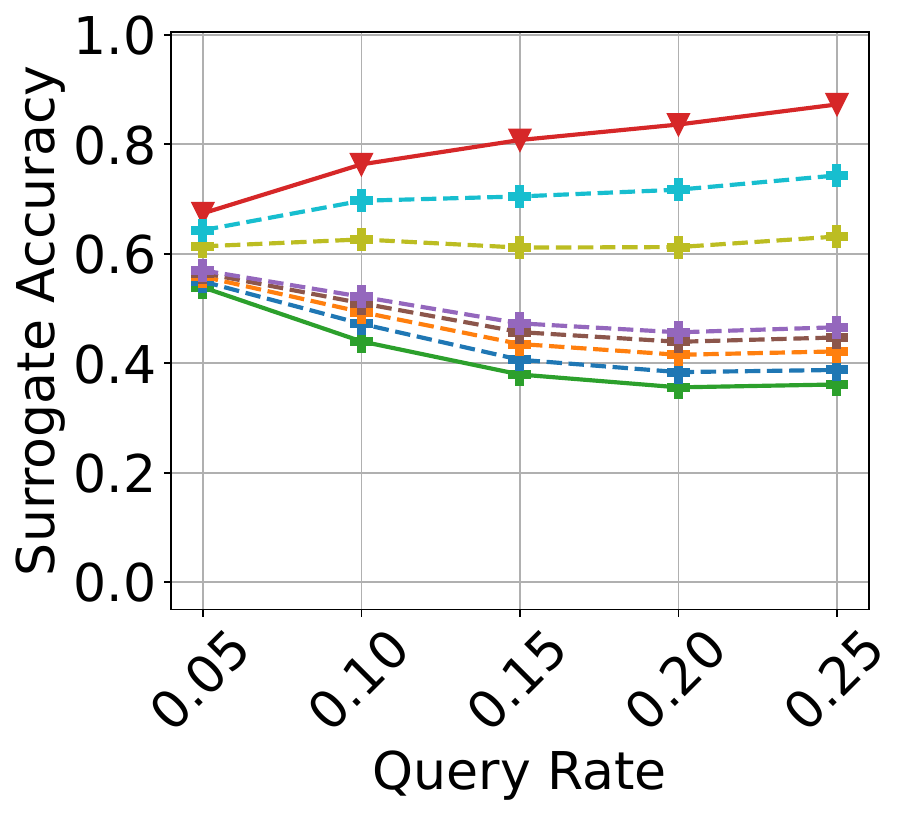}
     \caption{Projections.}
 \end{subfigure}
\caption{\textbf{Noise-averaging adaptive attacker (ACM, GAT).} We present the improvement of surrogate accuracy if the attacker repeats (REP) each query multiple times to average out the noise over the Baseline, where the attacker does not repeat any query.
Overall, with less than $200$ repeated queries, our defense can still degrade the stealing performance substantially (more than 40\%).
Only at a  very high query cost, \ie $2000$ REP, the attacker can improve surrogate performance slightly. 
However, the performance drop is still around $14\%$ with respect to the undefended target model (w/o \ours).
}
\label{fig:average_noise}
\end{figure}

\begin{table}[t]
\small
 \centering
 \caption{\textbf{Attacker taxonomy.} 
 PA denotes a \textit{perfect attacker} who knows the graph and its communities.
 KA refers to a \textit{knowledgeable attacker} who may know
 the number of communities ($K$) and/or the community detection algorithm ($Alg.$). \textbf{a} indicates that the attacker has access to this dimension of knowledge, whereas \textbf{b} indicates that the attacker does not have access to it. %
 }
 \label{tab:adaptive_attacker}
\begin{tabular}{cccc} 
 \hline
 \multirow{2}{*}{Attacker} & \multirow{2}{*}{$\mathbf{G}_{train}$} & \multicolumn{2}{c}{$\mathbf{G}_q$}\\
 \cline{3-4}
  & & $K$ & $Alg.$ \\
  \hline
 PA & \checkmark &  &  \\
 \hline
 \hline
 KA\_aa & \ding{55} & \checkmark & \checkmark \\
 KA\_ab & \ding{55} & \checkmark & \ding{55} \\
 KA\_ba & \ding{55} & \ding{55} & \checkmark \\
 KA\_bb & \ding{55} & \ding{55} & \ding{55} \\
 \hline
\end{tabular}
\end{table}

\begin{figure}[!h]
\centering
    \begin{subfigure}[t]{0.35\textwidth}
         \centering
         \includegraphics[width=\textwidth]{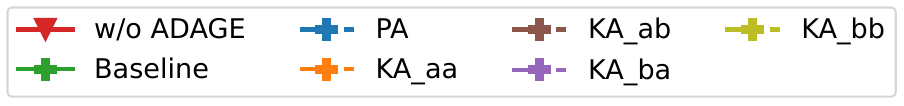}
     \end{subfigure}
     \vfill  
     \begin{subfigure}[t]{0.15\textwidth}
         \centering
         \includegraphics[width=\textwidth]{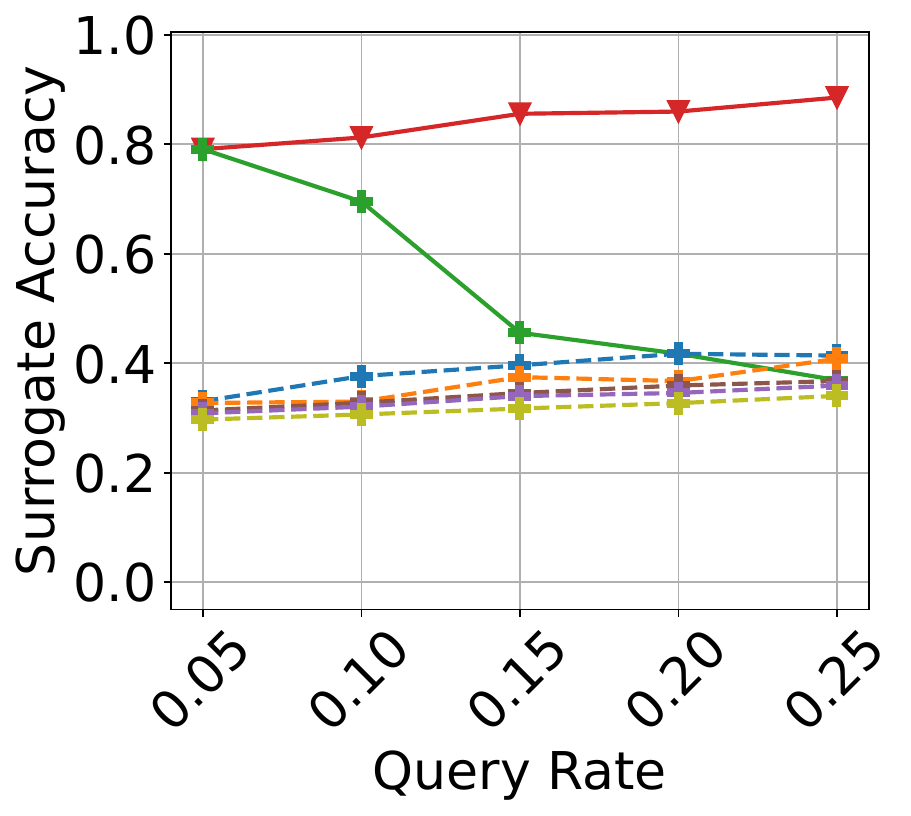}
         \caption{Probabilities.}
     \end{subfigure}
     \begin{subfigure}[t]{0.15\textwidth}
         \centering
         \includegraphics[width=\textwidth]{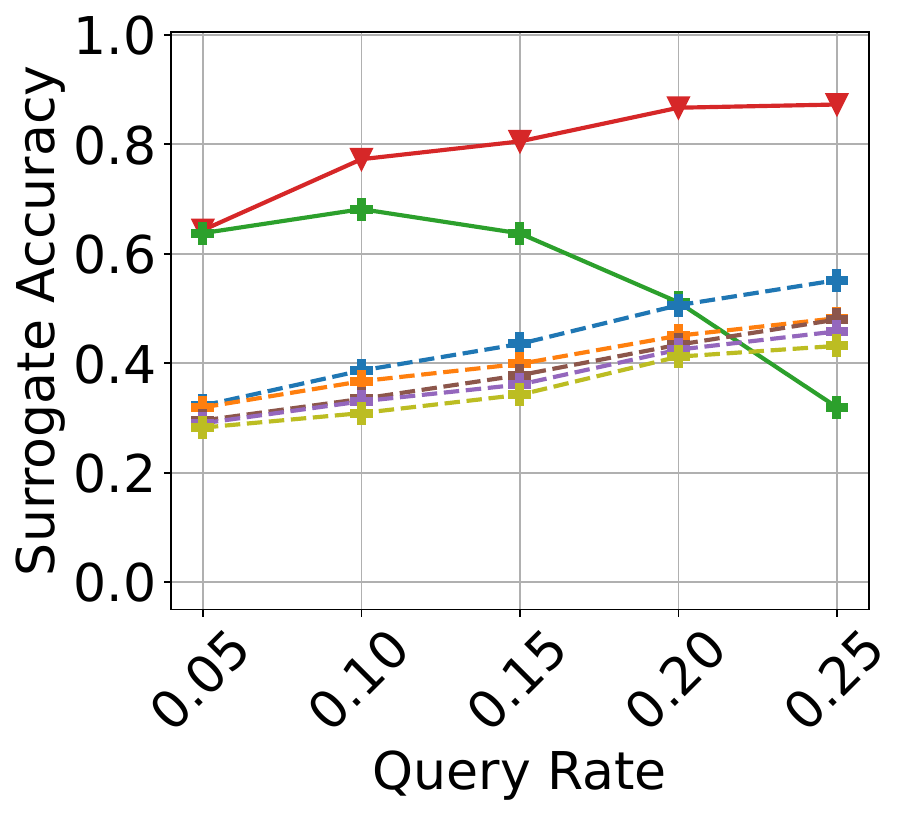}
         \caption{Embeddings.}
     \end{subfigure}
     \begin{subfigure}[t]{0.15\textwidth}
         \centering
         \includegraphics[width=\textwidth]{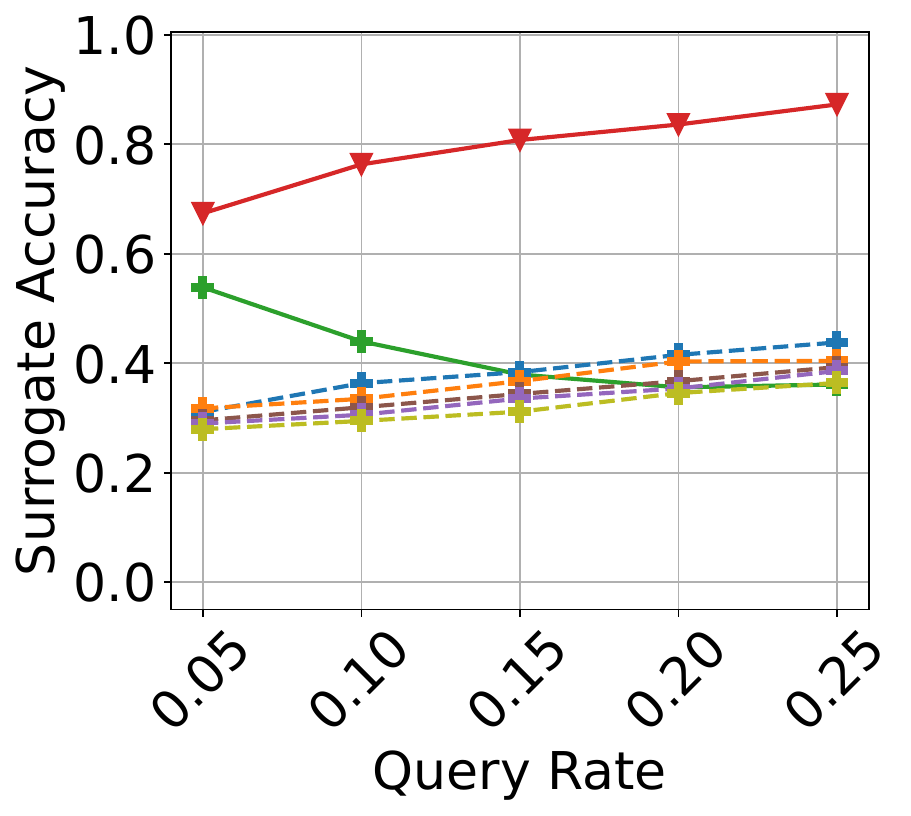}
         \caption{Projections.}
     \end{subfigure}
\caption{\textbf{Performance of the surrogate model with the second adaptive attack (ACM, GAT).} Overall, with knowledge of communities in the underlying graph, the surrogate accuracy increases as the query rate rises, but it is still low.
}
\label{fig:adaptive_attacker_community_knowledge}
\end{figure}

\begin{table*}[htpb]
\small
 \centering
 \caption{\textbf{Impact of transformations on the performance of downstream tasks in B) and C) setups (ACM).} We show the results on the $c_1$ community. Overall, the transformations applied per-account do not harm the performance of downstream tasks.}
 \label{table:results_with transformation_acm}
\begin{tabular}{ccccccc} 
 \hline
Attack setup & Transformation & GAT & GIN & GraphSAGE & Graph Transformer \\
 \hline
 \hline

\multirow{4}{*}{Embeddings} &  N/A & {$89.50\pm1.16$} & {$89.35\pm0.76$} & {$88.99\pm0.93$} & {$96.72\pm0.30$}\\
 &  Affine & {$88.29\pm2.09$} & {$85.12\pm4.78$} & {$87.34\pm1.93$} & {$95.76\pm0.52$}\\
 &  Shuffle & {$89.47\pm1.15$} & {$89.09\pm0.94$} & {$88.73\pm1.11$} & {$96.61\pm0.17$}\\
 &  Affine + Shuffle & {$88.77\pm1.90$} & {$86.81\pm1.35$} & {$88.37\pm1.03$} & {$96.20\pm0.56$}\\
 \hline
 \hline

\multirow{4}{*}{Projections} &  N/A & {$88.56\pm0.93$} & {$89.03\pm1.22$} & {$90.08\pm2.26$} & {$95.79\pm0.79$}\\
 &  Affine & {$87.82\pm1.96$} & {$89.03\pm1.22$} & {$89.35\pm2.60$} & {$95.24\pm1.11$}\\
 &  Shuffle & {$88.19\pm1.44$} & {$88.99\pm1.23$} & {$90.08\pm2.26$} & {$95.50\pm0.76$}\\
 &  Affine + Shuffle & {$88.08\pm1.60$} & {$88.96\pm1.12$} & {$89.17\pm2.58$} & {$95.46\pm0.84$}\\
 \hline
 \hline
\end{tabular}
\end{table*}

\begin{figure}[!h]
    \centering
     \begin{subfigure}[t]{0.116\textwidth}
         \centering
         \includegraphics[width=\textwidth]{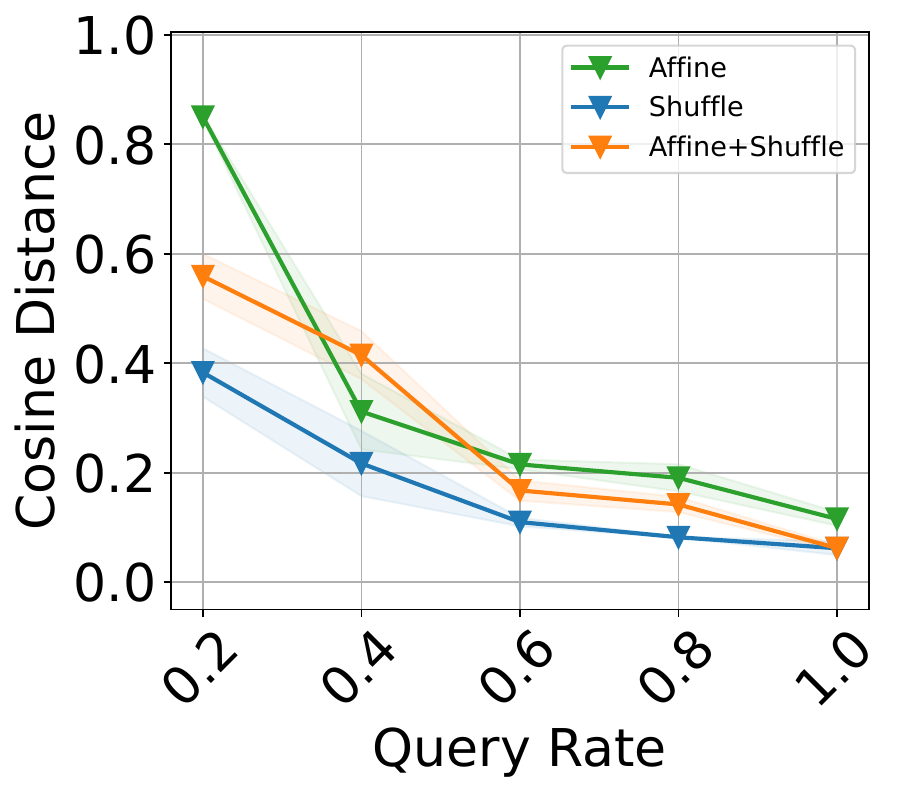}
         \caption{GAT.}
     \end{subfigure}
     \hfill
     \begin{subfigure}[t]{0.116\textwidth}
         \centering
         \includegraphics[width=\textwidth]{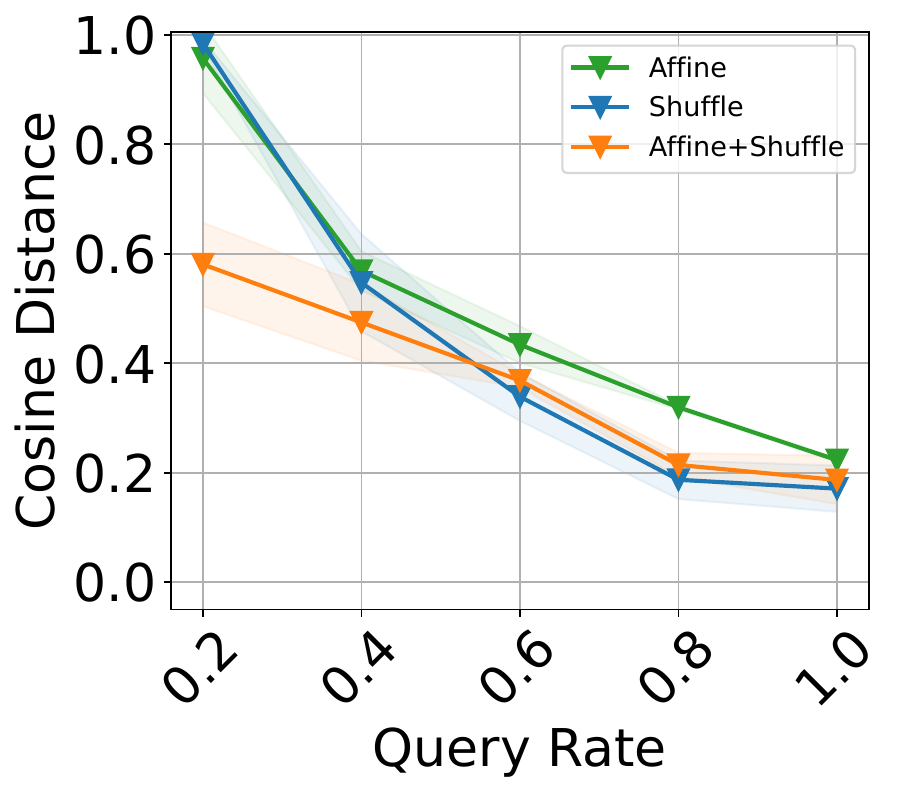}
         \caption{GIN.}
     \end{subfigure}
     \hfill
     \begin{subfigure}[t]{0.116\textwidth}
         \centering
         \includegraphics[width=\textwidth]{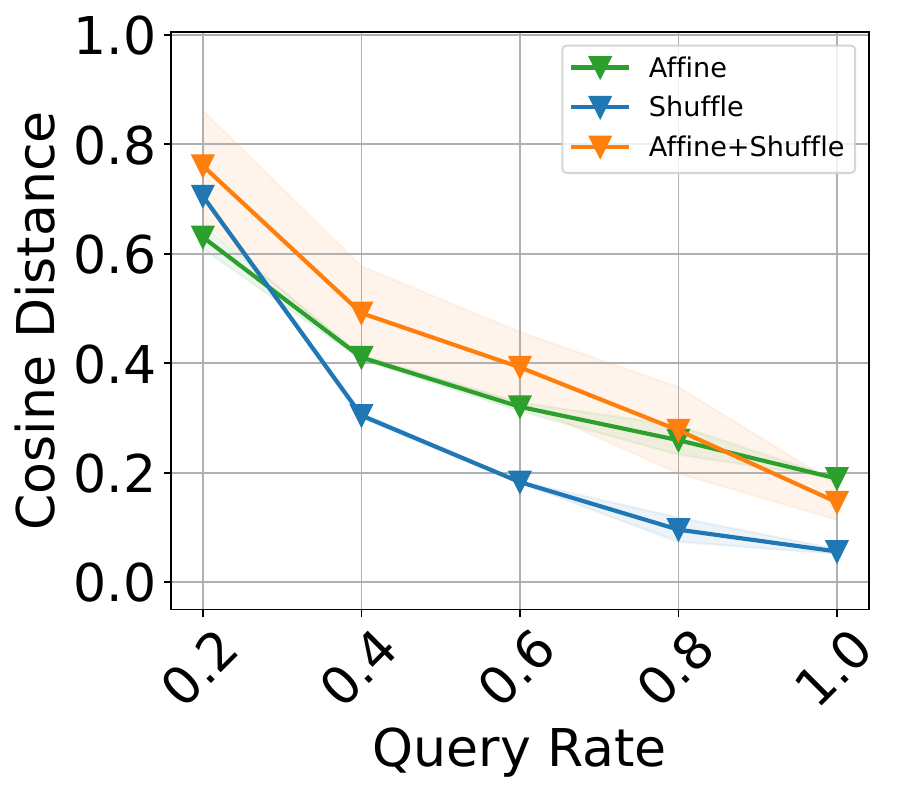}
         \caption{SAGE.}
     \end{subfigure}
     \hfill
     \begin{subfigure}[t]{0.116\textwidth}
         \centering
         \includegraphics[width=\textwidth]{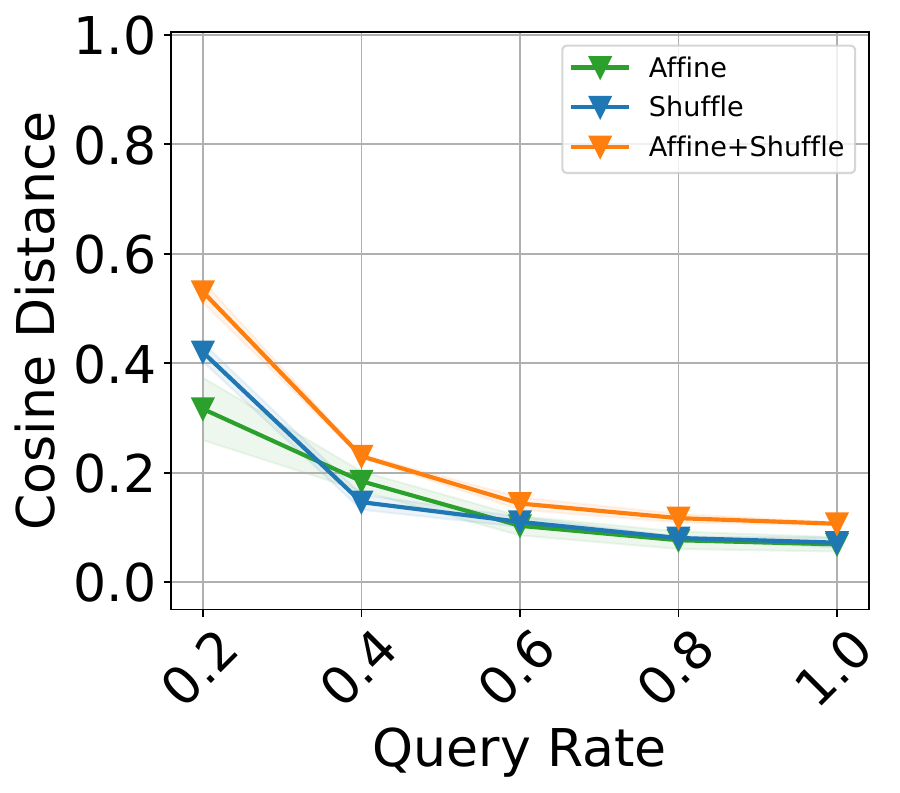}
         \caption{GT.}
     \end{subfigure}
    \caption{\textbf{Remapping quality (ACM, Projections setup, GT-Graph Transformer).} It is difficult and costly for attackers to perfectly remap representations over Sybil accounts.}
    \label{fig:remapping_distance_acm}
\end{figure}

\subsection{Sybil-Attacker}
Finally, we further consider Sybil attacks~\citep{Douceur2002sybilattacks}. 
A Sybil attack is a type of attack in which an attacker subverts the service's system by creating a large number of pseudonymous identities and uses them to gain benefits~\citep{Douceur2002sybilattacks,neary2011real}. 
In our threat model, since \ours analyses samples queried by a single user, an attacker may distribute its queries among several users to avoid detection. 
However, there are many general countermeasures against Sybil attacks. 
For instance, validation techniques can be used to prevent Sybil attacks~\citep{maheswaran2016building}, where a user who wants to query the target GNN model through an API has to establish a remote identity based on a trusted third party that ensures a one-to-one correspondence between an identity and a user. 
In addition, imposing economic costs can be used to make Sybil attacks more expensive. Proof-of-work-based defense, for instance, requires a user to prove that they expended a certain amount of computation effort to solve a cryptographic puzzle~\citep{powDefense}. With an increasing number of users, more computation effort is required to solve the puzzles. 
Investments in other resources, such as storage or a stake in an existing cryptocurrency, can also be used to impose such economic costs. 

To defend the Sybil attackers in the B) node embedding and C) projection setups, we leverage the performance preserving per-user transformations of embeddings from~\citep{dubinski2023bucks}. The transformations should follow two requirements: 1) they should not harm the performance of downstream tasks, and 2) they should be costly to reverse for the attackers. 
We present the performance of downstream tasks with different transformations in \Cref{table:results_with transformation_acm}, taking the ACM dataset as an example. %
As we can observe, with transformations, the downstream accuracy in all cases has a negligible drop, \ie less than $3\%$, which indicates that the transformations preserve the performance of downstream tasks. 

Furthermore, to evaluate the remapping cost for the attackers, we assess the fidelity of remapped representations as a function of the number of overlapping queries between the accounts. 
Specifically, we assume an attacker who queries from two Sybil accounts and aims to learn a remapping function that transforms the representations from account \#2 to the representation space of account \#1. Using more accounts for the attacker potentially leads to more remapping cost. Thus, our evaluation here represents a lower bound on the cost caused to the attacker through our transformations. 
A two-layer linear model is trained on overlapping representations between the accounts to learn the mapping between two accounts' representations. 
The number of overlapping representations is decided by the query rate within the query graph $G_Q$, from $0.2$ to $1.0$. 
When the remapping model is learned, we query the test data through two accounts. Then, we apply the learned remapping model to the representations of account \#2 and compute the pairwise cosine distances between the representations from account \#1 and their remapped counterparts from account \#2. 
In the projection setup, the dimension of output representation is smallest, \ie 2, which potentially leads to the least cost for the Sybil attackers to remap between different accounts' representations. Thus, we evaluate the remapping quality in the projection setup, as shown in \Cref{fig:remapping_distance_acm}, also on ACM dataset. %
We show that with increasing query rate, the remapping quality increases for all transformations. 
However, generally, it is difficult and costly for the attackers to perfectly remap representations over different accounts, \eg for the GIN model, the cosine distance is less than $0.2$ until the query rate is more than $0.8$, which is much higher than the query rate of stealing (up to $0.25$). 
In the case of setup A) with labels or output probabilities, we suggest to measure the privacy leakage per query as in~\citep{powDefense} to increase the cost of queries that incur more information about the target model. We further evaluate the performance of the Sybil attack under increasing numbers of accounts, as presented in \Cref{app:subsec_adaptive_attacks}. The results demonstrate that our defense disincentivizes Sybil attackers under our threat model.

\section{Related Work}
\label{sec:related_work}

\noindent \textbf{Model Stealing Attacks against ML.} 
There are also existing works on stealing the link or underlying graph training data from GNNs. \citet{guan2024large} proposed a novel link stealing attack method that takes advantage of cross-dataset and Large Language Models (LLMs). LinkThief \citep{zhang2024linkthief} combines generalized structure knowledge with node similarity, to improve link stealing attack. 
There is also a new threat model that steals the underlying graph training data given a trained graph model \citep{lin2024stealingtraininggraphsgraph}. In contrast, our work considers the stealing of the graph model itself. 
Byond GNNs, Model stealing attacks against supervised learning (SL) models involve an attacker querying the victim model to obtain labels for the attacker's own training data~\citep{tramer2016stealing}. The primary objectives of such an attack are for the adversary to either attain a specified level of accuracy on a task using their extracted model~\citep{orekondy2019knockoff} or recreate a high-fidelity replica model that can facilitate further attacks~\citep{jagielski2020highfidelity}. An example of a follow-up (reconnaissance) on the high-fidelity stealing is the construction of adversarial examples to fool the victim model~\citep{szegedy2013intriguing,biggio2013adv,goodfellow2014explaining}. A key goal for the attacker is to minimize the number of queries to the victim model required to successfully steal a model that meets their intended purpose.
In the self-supervised learning (SSL) setting, the goal of an attacker is to learn
high-quality representations that can be used to achieve high performance on many downstream tasks~\citep{SSLextraction}. 
Contrastive learning is used in the model stealing attacks against encoders trained in self-supervised setting~\citep{liu2022stolenencoder}.

\noindent \textbf{Defending Against GNN Stealing.} 
To protect the training graph from link stealing attacks on GNNs, GRID \citep{gridlou} adds carefully crafted noises to the nodes' prediction vectors for disguising adjacent nodes as n-hop indirect neighboring nodes. 
Regarding defending against model stealing attacks on GNNs, \citet{zhao2021watermarking} proposed a GNN watermark for inductive node classification GNNs based on an Erdős–Rényi random graph with random node feature vectors and labels. %
\citet{xu2023watermarking} further extended that work to transductive GNNs and graph classification tasks by proposing a watermarking method for GNNs based on backdoor attacks. 
In a similar vein, \citet{waheed2023grove} presented a GNN model fingerprinting scheme for inductive GNNs. Their approach identifies GNN embeddings as a potential fingerprint and, given a target model and a suspect model, can determine if the suspect model was stolen or derived from the target model. %
All these defenses focus on one particular stealing setup, \ie the GNN model outputs either node embeddings or prediction probabilities. Additionally, they are limited to detecting stolen models, \ie they operate after the harm has already been incurred.
In contrast, our \ours is general and can be applied to protect GNNs in multiple stealing setups with different types of model outputs. Moreover, \ours actively prevents the stealing while it is happening. 
Similarly to \citet{9157021}, we output incorrect predictions with a calibrated probability to impede the stealing process. However, our method operates on GNNs instead of standard vision models. 

\noindent \textbf{Defenses Against Model Stealing.} Defenses against stealing machine learning models can be categorized based on when they are used in the stealing~\citep{powDefense}. There are \textit{active} defenses that aim to prevent model theft before it occurs by increasing the cost of stealing or by introducing perturbations to outputs to poison the training objective of an attacker, \textit{passive} defenses that try to detect attacks, and \textit{reactive} defenses that try to determine if a model was stolen. 

Active defenses like proof-of-work~\citep{powDefense}
require API users to solve puzzles before accessing model outputs, with the puzzle difficulty calibrated based on deviations from expected legitimate users' behavior. 
Another active defense~\citep{10049136} disables the usable functionality of the stolen model by constructively minimizing the diverged confidence information that is essential to train the surrogate model. 
Other active defenses add noise to outputs or truncate them, lowering result quality~\citep{dubinski2023bucks}. 

Passive defenses monitor for signs of an attack in progress. For example, they analyze the distribution of the users' queries and try to identify if there is a deviation of a given query distribution from the assumed normal distribution~\citep{juuti2019prada}. 

Finally, reactive defenses, \eg watermarking~\citep{entangled_wm, xu2023watermarking}, dataset inference~\citep{maini2021dataset,datasetinference2022neurips}, and Proof-of-Learning~\citep{jia2021proof}, attempt to enable model owners to prove ownership after the fact if theft is suspected. For example, dataset inference detects if a signal from the private training data of the model owner is present in a suspect copy, while Proof-of-Learning shows ownership by demonstrating incremental updates from model training.

\section{Discussion}
\label{discussion}
Our work introduces an active defense against GNN model stealing, which dynamically adjusts perturbations in the model output based on the accumulated query diversity. The experiments with three adaptive attackers demonstrate that \ours substantially increases the cost of successful stealing while maintaining high downstream utility. We highlight several discussion points around the broader implications and design trade-offs of our defense.

\textbf{Query Diversity.} The intuitive idea of \ours is based on the use of query diversity as an indicator for suspicious behavior. While this successfully captures the behavior of attackers, it also means that normal users with highly diverse queries may be subject to stronger penalties. Notably, our method never attempts to classify queries as \textit{benign} or \textit{malicious}; rather, it adaptively penalizes based on accumulated query diversity. In practice, this ensures that such users still experience much smaller penalties than adversaries. Future work could extend this design by combining query diversity with explicit user modeling to better accommodate diverse but normal behaviors. 
Current query diversity metric is designed to be simple, efficient to compute in real-time. This metric could be extended by integrating query distribution and community size to design a richer diversity metric, in the future, as discussed in \Cref{app:query_diversity_metric}.

\textbf{Penalty Mechanisms.} In this work, we design the penalty via adaptive noise addition to the model outputs. However, our framework is not restricted to noise. Alternative mechanisms, \eg requiring additional computational work~\citep{powDefense}, could be integrated to also achieve similar goals. The general principle remains the same, \ie defenses cannot guarantee absolute prevention of model stealing, but they can make it less attractive for the attacker by raising the attacker's cost beyond the resources required to train a similar high-performance model from scratch.

\section{Conclusions}
This paper proposes \ours, the first general and active defenses against GNN model stealing. 
\ours analyzes the diversity of queries to the target models with respect to the communities in the underlying graph and calibrates the defense strength accordingly. 
We show that \ours can be applied in all common stealing attack setups, where attackers query for labels (posterior probabilities), node embeddings, or projections.
We conduct extensive experiments on four popular inductive GNN models, six benchmark datasets, and with three adaptive attackers. Our empirical results show that our defense can prevent model stealing in all attack setups while maintaining the performance on downstream tasks.

\begin{acks}
This work was also supported by the German Research Foundation (DFG) within the framework of the Weave Programme under the project titled "Protecting Creativity: On the Way to Safe Generative Models" with number 545047250.
\end{acks}

\bibliographystyle{ACM-Reference-Format}
\bibliography{main}

@InProceedings{biggio2013adv,
author="Biggio, Battista
and Corona, Igino
and Maiorca, Davide
and Nelson, Blaine
and {\v{S}}rndi{\'{c}}, Nedim
and Laskov, Pavel
and Giacinto, Giorgio
and Roli, Fabio",
editor="Blockeel, Hendrik
and Kersting, Kristian
and Nijssen, Siegfried
and {\v{Z}}elezn{\'y}, Filip",
title="Evasion Attacks against Machine Learning at Test Time",
booktitle="Machine Learning and Knowledge Discovery in Databases",
year="2013",
publisher="Springer Berlin Heidelberg",
address="Berlin, Heidelberg",
pages="387--402"
}

@article{goodfellow2014explaining,
  title={Explaining and harnessing adversarial examples},
  author={Goodfellow, Ian J and Shlens, Jonathon and Szegedy, Christian},
  journal={arXiv preprint arXiv:1412.6572},
  year={2014}
}

@article{entangled_wm,
    author={Hengrui Jia and C. A. Choquette-Choo and V. Chandrasekaran and N. Papernot.},
    title={Entangled Watermarks as a Defense against Model Extraction.},
    journal={USENIX Security Symposium},
    year={2021},
}

@inproceedings{orekondy2019knockoff,
  title={Knockoff nets: Stealing functionality of black-box models},
  author={Orekondy, Tribhuvanesh and Schiele, Bernt and Fritz, Mario},
  booktitle={Proceedings of the IEEE/CVF Conference on Computer Vision and Pattern Recognition},
  pages={4954--4963},
  year={2019}
}

@inproceedings{maini2021dataset,
  title={Dataset inference: Ownership resolution in machine learning},
  author={Maini, Pratyush and Yaghini, Mohammad and Papernot, Nicolas},
  booktitle={Proceedings of ICLR 2021: 9th International Conference on Learning Representationsn},
  year={2021}
}

@inproceedings{juuti2019prada,
  title={PRADA: protecting against DNN model stealing attacks},
  author={Juuti, Mika and Szyller, Sebastian and Marchal, Samuel and Asokan, N},
  booktitle={2019 IEEE European Symposium on Security and Privacy (EuroS\&P)},
  pages={512--527},
  year={2019},
  organization={IEEE}
}

@inproceedings{powDefense,
title={Increasing the Cost of Model Extraction with Calibrated Proof of Work},
author={Dziedzic, Adam and Kaleem, Muhammad Ahmad and Lu, Yu Shen and Papernot, Nicolas},
booktitle={International Conference on Learning Representations},
year={2022},
url={https://arxiv.org/abs/2201.09243},
}

@inproceedings{orekondy2019prediction,
title={Prediction Poisoning: Towards Defenses Against DNN Model Stealing Attacks},
author={Tribhuvanesh Orekondy and Bernt Schiele and Mario Fritz},
booktitle={International Conference on Learning Representations},
year={2020},
url={https://openreview.net/forum?id=SyevYxHtDB}
}

@article{jia2021proof,
  title={Proof-of-Learning: Definitions and Practice},
  author={Jia, Hengrui and Yaghini, Mohammad and Choquette-Choo, Christopher A and Dullerud, Natalie and Thudi, Anvith and Chandrasekaran, Varun and Papernot, Nicolas},
  journal={arXiv preprint arXiv:2103.05633},
  year={2021}
}

@article{szegedy2013intriguing,
  title={Intriguing properties of neural networks},
  author={Szegedy, Christian and Zaremba, Wojciech and Sutskever, Ilya and Bruna, Joan and Erhan, Dumitru and Goodfellow, Ian and Fergus, Rob},
  booktitle={International Conference on Learning Representations},
  url={https://openreview.net/forum?id=kklr_MTHMRQjG},
  year={2014}
}

@inproceedings{papernot2017practical,
  title={Practical black-box attacks against machine learning},
  author={Papernot, Nicolas and McDaniel, Patrick and Goodfellow, Ian and Jha, Somesh and Celik, Z Berkay and Swami, Ananthram},
  booktitle={Proceedings of the 2017 ACM on Asia conference on computer and communications security},
  pages={506--519},
  year={2017}
}

@inproceedings{ContSteal,
  title={Can't steal? Cont-steal! Contrastive stealing attacks against image encoders},
  author={Sha, Zeyang and He, Xinlei and Yu, Ning and Backes, Michael and Zhang, Yang},
  booktitle={Proceedings of the IEEE/CVF Conference on Computer Vision and Pattern Recognition},
  pages={16373--16383},
  year={2023}
}

@inproceedings{SSLextraction,
  title = {On the Difficulty of Defending Self-Supervised Learning against Model Extraction},
  author = {Dziedzic, Adam and Dhawan, Nikita and Kaleem, Muhammad Ahmad and Guan, Jonas and Papernot, Nicolas},
  booktitle = {International Conference on Machine Learning},
  year = {2022}
}

@inproceedings{dubinski2023bucks,
  title={Bucks for Buckets (B4B): Active Defenses Against Stealing Encoders},
  author={Dubi{\'n}ski, Jan and Pawlak, Stanis{\l}aw and Boenisch, Franziska and Trzcinski, Tomasz and Dziedzic, Adam},
  booktitle={Thirty-seventh Conference on Neural Information Processing Systems},
  year={2023}
}

@inproceedings{shen2022model,
  title={Model stealing attacks against inductive graph neural networks},
  author={Shen, Yun and He, Xinlei and Han, Yufei and Zhang, Yang},
  booktitle={2022 IEEE Symposium on Security and Privacy (SP)},
  pages={1175--1192},
  year={2022},
  organization={IEEE}
}

@article{kipf2016semi,
  title={Semi-supervised classification with graph convolutional networks},
  author={Kipf, Thomas N and Welling, Max},
  journal={arXiv preprint arXiv:1609.02907},
  year={2016}
}

@article{hamilton2017inductive,
  title={Inductive representation learning on large graphs},
  author={Hamilton, Will and Ying, Zhitao and Leskovec, Jure},
  journal={Advances in neural information processing systems},
  volume={30},
  year={2017}
}

@article{xu2018powerful,
  title={How powerful are graph neural networks?},
  author={Xu, Keyulu and Hu, Weihua and Leskovec, Jure and Jegelka, Stefanie},
  journal={arXiv preprint arXiv:1810.00826},
  year={2018}
}

@article{velivckovic2017graph,
  title={Graph attention networks},
  author={Veli{\v{c}}kovi{\'c}, Petar and Cucurull, Guillem and Casanova, Arantxa and Romero, Adriana and Lio, Pietro and Bengio, Yoshua},
  journal={arXiv preprint arXiv:1710.10903},
  year={2017}
}

@article{vishwanathan2010graph,
  title={Graph kernels},
  author={Vishwanathan, S Vichy N and Schraudolph, Nicol N and Kondor, Risi and Borgwardt, Karsten M},
  journal={Journal of Machine Learning Research},
  volume={11},
  pages={1201--1242},
  year={2010},
  publisher={MIT Press}
}

@article{shervashidze2011weisfeiler,
  title={Weisfeiler-lehman graph kernels.},
  author={Shervashidze, Nino and Schweitzer, Pascal and Van Leeuwen, Erik Jan and Mehlhorn, Kurt and Borgwardt, Karsten M},
  journal={Journal of Machine Learning Research},
  volume={12},
  number={9},
  year={2011}
}

@article{zhang2018link,
  title={Link prediction based on graph neural networks},
  author={Zhang, Muhan and Chen, Yixin},
  journal={Advances in neural information processing systems},
  volume={31},
  year={2018}
}

@article{defazio2019adversarial,
  title={Adversarial model extraction on graph neural networks},
  author={DeFazio, David and Ramesh, Arti},
  journal={arXiv preprint arXiv:1912.07721},
  year={2019}
}

@inproceedings{fan2019graph,
  title={Graph neural networks for social recommendation},
  author={Fan, Wenqi and Ma, Yao and Li, Qing and He, Yuan and Zhao, Eric and Tang, Jiliang and Yin, Dawei},
  booktitle={The world wide web conference},
  pages={417--426},
  year={2019}
}

@inproceedings{wu2022model,
  title={Model extraction attacks on graph neural networks: Taxonomy and realisation},
  author={Wu, Bang and Yang, Xiangwen and Pan, Shirui and Yuan, Xingliang},
  booktitle={Proceedings of the 2022 ACM on Asia Conference on Computer and Communications Security},
  pages={337--350},
  year={2022}
}

@inproceedings{zhao2021watermarking,
  title={Watermarking graph neural networks by random graphs},
  author={Zhao, Xiangyu and Wu, Hanzhou and Zhang, Xinpeng},
  booktitle={2021 9th International Symposium on Digital Forensics and Security (ISDFS)},
  pages={1--6},
  year={2021},
  organization={IEEE}
}

@inproceedings{xu2023watermarking,
  title={Watermarking graph neural networks based on backdoor attacks},
  author={Xu, Jing and Koffas, Stefanos and Ersoy, O{\u{g}}uzhan and Picek, Stjepan},
  booktitle={2023 IEEE 8th European Symposium on Security and Privacy (EuroS\&P)},
  pages={1179--1197},
  year={2023},
  organization={IEEE}
}

@article{waheed2023grove,
  title={GrOVe: Ownership Verification of Graph Neural Networks using Embeddings},
  author={Waheed, Asim and Duddu, Vasisht and Asokan, N},
  journal={arXiv preprint arXiv:2304.08566},
  year={2023}
}

@inproceedings{kipf2017semi,
  title={Semi-Supervised Classification with Graph Convolutional Networks},
  author={Kipf, Thomas N. and Welling, Max},
  booktitle={ICLR},
  year={2017}
}

@article{velickovic2018graph,
  title="{Graph Attention Networks}",
  author={Veli{\v{c}}kovi{\'{c}}, Petar and Cucurull, Guillem and Casanova, Arantxa and Romero, Adriana and Li{\`{o}}, Pietro and Bengio, Yoshua},
  journal={ICLR},
  year={2018},
  url={https://openreview.net/forum?id=rJXMpikCZ},
}

@article{ying2018hierarchical,
  title={Hierarchical graph representation learning with differentiable pooling},
  author={Ying, Zhitao and You, Jiaxuan and Morris, Christopher and Ren, Xiang and Hamilton, Will and Leskovec, Jure},
  journal={Advances in neural information processing systems},
  volume={31},
  year={2018}
}

@inproceedings{zhang2018end,
  title={An end-to-end deep learning architecture for graph classification},
  author={Zhang, Muhan and Cui, Zhicheng and Neumann, Marion and Chen, Yixin},
  booktitle={Proceedings of the AAAI conference on artificial intelligence},
  volume={32},
  year={2018}
}

@inproceedings{tramer2016stealing,
  title={Stealing machine learning models via prediction $\{$APIs$\}$},
  author={Tram{\`e}r, Florian and Zhang, Fan and Juels, Ari and Reiter, Michael K and Ristenpart, Thomas},
  booktitle={25th USENIX security symposium (USENIX Security 16)},
  pages={601--618},
  year={2016}
}

@inproceedings{huang2020gnnvis,
  title={GNNVIs: Visualize large-scale data by learning a graph neural network representation},
  author={Huang, Yajun and Zhang, Jingbin and Yang, Yiyang and Gong, Zhiguo and Hao, Zhifeng},
  booktitle={Proceedings of the 29th ACM International Conference on Information \& Knowledge Management},
  pages={545--554},
  year={2020}
}

@article{zhu2021transfer,
  title={Transfer learning of graph neural networks with ego-graph information maximization},
  author={Zhu, Qi and Yang, Carl and Xu, Yidan and Wang, Haonan and Zhang, Chao and Han, Jiawei},
  journal={Advances in Neural Information Processing Systems},
  volume={34},
  pages={1766--1779},
  year={2021}
}

@article{he2021fedgraphnn,
  title={Fedgraphnn: A federated learning system and benchmark for graph neural networks},
  author={He, Chaoyang and Balasubramanian, Keshav and Ceyani, Emir and Yang, Carl and Xie, Han and Sun, Lichao and He, Lifang and Yang, Liangwei and Yu, Philip S and Rong, Yu and others},
  journal={arXiv preprint arXiv:2104.07145},
  year={2021}
}

@article{hu2019strategies,
  title={Strategies for pre-training graph neural networks},
  author={Hu, Weihua and Liu, Bowen and Gomes, Joseph and Zitnik, Marinka and Liang, Percy and Pande, Vijay and Leskovec, Jure},
  journal={arXiv preprint arXiv:1905.12265},
  year={2019}
}

@inproceedings{song2019auditing,
  title={Auditing data provenance in text-generation models},
  author={Song, Congzheng and Shmatikov, Vitaly},
  booktitle={Proceedings of the 25th ACM SIGKDD International Conference on Knowledge Discovery \& Data Mining},
  pages={196--206},
  year={2019}
}

@inproceedings{wang2019heterogeneous,
  title={Heterogeneous graph attention network},
  author={Wang, Xiao and Ji, Houye and Shi, Chuan and Wang, Bai and Ye, Yanfang and Cui, Peng and Yu, Philip S},
  booktitle={The world wide web conference},
  pages={2022--2032},
  year={2019}
}

@inproceedings{pan2016tri,
  title={Tri-party deep network representation},
  author={Pan, Shirui and Wu, Jia and Zhu, Xingquan and Zhang, Chengqi and Wang, Yang},
  booktitle={International Joint Conference on Artificial Intelligence 2016},
  pages={1895--1901},
  year={2016},
  organization={Association for the Advancement of Artificial Intelligence (AAAI)}
}

@article{sen2008collective,
  title={Collective classification in network data},
  author={Sen, Prithviraj and Namata, Galileo and Bilgic, Mustafa and Getoor, Lise and Galligher, Brian and Eliassi-Rad, Tina},
  journal={AI magazine},
  volume={29},
  number={3},
  pages={93--93},
  year={2008}
}

@inproceedings{giles1998citeseer,
  title={CiteSeer: An automatic citation indexing system},
  author={Giles, C Lee and Bollacker, Kurt D and Lawrence, Steve},
  booktitle={Proceedings of the third ACM conference on Digital libraries},
  pages={89--98},
  year={1998}
}

@inproceedings{mcauley2015image,
  title={Image-based recommendations on styles and substitutes},
  author={McAuley, Julian and Targett, Christopher and Shi, Qinfeng and Van Den Hengel, Anton},
  booktitle={Proceedings of the 38th international ACM SIGIR conference on research and development in information retrieval},
  pages={43--52},
  year={2015}
}

@article{shchur2018pitfalls,
  title={Pitfalls of graph neural network evaluation. arXiv 2018},
  author={Shchur, Oleksandr and Mumme, Maximilian and Bojchevski, Aleksandar and G{\"u}nnemann, Stephan},
  journal={arXiv preprint arXiv:1811.05868},
  year={2018}
}

@inproceedings{he2021stealing,
  title={Stealing links from graph neural networks},
  author={He, Xinlei and Jia, Jinyuan and Backes, Michael and Gong, Neil Zhenqiang and Zhang, Yang},
  booktitle={30th USENIX security symposium (USENIX security 21)},
  pages={2669--2686},
  year={2021}
}

@article{blondel2008fast,
  title={Fast unfolding of communities in large networks},
  author={Blondel, Vincent D and Guillaume, Jean-Loup and Lambiotte, Renaud and Lefebvre, Etienne},
  journal={Journal of statistical mechanics: theory and experiment},
  volume={2008},
  number={10},
  pages={P10008},
  year={2008},
  publisher={IOP Publishing}
}

@article{rustamaji2024community,
  title={Community detection with greedy modularity disassembly strategy},
  author={Rustamaji, Heru Cahya and Kusuma, Wisnu Ananta and Nurdiati, Sri and Batubara, Irmanida},
  journal={Scientific Reports},
  volume={14},
  number={1},
  pages={4694},
  year={2024},
  publisher={Nature Publishing Group UK London}
}

@article{neary2011real,
  title={Real ‘Sybil’Admits Multiple Personalities Were Fake},
  author={Neary, Lynn},
  journal={National Public Radio. NPR},
  volume={20},
  year={2011}
}

@inproceedings{maheswaran2016building,
  title={Building privacy-preserving cryptographic credentials from federated online identities},
  author={Maheswaran, John and Jackowitz, Daniel and Zhai, Ennan and Wolinsky, David Isaac and Ford, Bryan},
  booktitle={Proceedings of the Sixth ACM Conference on Data and Application Security and Privacy},
  pages={3--13},
  year={2016}
}

@article{clauset2004finding,
  title={Finding community structure in very large networks},
  author={Clauset, Aaron and Newman, Mark EJ and Moore, Cristopher},
  journal={Physical review E},
  volume={70},
  number={6},
  pages={066111},
  year={2004},
  publisher={APS}
}

@inproceedings{jagielski2020highfidelity,
author = {Jagielski, Matthew and Carlini, Nicholas and Berthelot, David and Kurakin, Alex and Papernot, Nicolas},
title = {High accuracy and high fidelity extraction of neural networks},
year = {2020},
isbn = {978-1-939133-17-5},
publisher = {USENIX Association},
address = {USA},
booktitle = {Proceedings of the 29th USENIX Conference on Security Symposium},
articleno = {76},
numpages = {18},
series = {SEC'20}
}

@inproceedings{datasetinference2022neurips,
  title = {Dataset Inference for Self-Supervised Models},
  author = {Dziedzic, Adam and Duan, Haonan and Kaleem, Muhammad Ahmad and Dhawan, Nikita and Guan, Jonas and Cattan, Yannis and Boenisch, Franziska and Papernot, Nicolas},
  booktitle = {NeurIPS (Neural Information Processing Systems)},
  year = {2022}
}

@inproceedings{Douceur2002sybilattacks,
  title={The Sybil Attack},
  author={John R. Douceur},
  booktitle={International Workshop on Peer-to-Peer Systems},
  year={2002},
  url={https://www.cs.cornell.edu/people/egs/714-spring05/sybil.pdf}
}

@inproceedings{liu2022stolenencoder,
  title={Stolenencoder: stealing pre-trained encoders in self-supervised learning},
  author={Liu, Yupei and Jia, Jinyuan and Liu, Hongbin and Gong, Neil Zhenqiang},
  booktitle={Proceedings of the 2022 ACM SIGSAC Conference on Computer and Communications Security},
  pages={2115--2128},
  year={2022}
}

@ARTICLE{10049136,
  author={Zhang, Jiliang and Peng, Shuang and Gao, Yansong and Zhang, Zhi and Hong, Qinghui},
  journal={IEEE Transactions on Information Forensics and Security}, 
  title={APMSA: Adversarial Perturbation Against Model Stealing Attacks}, 
  year={2023},
  volume={18},
  number={},
  pages={1667-1679},
  doi={10.1109/TIFS.2023.3246766}}

@inproceedings{ijcai2021p214,
  title     = {Masked Label Prediction: Unified Message Passing Model for Semi-Supervised Classification},
  author    = {Shi, Yunsheng and Huang, Zhengjie and Feng, Shikun and Zhong, Hui and Wang, Wenjing and Sun, Yu},
  booktitle = {Proceedings of the Thirtieth International Joint Conference on
               Artificial Intelligence, {IJCAI-21}},
  publisher = {International Joint Conferences on Artificial Intelligence Organization},
  editor    = {Zhi-Hua Zhou},
  pages     = {1548--1554},
  year      = {2021},
  month     = {8},
  note      = {Main Track},
  doi       = {10.24963/ijcai.2021/214},
  url       = {https://doi.org/10.24963/ijcai.2021/214},
}

@article{guan2024large,
  title={Large Language Models Merging for Enhancing the Link Stealing Attack on Graph Neural Networks},
  author={Guan, Faqian and Zhu, Tianqing and Chang, Wenhan and Ren, Wei and Zhou, Wanlei},
  journal={arXiv preprint arXiv:2412.05830},
  year={2024}
}

@inproceedings{zhang2024linkthief,
  title={LinkThief: Combining Generalized Structure Knowledge with Node Similarity for Link Stealing Attack against GNN},
  author={Zhang, Yuxing and Meng, Siyuan and Chen, Chunchun and Peng, Mengyao and Gu, Hongyan and Huang, Xinli},
  booktitle={Proceedings of the 32nd ACM International Conference on Multimedia},
  pages={4947--4956},
  year={2024}
}

@inproceedings{lin2024stealingtraininggraphsgraph,
      title={Stealing Training Graphs from Graph Neural Networks}, 
      author={Minhua Lin and Enyan Dai and Junjie Xu and Jinyuan Jia and Xiang Zhang and Suhang Wang},
      booktitle={Proceedings of the 31st ACM SIGKDD Conference on Knowledge Discovery and Data Mining},
      year={2025}
}

@INPROCEEDINGS{gridlou,
      author = { Lou, Jiadong and Yuan, Xu and Zhang, Rui and Yuan, Xingliang and Gong , Neil and Tzeng , Nian-Feng },
      booktitle = { 2025 IEEE Symposium on Security and Privacy (SP) },
      title = { GRID: Protecting Training Graph from Link Stealing Attacks on GNN Models },
      year = {2025},
      ISSN = {2375-1207},
      pages = {59-59},
      doi = {10.1109/SP61157.2025.00059},
      publisher = {IEEE Computer Society},
      address = {Los Alamitos, CA, USA},
}

@INPROCEEDINGS {9157021,
      author = { Kariyappa, Sanjay and Qureshi, Moinuddin K. },
      booktitle = { 2020 IEEE/CVF Conference on Computer Vision and Pattern Recognition (CVPR) },
      title = {{ Defending Against Model Stealing Attacks With Adaptive Misinformation }},
      year = {2020},
      pages = {767-775},
      doi = {10.1109/CVPR42600.2020.00085},
      url = {https://doi.ieeecomputersociety.org/10.1109/CVPR42600.2020.00085},
      publisher = {IEEE Computer Society},
      address = {Los Alamitos, CA, USA},
      month =Jun
}

@incollection{steinhaeuser2008community,
  title={Community detection in a large real-world social network},
  author={Steinhaeuser, Karsten and Chawla, Nitesh V},
  booktitle={Social computing, behavioral modeling, and prediction},
  pages={168--175},
  year={2008},
  publisher={Springer}
}

@inproceedings{chintalapudi2015survey,
  title={A survey on community detection algorithms in large scale real world networks},
  author={Chintalapudi, S Rao and Prasad, MHM Krishna},
  booktitle={2015 2nd international conference on computing for sustainable global development (INDIACom)},
  pages={1323--1327},
  year={2015},
  organization={IEEE}
}

@inproceedings{ciglan2013community,
  title={On community detection in real-world networks and the importance of degree assortativity},
  author={Ciglan, Marek and Laclav{\'\i}k, Michal and N{\o}rv{\aa}g, Kjetil},
  booktitle={Proceedings of the 19th ACM SIGKDD international conference on Knowledge discovery and data mining},
  pages={1007--1015},
  year={2013}
}

@misc{facebook_ad_relevance,
  author       = {Facebook},
  title        = {About looklike audiences},
  howpublished = {\url{https://www.facebook.com/business/help/164749007013531}},
}

@inproceedings{dou2020enhancing,
  title={Enhancing graph neural network-based fraud detectors against camouflaged fraudsters},
  author={Dou, Yingtong and Liu, Zhiwei and Sun, Li and Deng, Yutong and Peng, Hao and Yu, Philip S},
  booktitle={Proceedings of the 29th ACM international conference on information \& knowledge management},
  pages={315--324},
  year={2020}
}

\crefalias{section}{appendix}
\appendix
\section{Broader Impacts}
\label{app:broader_impacts}
Our research aims to actively defend graph neural networks against various model-stealing attacks. The primary positive social impact of our work is protecting the intellectual property of organizations and researchers who develop GNN models, which contributes to enhancing the fairness of the ML community and society. 
One potentially negative impact of our work could be the degradation of performance on downstream tasks. However, our experimental results indicate that our defense can still maintain the downstream task performance, therefore mitigating this concern.

\section{Ethics Considerations}
\label{sec:ethics_considerations}

There is no human subjects involved in this research, and no personal data or identifiable information was collected or processed. 
The aim of our method is to enhance the security of valuable GNN models by defending against model stealing attacks, aligning with ethical objectives of protecting intellectual property and promoting responsible usage of machine learning. 
The effectiveness of our defense mechanism has been evaluated through comprehensive experiments. 
To further ensure ethical compliance, we have adhered to principles of transparency and fairness throughout the research process. 
All experiments were conducted using publicly available open datasets, models, and open-source frameworks, ensuring transparency, accessibility, and reproducibility. 

\section{Hyperparameter Configuration}
\label{app:hyperparamters}

Here, we summarize the hyperparameters used for training target and surrogate models. And we explore the impact of the number of communities $K$ on query diversity estimation. 
What's more, the goal of the penalty design in our defense is that we add a low penalty to the model outputs for the target downstream tasks, while a high penalty to those of the attackers. To achieve this goal, we need to calibrate the penalty functions as described in \Cref{sub:penalty_design} so that the value of the \Cref{equ:cost_label_stealing} and \Cref{equ:calibration}is low for low-diversity query and high for high-diversity query. 

\subsection{Hyperparameter of Target/Surrogate models}
\label{app:hyperparameter_models}
The default hyperparameters used for training target and surrogate models are presented in \Cref{Table:parameter_setting_target} and \Cref{Table:parameter_setting_surrogate}, respectively. 

\begin{table}[!h]
\centering
\scriptsize
\begin{threeparttable}
 \caption{Default hyperparameter setting for target model training.}
\begin{tabular}{lll}
 \hline
 Type & Hyperparameter & Setting\\
 \hline
 \hline
\multirow{3}{*}{GAT} & Architecture & 3 layers \\
 & Hidden unit size & 256 \\
 & \# Heads & 4 \\
 \hline
 \multirow{2}{*}{GIN} & Architecture & 3 layers \\
 & Hidden unit size & 256 \\
 \hline
 \multirow{2}{*}{GraphSAGE} & Architecture & 3 layers \\
  & Hidden unit size & 256 \\
 \hline
 \multirow{3}{*}{Graph Transformer} & Architecture & 3 layers \\
  & Hidden unit size & 256 \\
  & \# Heads & 4 \\
  \hline
  \hline
\multirow{4}{*}{Training} & Learning rate & $0.001$ \\
 & Optimizer & Adam\\
 & Epochs & 200\\
 & Batch size & 32\\
 \hline
\end{tabular}
\label{Table:parameter_setting_target}
\end{threeparttable}
\end{table}

\begin{table}[!h]

\centering
\scriptsize
\begin{threeparttable}
 \caption{Default hyperparameter setting for surrogate model training. BE: Backbone Encoder, CH: Classification Head (optional), GT: Graph Transformer.}
\begin{tabular}{c|lll} 
 \hline
 & Type & Hyperparameter & Setting\\
 \hline
 \hline
 \multirow{10}{*}{\parbox{0.5cm}{BE}} & \multirow{3}{*}{\parbox{1.4cm}{GAT}} & Architecture & 2 layers \\
 & & Hidden unit size & 256 \\
 & & \#Heads & 4 \\
 \cline{2-4}
 & \multirow{2}{*}{\parbox{1.4cm}{GIN}} & Architecture & 2 layers \\
 & & Hidden unit size & 256 \\
 \cline{2-4}
 & \multirow{2}{*}{\parbox{1.4cm}{GraphSAGE}} & Architecture & 2 layers \\
 & & Hidden unit size & 256 \\
 \cline{2-4}
 & \multirow{3}{*}{\parbox{1.4cm}{GT}} & Architecture & 2 layers \\
 & & Hidden unit size & 256 \\
 & & \# Heads & 4 \\
 \hline
 \multirow{2}{*}{\parbox{0.5cm}{*CH}} & \multirow{2}{*}{\parbox{1.4cm}{MLP}} & Architecture & 2 layers \\
 & & Hidden unit size & 100 \\
 \hline
 \hline
& \multirow{4}{*}{\parbox{1.4cm}{Training}} & Learning rate & $0.001$ \\
& & Optimizer & Adam\\
& & Epochs & 200 (BE), 300 (CH)\\
& & Batch size & 32\\
 \hline
\end{tabular}
\label{Table:parameter_setting_surrogate}
\end{threeparttable}
\end{table}

\begin{table}[htpb]
     \centering
     \scriptsize
     \caption{\textbf{Setting of $K$ and $\beta$ for different datasets.}}
     \label{table:hyperparameter}
    \begin{tabular}{ccccccc} 
     \hline
     Dataset & ACM & DBLP & Pubmed & Citeseer & Amazon & Coauthor\\
     \hline
     $K$ & $300$ & $150$ & $300$ & $250$ & $150$ & $300$\\
     $\beta$ & $40$ & $90$ & $90$ & $70$ & $80$ & $90$\\
     \hline
    \end{tabular}
\end{table}

\begin{figure*}[!h]
    \centering
    \begin{subfigure}[t]{0.45\textwidth}
         \centering
         \includegraphics[width=\textwidth]{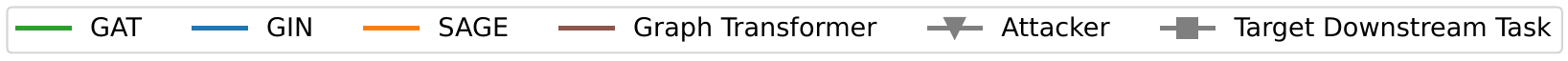}
     \end{subfigure}
     \vfill
    \begin{subfigure}[b]{0.2\textwidth}
        \centering
        \includegraphics[width=\textwidth]{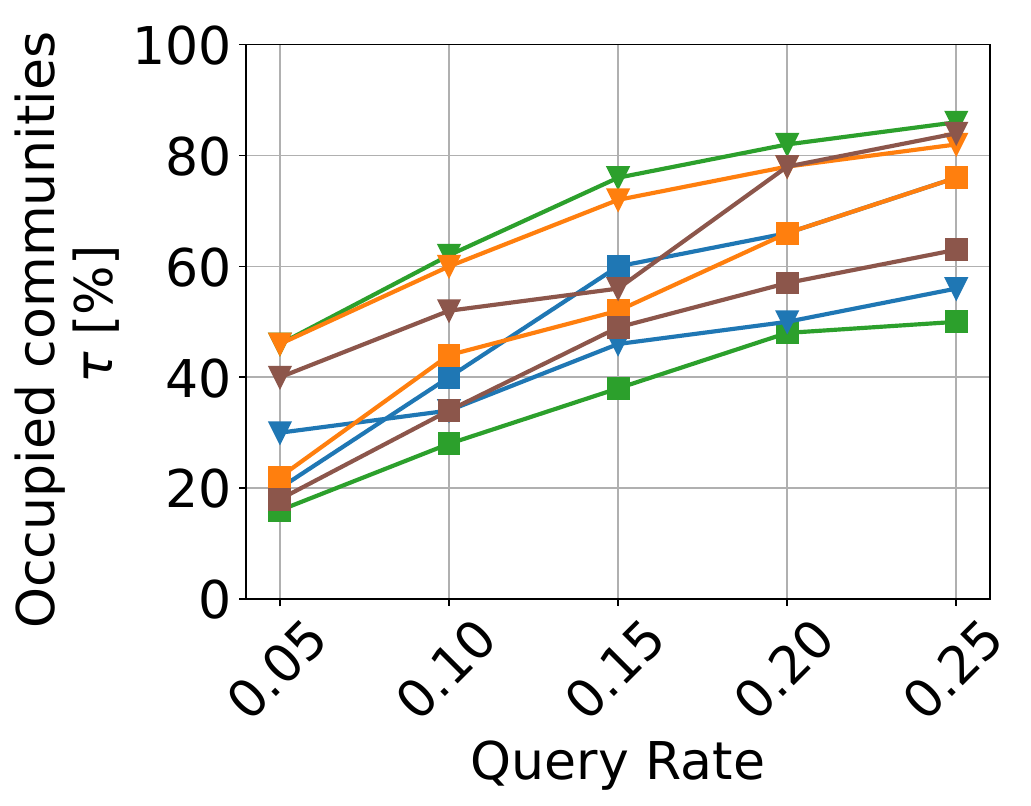}
        \caption{50 communities}
    \end{subfigure}
    \begin{subfigure}[b]{0.2\textwidth}
        \centering
        \includegraphics[width=\textwidth]{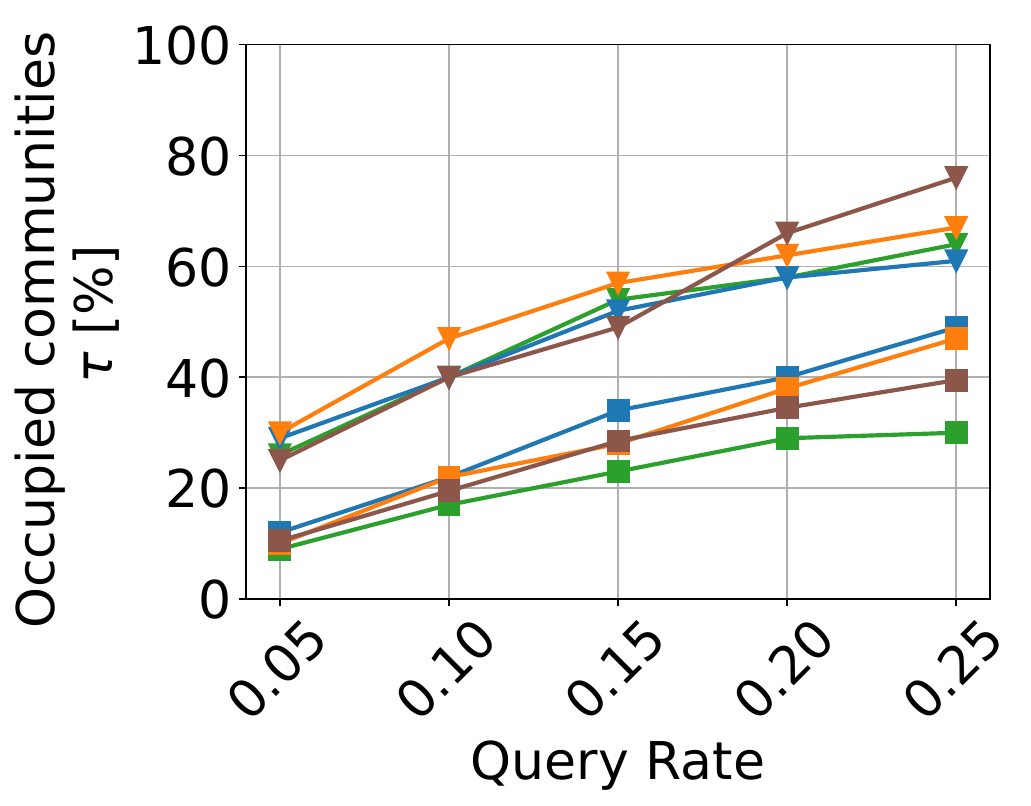}
        \caption{100 communities}
    \end{subfigure}
    \begin{subfigure}[b]{0.2\textwidth}
        \centering
        \includegraphics[width=\textwidth]{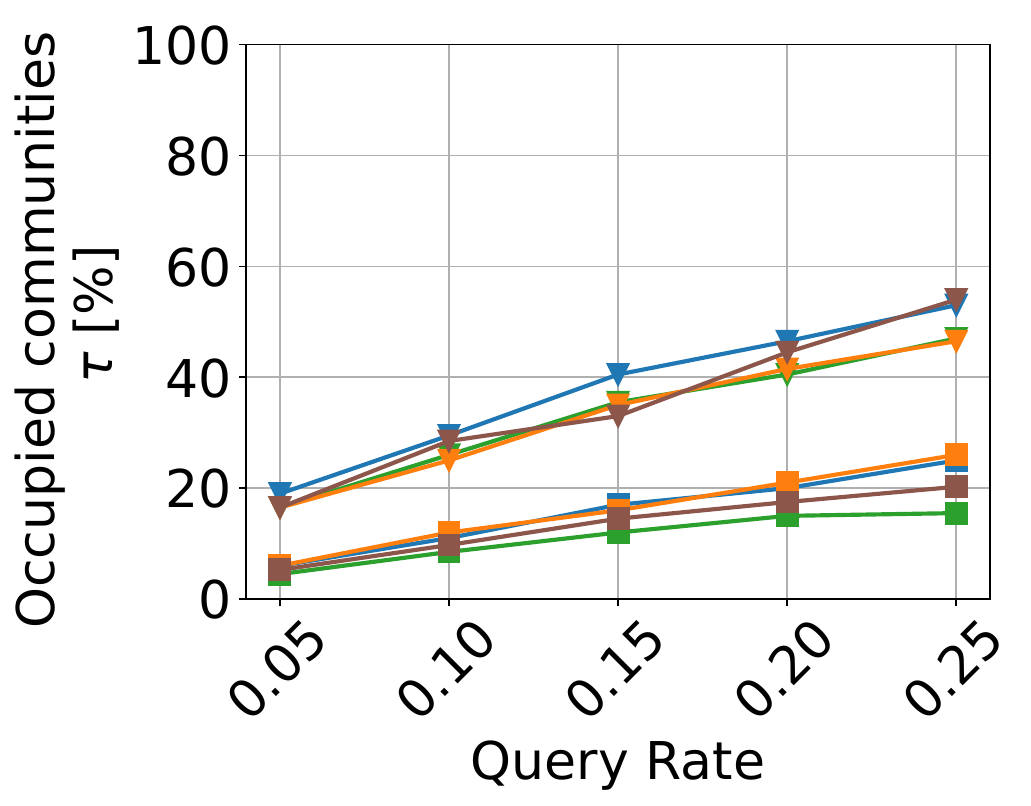}
        \caption{200 communities}
    \end{subfigure}
    \begin{subfigure}[b]{0.2\textwidth}
        \centering
        \includegraphics[width=\textwidth]{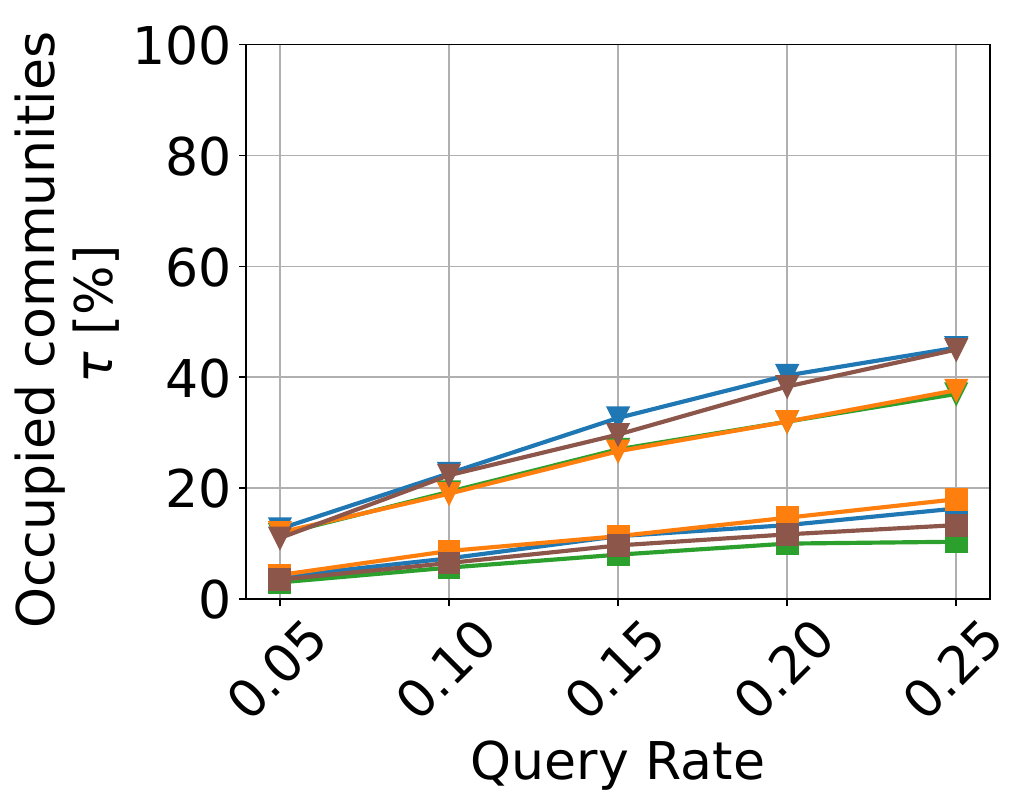}
        \caption{300 communities}
    \end{subfigure}
    \caption{\textbf{Query diversity between the attacker and a target downstream task with different $K$ (ACM).} Generally, with $K=300$, the relative difference is the largest for all three models.}
    \label{fig:different_k_acm}
\end{figure*}

\subsection{Impact of the Number of Communities $K$}
\label{app:different_k}
We experiment with different numbers of communities.
We aim to optimize the value of $K$ for each dataset such that we obtain the largest relative difference in query diversity between attackers and target downstream tasks. As explained in \Cref{sub:query_diversity}, the query diversity can be quantified as fractions of occupied communities.
For example, \Cref{fig:different_k_acm} shows the query diversity for attackers and target downstream tasks with different $K$'s on the ACM dataset. The largest relative difference is obtained with $K=300$. 
Yet, for all other values of $K$ alike, there is a significant difference between the curve for target downstream tasks and attackers. This highlights that under all these different setups for $K$, we are still able to distinguish between the query diversity of these two, which means the effectiveness of our approach is not significantly affected by the choice of $K$. 
The final chosen values for $K$ over all datasets' results of this paper are shown in \Cref{table:hyperparameter}.

\begin{figure}[!h]
    \centering
    \includegraphics[width=0.2\textwidth]{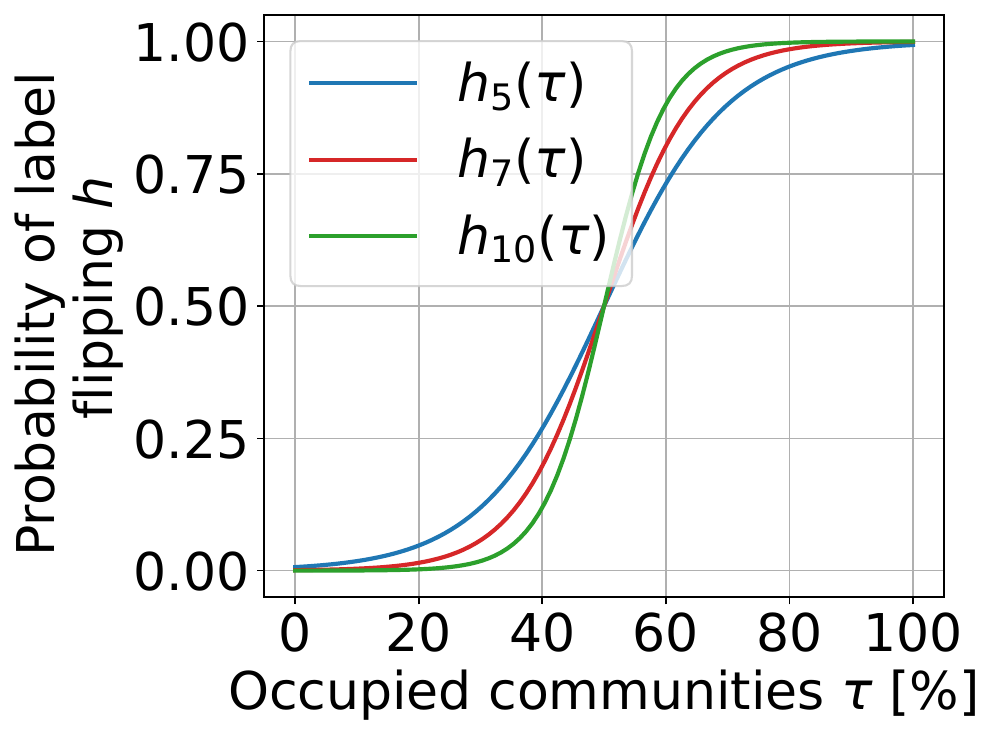}
    \captionof{figure}{\textbf{Calibration function of label flipping with different $\eta$ values (ACM).}}
    \label{fig:cost_label_stealing}
\end{figure}

\subsection{Hyperparameter $\eta$ in \Cref{equ:cost_label_stealing}}
The label flipping probability for the A) attack setup is returned by the calibration function in \Cref{equ:cost_label_stealing}, and the behavior of this function is controlled by the hyperparameter $\eta$ which specifies the level of squeezing the curve. 
With a larger $\eta$, we can obtain lower penalties for a small fraction of occupied communities and higher penalties for large fractions. 
The calibration functions with different $\eta$ values are shown in \Cref{fig:cost_label_stealing} for the ACM dataset. 
We can indeed observe that with a larger $\eta$ (e.g., 10), $h(\tau)$ can output a smaller value for a low fraction of occupied communities and a larger value for a high percentage of occupied communities. Thus, we set $\eta=10$.

\begin{figure}[!h]
    \centering
    \includegraphics[width=0.3\textwidth]{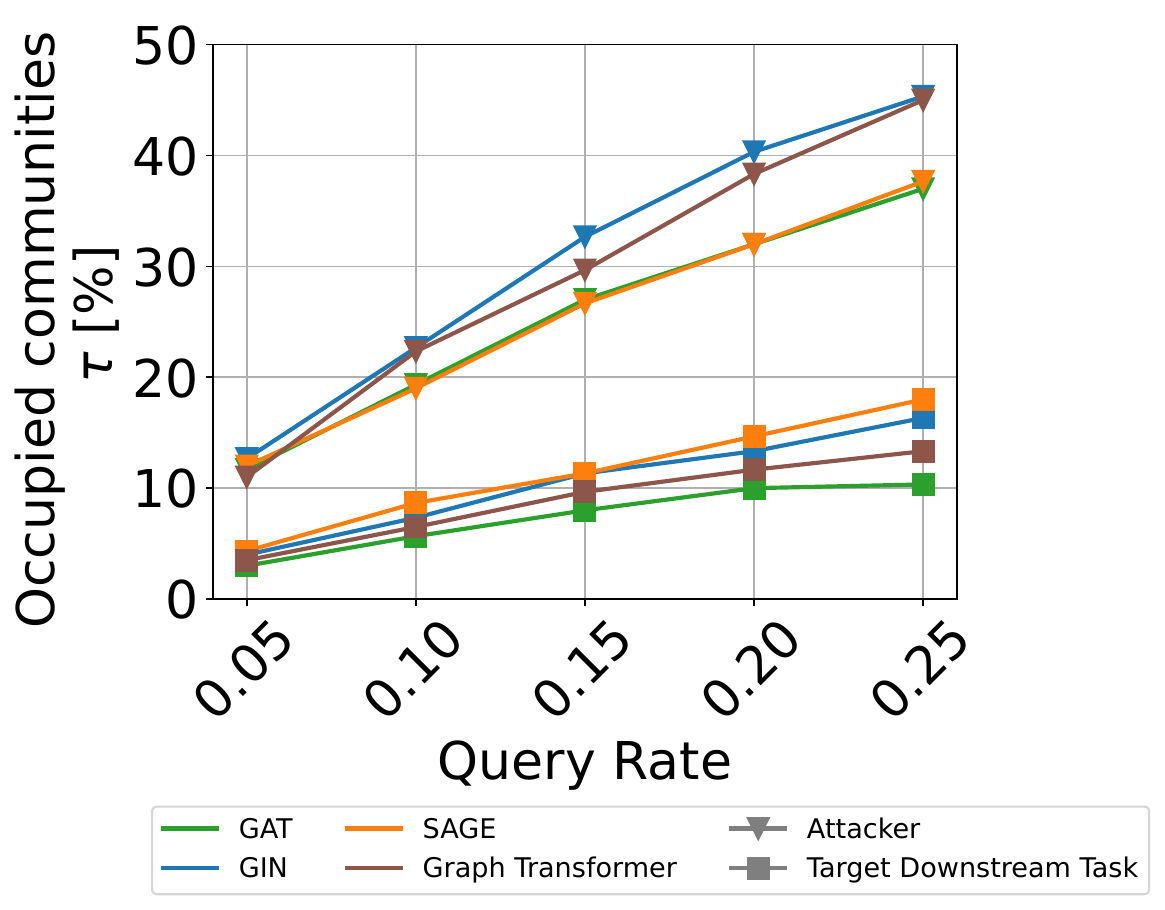}
    \captionof{figure}{\textbf{An example of query diversity of target downstream tasks and attackers (ACM, $K=300$).}}
    \label{fig:query_diversity_example}
\end{figure}

\subsection{Hyperparameters in \Cref{equ:calibration}}

For the penalty function in \Cref{equ:calibration}, hyperparameters $\beta, \alpha, \lambda$ should be calibrated. $\beta$ specifies how many occupied communities are considered safe and normal information leakage for target downstream tasks. Once the percentage of occupied communities is close or reaches $\beta$, a high penalty is necessary to be added to the model output to prevent the model from being stolen. 
Thus, we first present the query diversity of a target downstream task and attacker in \Cref{fig:query_diversity_example} and then, according to the percentages of occupied communities for attackers and target downstream tasks, $\beta$ is set as $40$ for ACM. The setting of $\beta$ for all datasets is shown in \Cref{table:hyperparameter}.

\begin{figure}
    \centering
\begin{subfigure}{0.2\textwidth}
    \includegraphics[width=\textwidth]{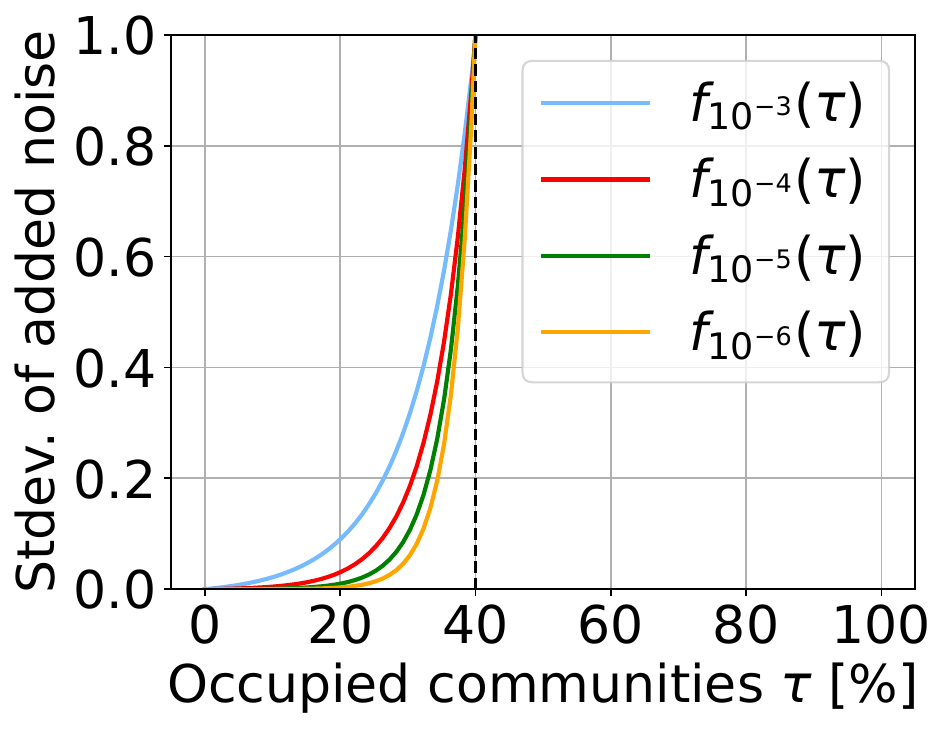}
    \caption{with different $\lambda$ values.}
    \label{fig:lambda_calibration_acm}
\end{subfigure}
\begin{subfigure}{0.2\textwidth}
    \includegraphics[width=\textwidth]{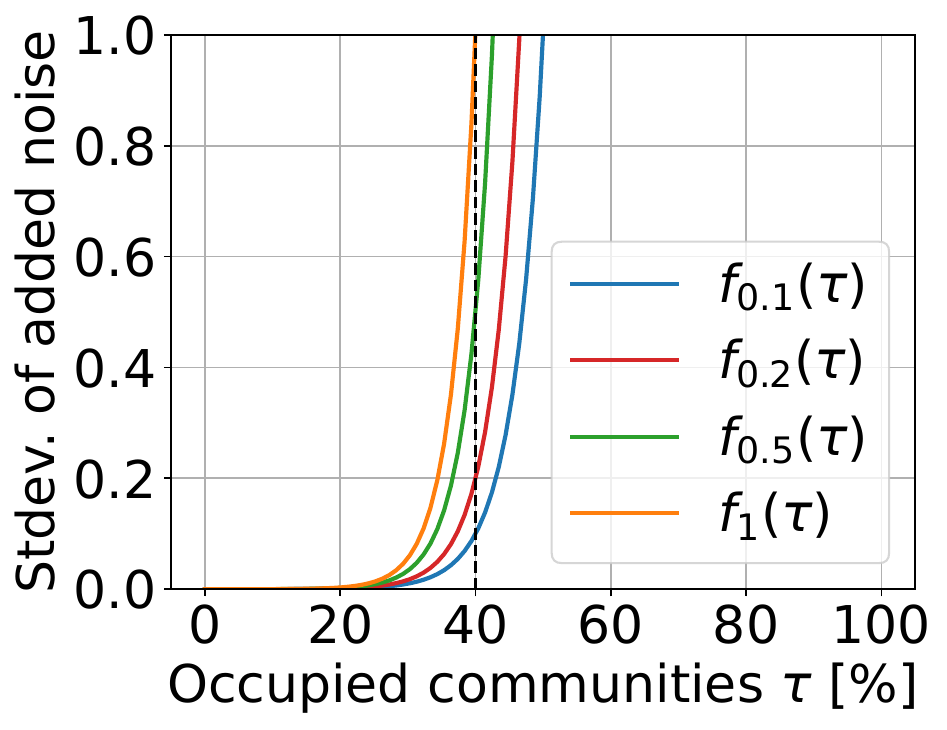}
    \caption{with different $\alpha$ values.}
    \label{fig:alpha_calibration_acm}
\end{subfigure}
\caption{\textbf{Calibration function for different hyperparameters (ACM).}}
\label{fig:defense_calibration}
\end{figure}

Hyperparameter $\lambda$ compresses the curve of the penalty function. As we can see from \Cref{fig:lambda_calibration_acm}, when the query diversity arrives $20\%$ which is the diversity level for the target downstream task (\Cref{fig:query_diversity_example}), the standard deviation of the added noise is decreasing with reduce of $\lambda$. Thus, we set $\lambda=10^{-6}$ to obtain a low $\sigma$ value for the target downstream task.

Hyperparameter $\alpha$ controls the level of penalty (\ie $\sigma$) once the information leakage specified by $\beta$ is reached. The calibration functions with different $\alpha$ values are shown in \Cref{fig:alpha_calibration_acm} for ACM dataset. As we can observe, when we set $\alpha=1$, the standard deviation of Gaussian noise can be maximized when the percentage of occupied communities reaches the pre-defined percentage of occupied communities, \ie $\beta$.

\section{Additional Experiments}
\label{app:experiments}

\subsection{Stealing Performance with/without \ours}
\label{app:adversary_performance}
The stealing performance under three attack setups, with and without applying \ours, on other datasets is illustrated in \Cref{fig:adversary_performance_dblp} to \Cref{fig:adversary_performance_coauthor}. Overall, after applying \ours, the stealing performance under all attack setups degrades dramatically, \ie below $40\%$ surrogate accuracy in most cases. 
The detailed stealing performance on these datasets is presented in \Cref{table:results_with_without_defenses_dblp} to \Cref{table:results_with_without_defenses_coauthor}. 
In general, our defense can significantly degrade the stealing performance while maintaining the performance of the downstream tasks. 

\begin{figure*}[htpb]
\centering
    \begin{subfigure}[t]{0.5\textwidth}
         \centering
         \includegraphics[width=\textwidth]{figures/adversary_performance_new/legend_with_transformer.pdf}
     \end{subfigure}
     \vfill
     \begin{subfigure}[b]{0.32\textwidth}
         \centering
         \includegraphics[width=\textwidth]{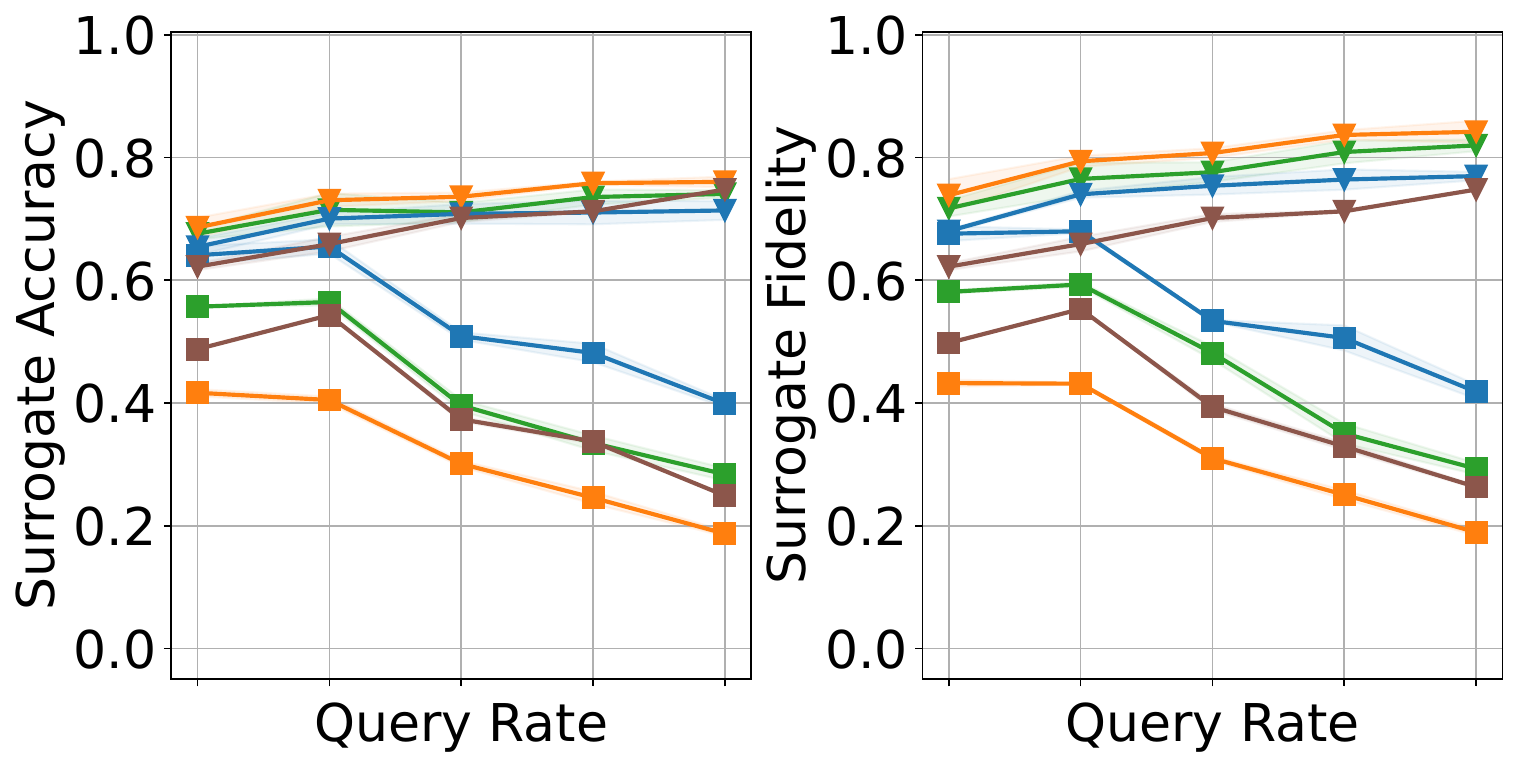}
         \caption{Probabilities.}
     \end{subfigure}
     \begin{subfigure}[b]{0.32\textwidth}
         \centering
         \includegraphics[width=\textwidth]{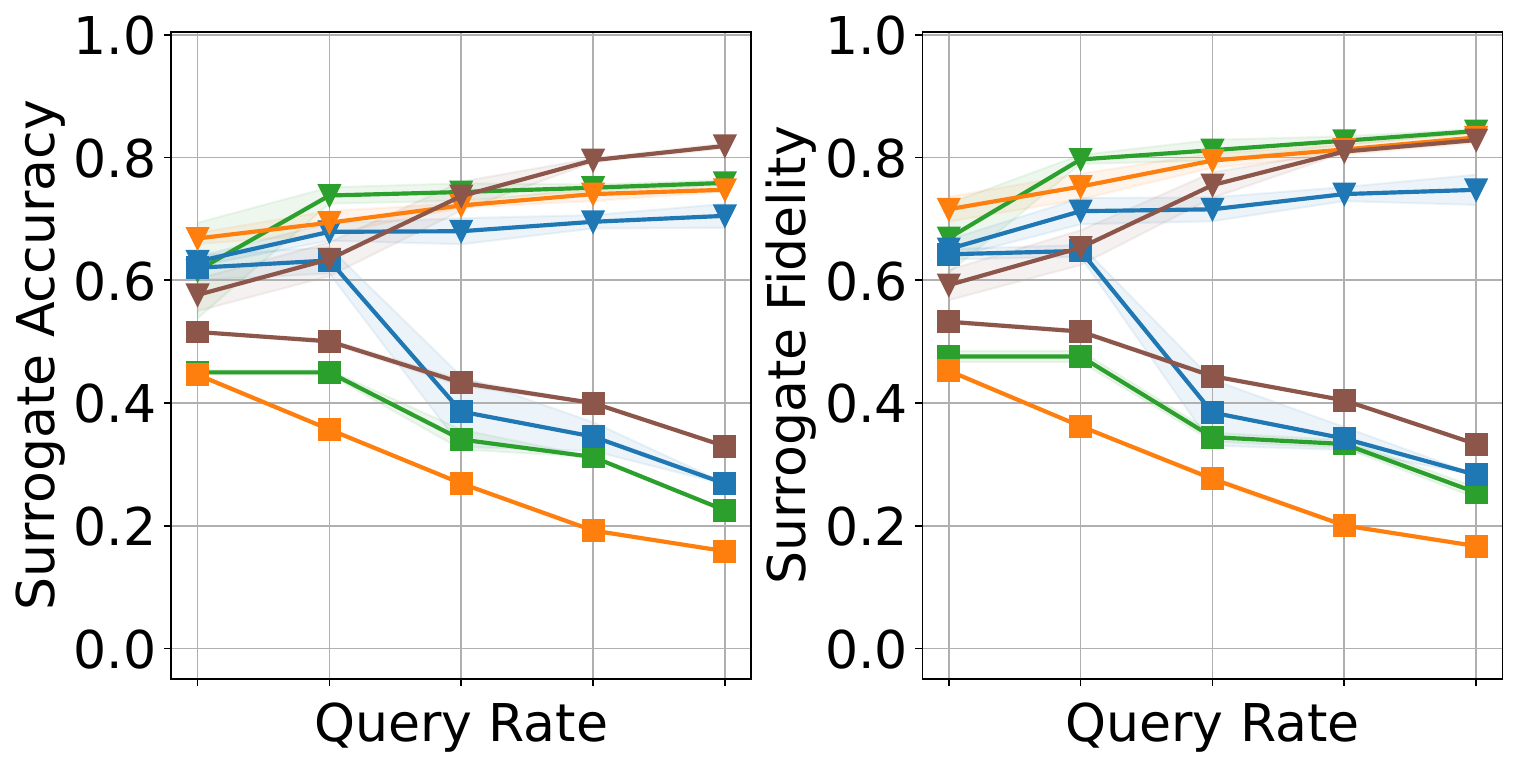}
         \caption{Embeddings.}
     \end{subfigure}
     \begin{subfigure}[b]{0.32\textwidth}
         \centering
         \includegraphics[width=\textwidth]{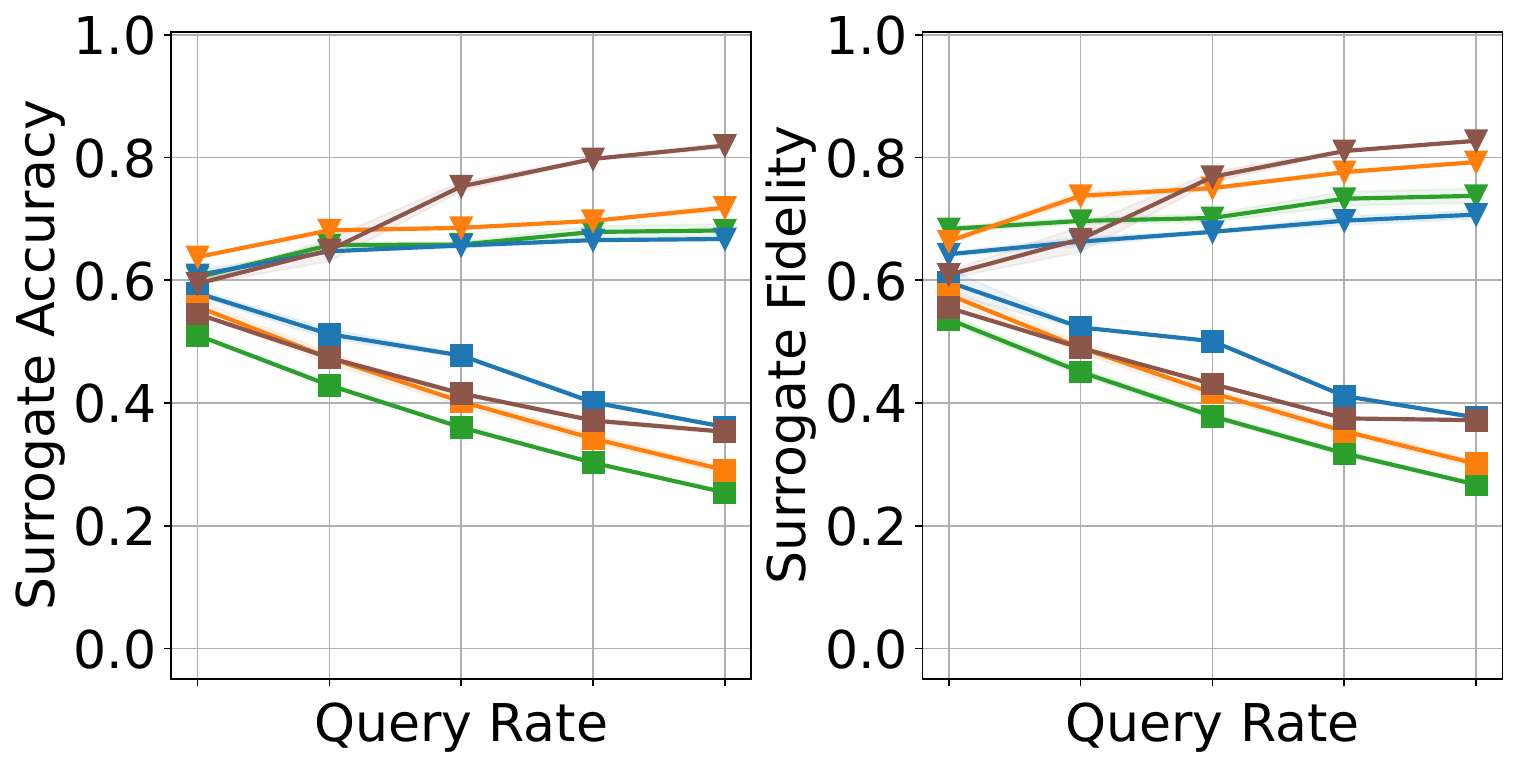}
         \caption{Projections.}
     \end{subfigure}
\caption{\textbf{Performance of the surrogate model with and without our defense (DBLP dataset).}}
\label{fig:adversary_performance_dblp}
\end{figure*}

\begin{figure*}[htpb]
\centering
    \begin{subfigure}[t]{0.5\textwidth}
         \centering
         \includegraphics[width=\textwidth]{figures/adversary_performance_new/legend_with_transformer.pdf}
     \end{subfigure}
     \vfill
     \begin{subfigure}[b]{0.32\textwidth}
         \centering
         \includegraphics[width=\textwidth]{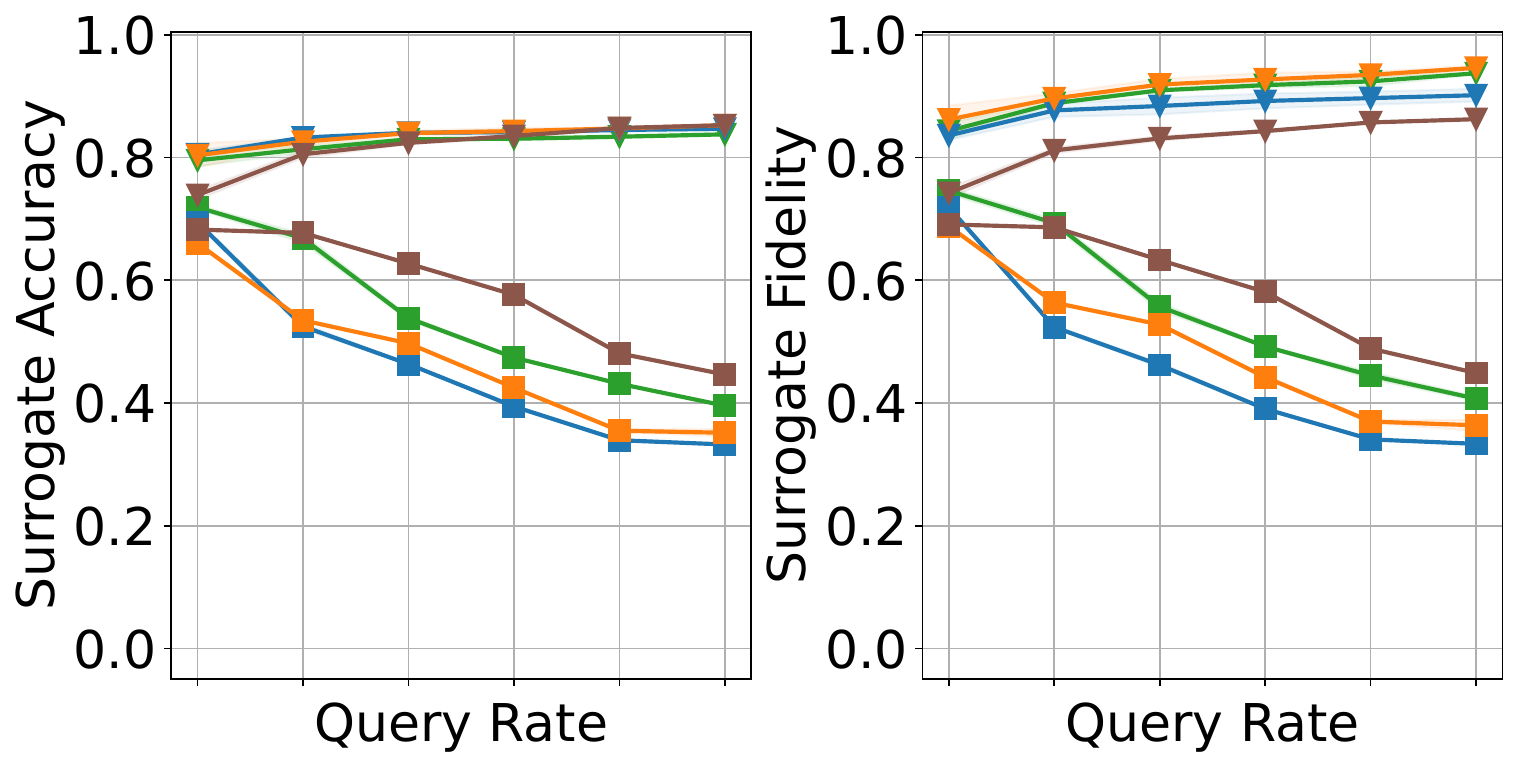}
         \caption{Probabilities.}
     \end{subfigure}
     \begin{subfigure}[b]{0.32\textwidth}
         \centering
         \includegraphics[width=\textwidth]{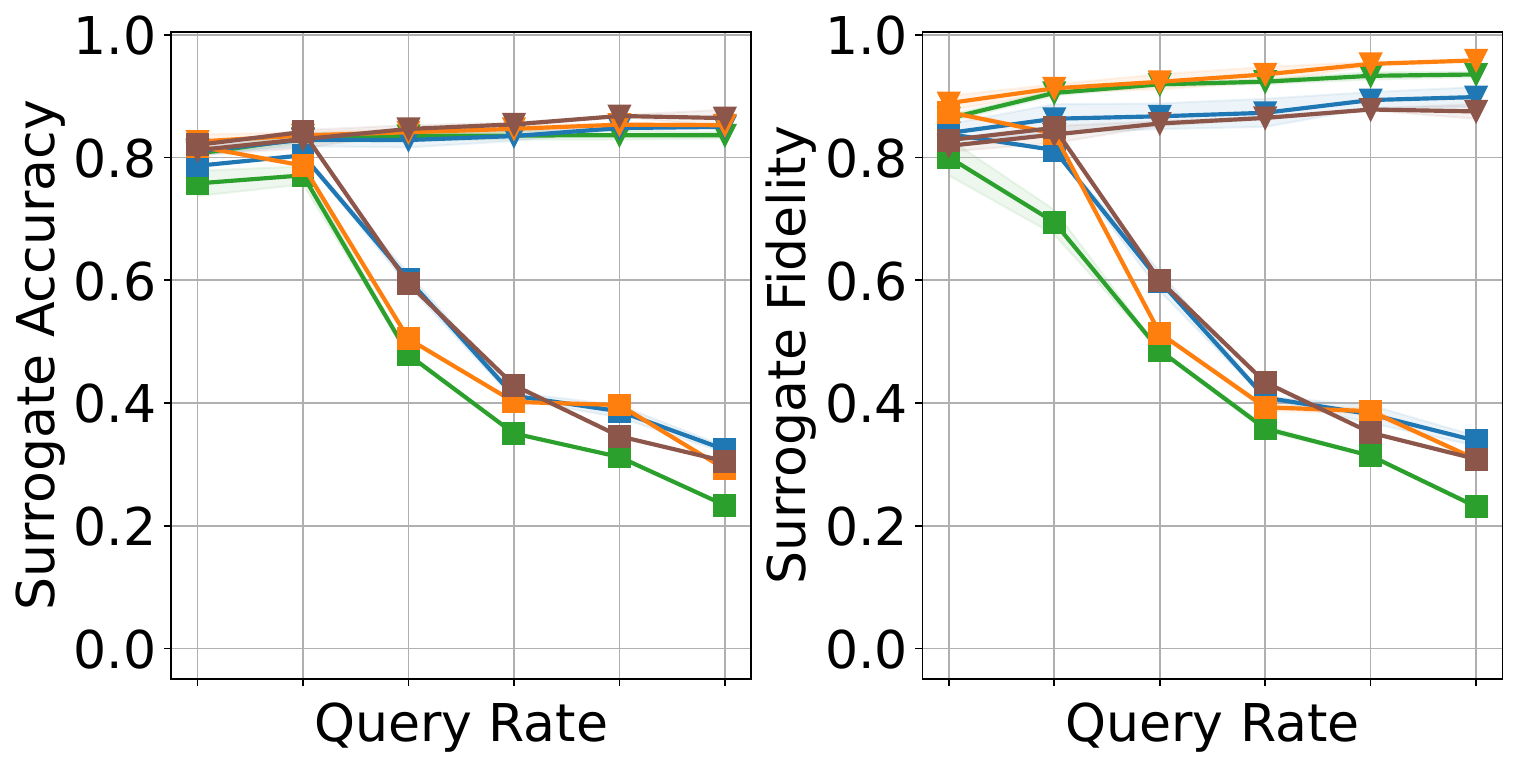}
         \caption{Embeddings.}
     \end{subfigure}
     \begin{subfigure}[b]{0.32\textwidth}
         \centering
         \includegraphics[width=\textwidth]{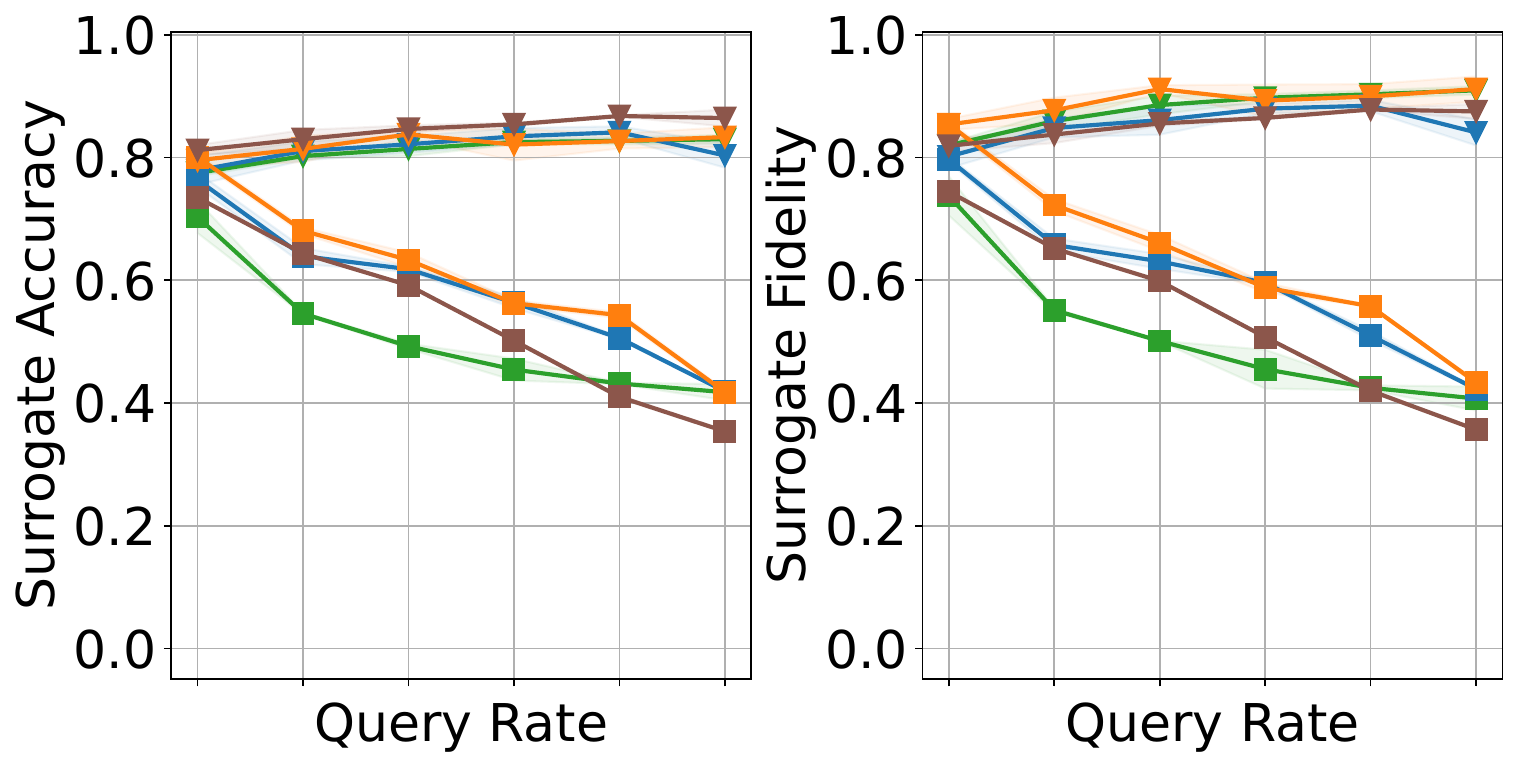}
         \caption{Projections.}
     \end{subfigure}
\caption{\textbf{Performance of the surrogate model with and without our defense (Pubmed dataset).}}
\label{fig:adversary_performance_pubmed}
\end{figure*}

\begin{figure*}[htpb]
\centering
    \begin{subfigure}[t]{0.5\textwidth}
         \centering
         \includegraphics[width=\textwidth]{figures/adversary_performance_new/legend_with_transformer.pdf}
     \end{subfigure}
     \vfill
     \begin{subfigure}[b]{0.32\textwidth}
         \centering
         \includegraphics[width=\textwidth]{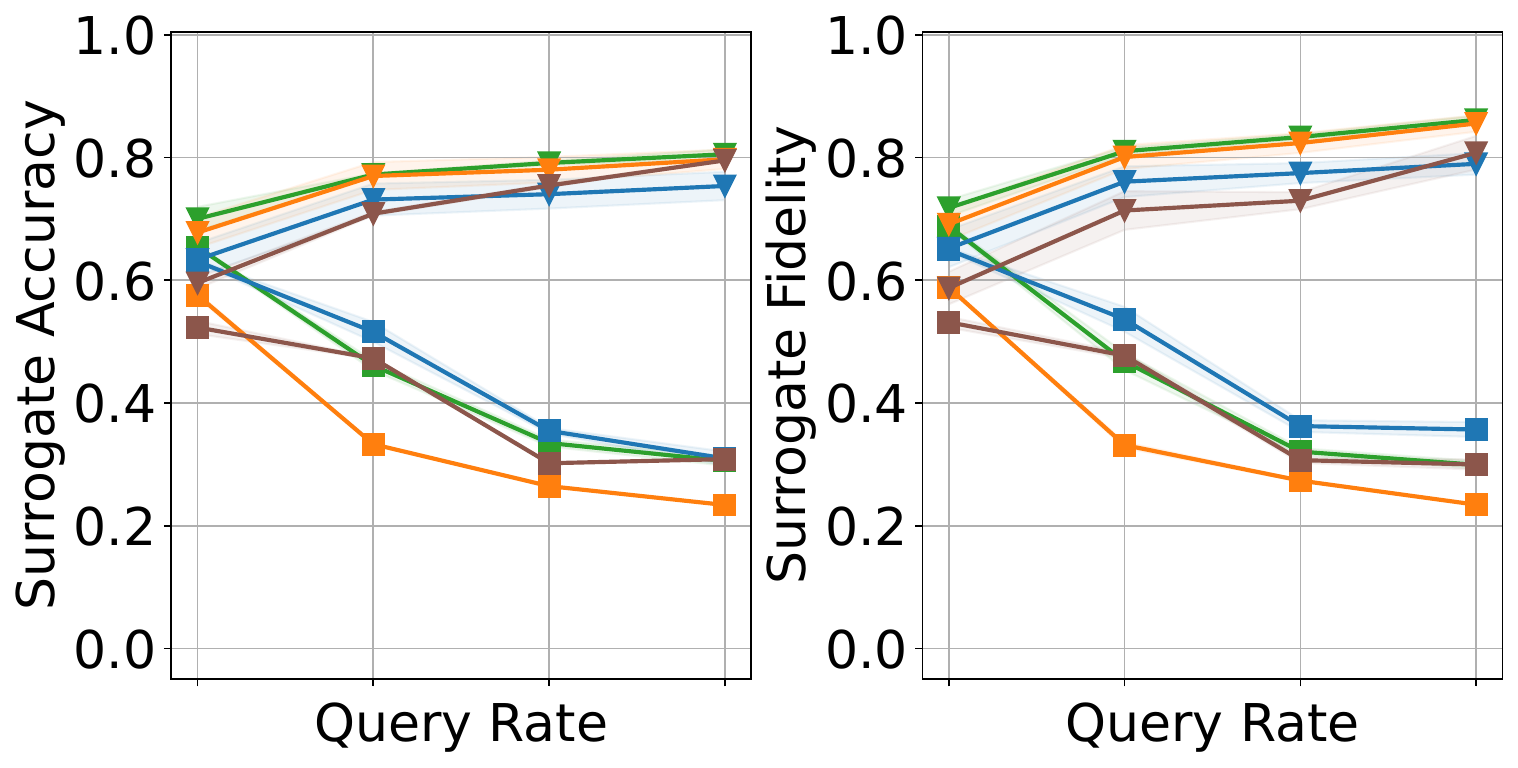}
         \caption{Probabilities.}
     \end{subfigure}
     \begin{subfigure}[b]{0.32\textwidth}
         \centering
         \includegraphics[width=\textwidth]{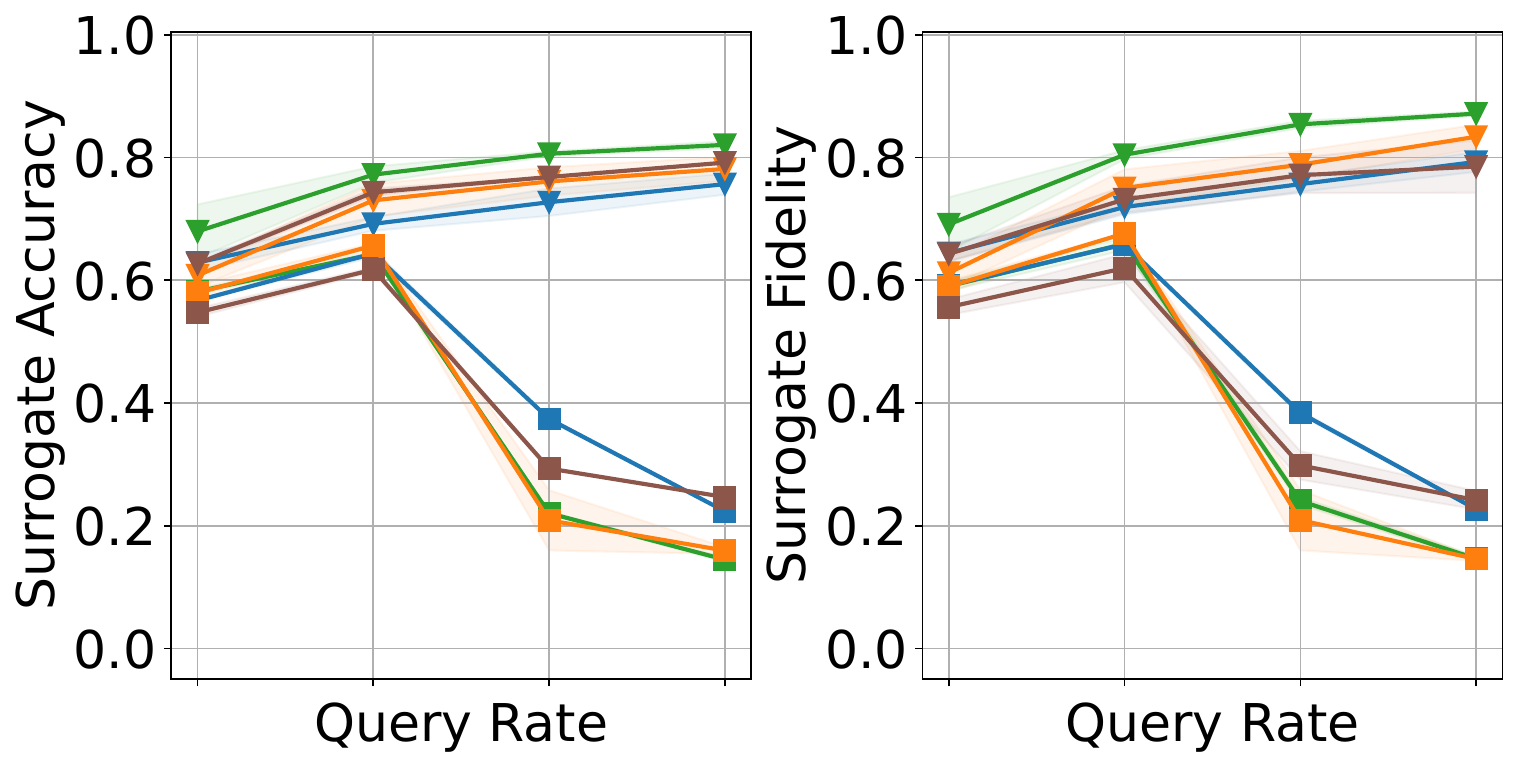}
         \caption{Embeddings.}
     \end{subfigure}
     \begin{subfigure}[b]{0.32\textwidth}
         \centering
         \includegraphics[width=\textwidth]{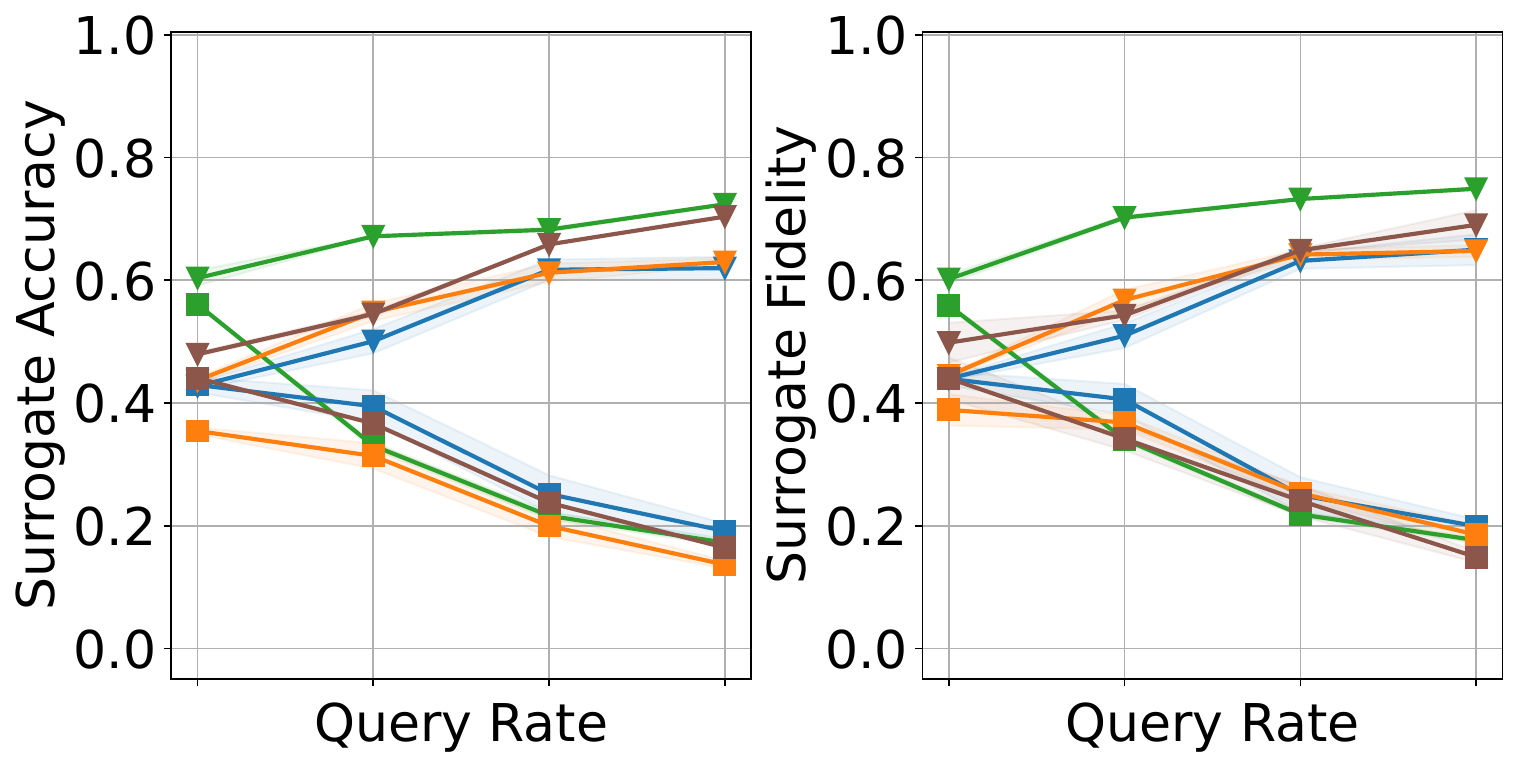}
         \caption{Projections.}
     \end{subfigure}
\caption{\textbf{Performance of the surrogate model with and without our defense (Citeseer dataset).}}
\label{fig:adversary_performance_citeseer}
\end{figure*}

\begin{figure*}[htpb]
\centering
    \begin{subfigure}[t]{0.5\textwidth}
         \centering
         \includegraphics[width=\textwidth]{figures/adversary_performance_new/legend_with_transformer.pdf}
     \end{subfigure}
     \vfill
     \begin{subfigure}[b]{0.32\textwidth}
         \centering
         \includegraphics[width=\textwidth]{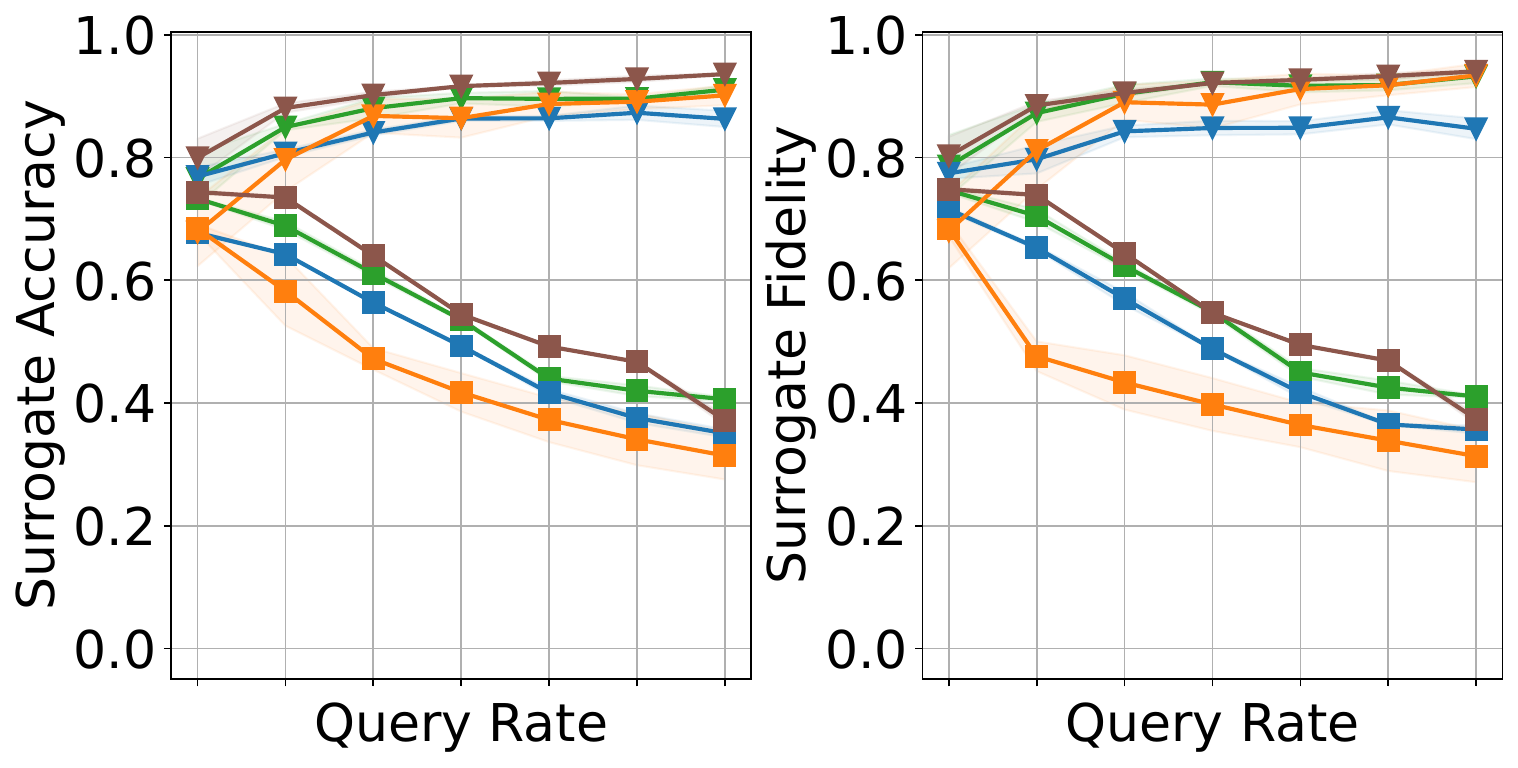}
         \caption{Probabilities.}
     \end{subfigure}
     \begin{subfigure}[b]{0.32\textwidth}
         \centering
         \includegraphics[width=\textwidth]{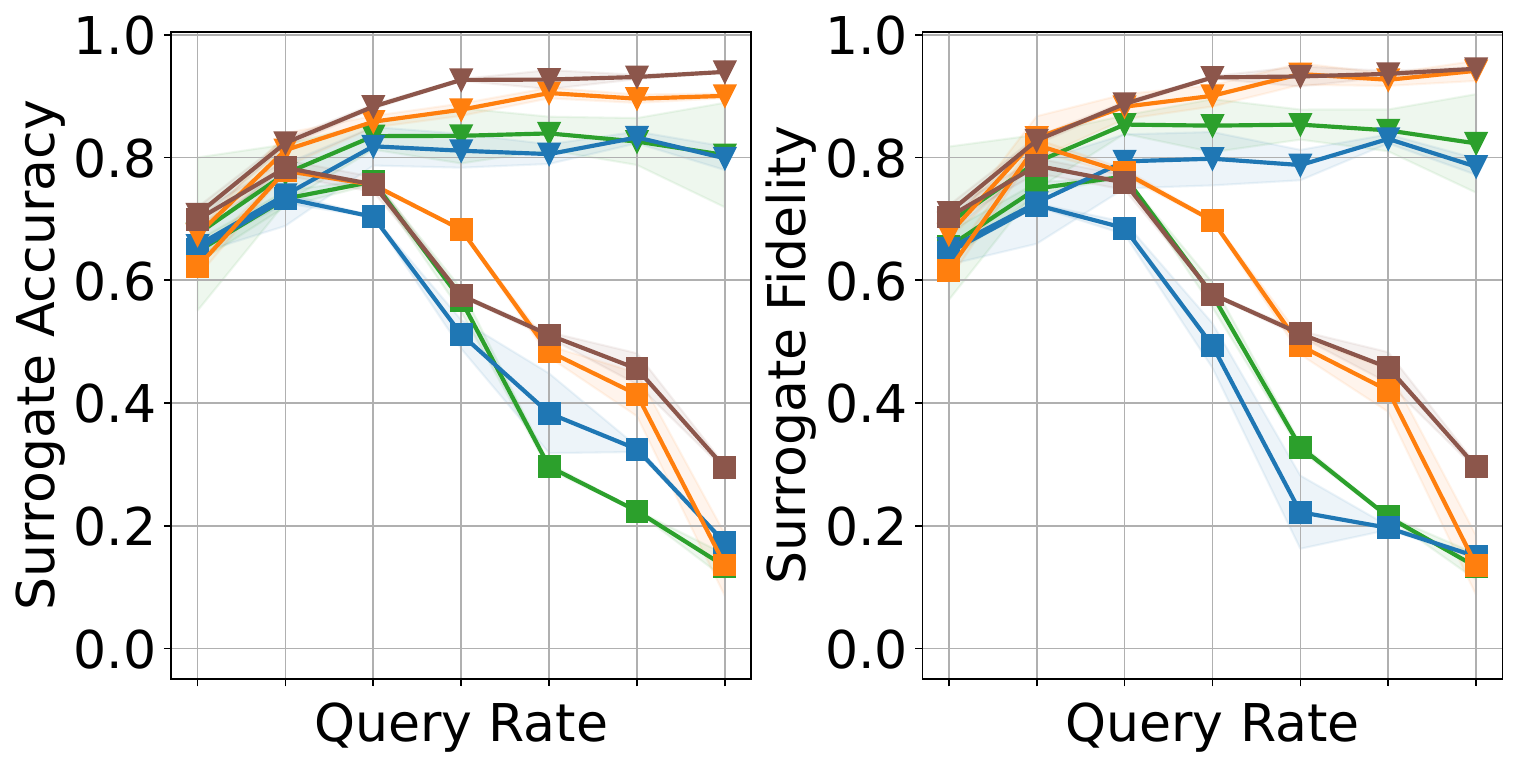}
         \caption{Embeddings.}
     \end{subfigure}
     \begin{subfigure}[b]{0.32\textwidth}
         \centering
         \includegraphics[width=\textwidth]{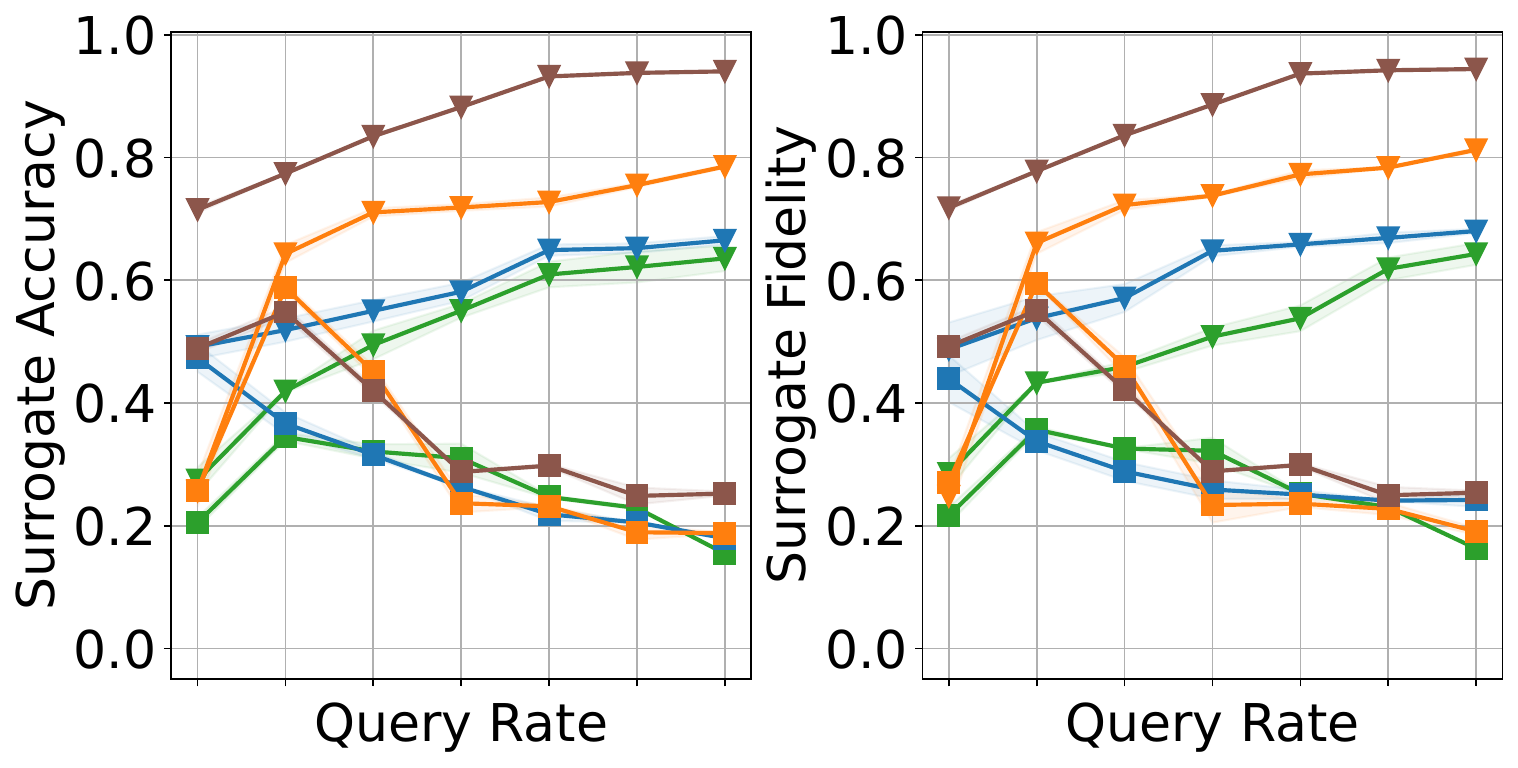}
         \caption{Projections.}
     \end{subfigure}
\caption{\textbf{Performance of the surrogate model with and without our defense (Amazon dataset).}}
\label{fig:adversary_performance_amazon}
\end{figure*}

\begin{figure*}[htpb]
\centering
    \begin{subfigure}[t]{0.5\textwidth}
         \centering
         \includegraphics[width=\textwidth]{figures/adversary_performance_new/legend_with_transformer.pdf}
     \end{subfigure}
     \vfill
     \begin{subfigure}[b]{0.32\textwidth}
         \centering
         \includegraphics[width=\textwidth]{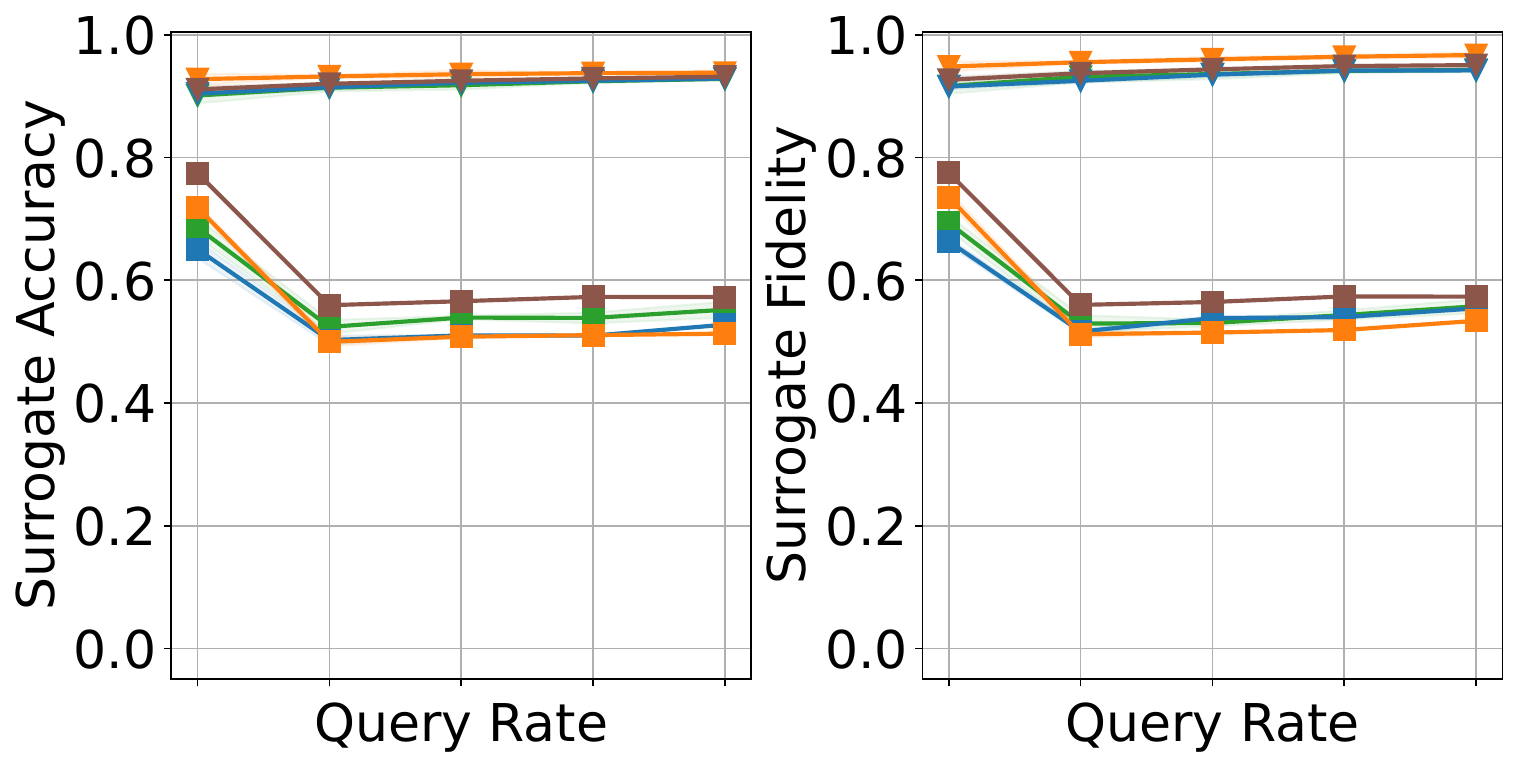}
         \caption{Probabilities.}
     \end{subfigure}
     \begin{subfigure}[b]{0.32\textwidth}
         \centering
         \includegraphics[width=\textwidth]{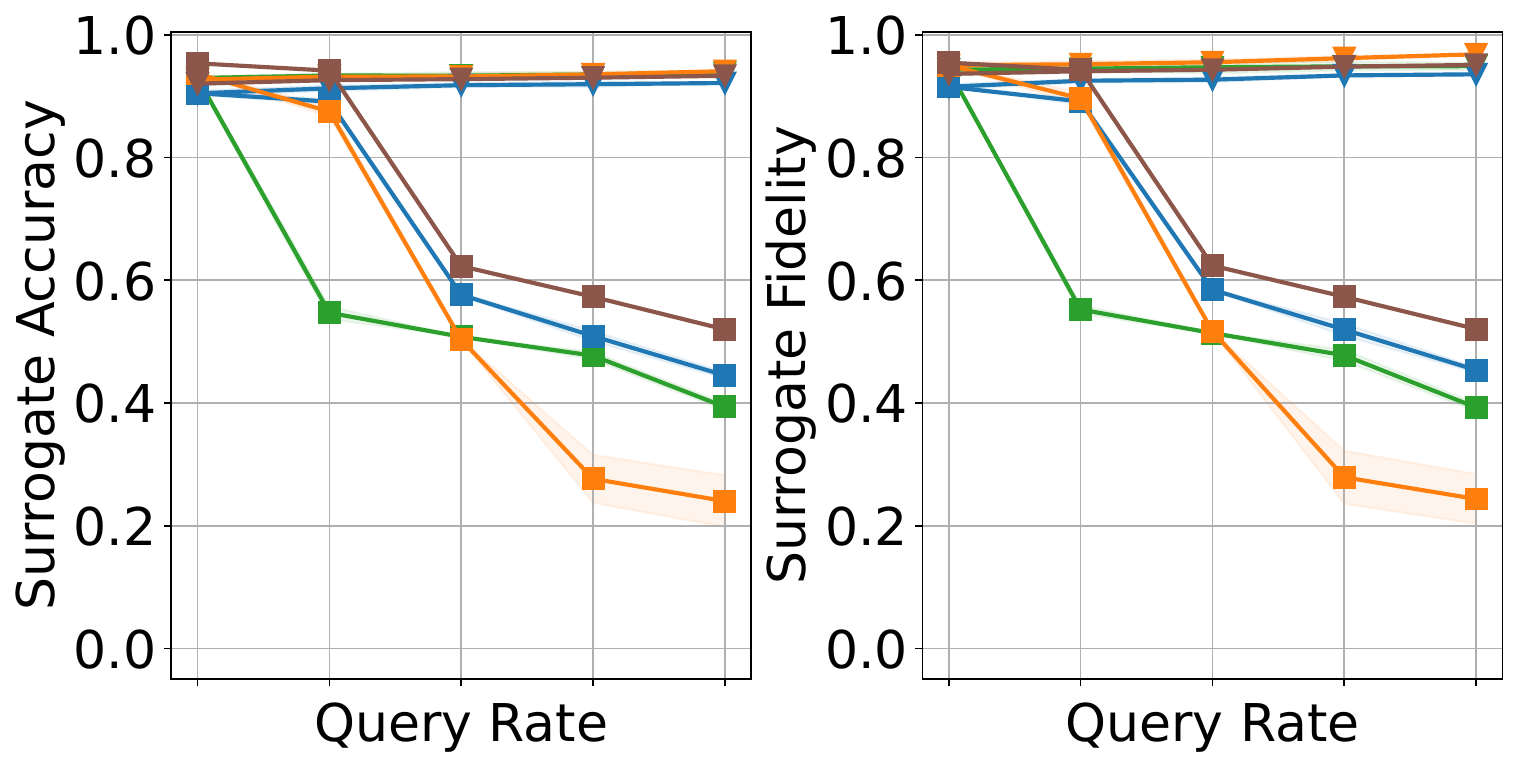}
         \caption{Embeddings.}
     \end{subfigure}
     \begin{subfigure}[b]{0.32\textwidth}
         \centering
         \includegraphics[width=\textwidth]{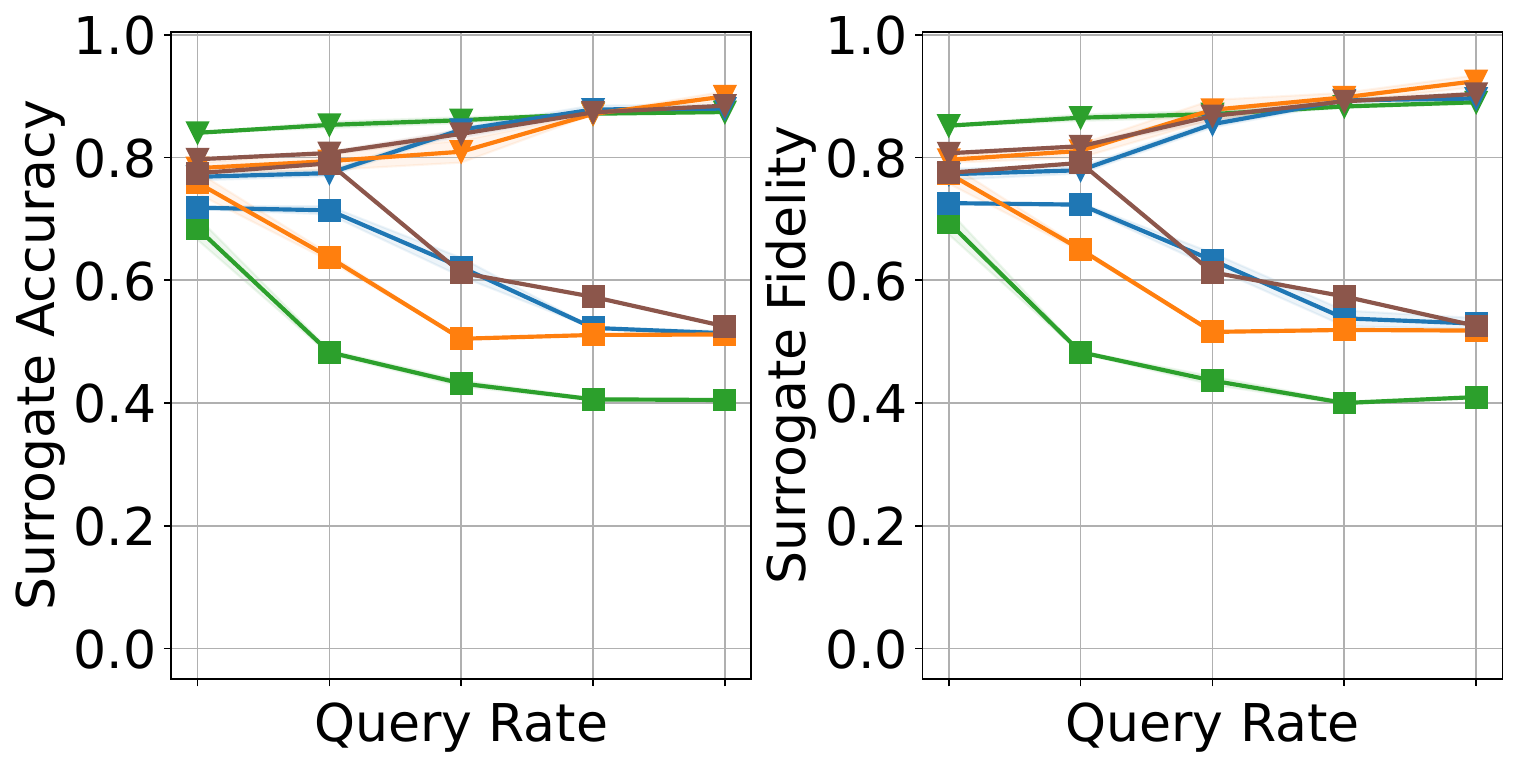}
         \caption{Projections.}
     \end{subfigure}
\caption{\textbf{Performance of the surrogate model with and without our defense (Coauthor dataset).}}
\label{fig:adversary_performance_coauthor}
\end{figure*}

\begin{table*}[htpb]
\small
 \centering
 \caption{\textbf{Performance for attacker and target downstream tasks with and without defense \ours in three attack setups (DBLP, $\delta=0.25$, $c_i$ represents a community).}
 }
 \label{table:results_with_without_defenses_dblp}
\begin{tabular}{cccccccc} 
 \hline
  & User & Dataset & Defense & GAT & GIN & GraphSAGE & Graph Transformer\\
 \hline
 & N/A & $\mathbf{G}_{test}$ & N/A & {$76.29\pm0.79$} & {$77.70\pm0.31$} & {$77.82\pm0.12$} & {$94.72\pm0.24$} \\
\hline
\hline
\multirow{5}{20mm}{\begin{center}Attack setup A (Probabilities)\end{center}} 
 & Attacker & $\mathbf{G}_{test}$ & NONE & {$74.07\pm0.60$}  & {$71.38\pm1.53$}  & {$76.04\pm0.86$} & {$73.89\pm0.24$} \\
 & Attacker & $\mathbf{G}_{test}$ & \ours & {$\textbf{28.37}\pm1.00$}  & {$\textbf{39.87}\pm0.58$}  & {$\textbf{18.71}\pm0.44$} & {$\textbf{24.90}\pm0.07$} \\
\cline{2-8}
 & Downstream Task 1 & $c_1$ & \ours & {$75.52\pm1.27$}  & {$73.64\pm2.04$}  & {$76.25\pm0.95$} & {$92.42\pm0.85$} \\
 & Downstream Task 2 & $c_2$ & \ours & {$73.98\pm2.11$}  & {$76.80\pm0.65$}  & {$77.24\pm1.16$} & {$93.62\pm0.73$} \\
 & Downstream Task 3 & $c_3$ & \ours & {$73.59\pm1.76$}  & {$75.09\pm0.67$}  & {$75.63\pm0.87$} & {$93.60\pm0.37$} \\
\hline
\hline

\multirow{5}{20mm}{\begin{center}Attack setup B (Embeddings)\end{center}}
 & Attacker & $\mathbf{G}_{test}$ & NONE & {$75.87\pm0.47$}  & {$70.50\pm1.95$}  & {$74.75\pm0.29$} & {$81.89\pm0.39$} \\
 & Attacker & $\mathbf{G}_{test}$ & \ours & {$\textbf{22.51}\pm0.17$}  & {$\textbf{26.87}\pm0.10$}  & {$\textbf{15.85}\pm0.01$} & {$\textbf{32.97}\pm0.03$} \\
\cline{2-8}
 & Downstream Task 1 & $c_1$ & \ours & {$75.60\pm0.23$}  & {$78.71\pm0.09$}  & {$74.25\pm0.18$} & {$93.34\pm1.15$} \\
 & Downstream Task 2 & $c_2$ & \ours & {$74.23\pm0.41$}  & {$75.71\pm0.02$}  & {$76.67\pm0.22$} & {$93.48\pm0.65$} \\
 & Downstream Task 3 & $c_3$ & \ours & {$75.65\pm0.75$}  & {$76.28\pm0.21$}  & {$77.23\pm0.16$} & {$93.69\pm0.90$} \\
\hline
\hline

\multirow{5}{20mm}{\begin{center}Attack setup C (Projections)\end{center}} 
 & Attacker & $\mathbf{G}_{test}$ & NONE & {$68.12\pm1.02$}  & {$66.72\pm0.18$}  & {$71.83\pm0.53$} & {$81.94\pm0.38$} \\
 & Attacker & $\mathbf{G}_{test}$ & \ours & {$\textbf{25.42}\pm0.12$}  & {$\textbf{36.09}\pm0.04$}  & {$\textbf{29.07}\pm0.56$} & {$\textbf{35.31}\pm0.27$} \\
\cline{2-8}
 & Downstream Task 1 & $c_1$ & \ours & {$72.84\pm1.73$}  & {$76.23\pm1.18$}  & {$76.17\pm1.33$} & {$93.35\pm0.82$} \\
 & Downstream Task 2 & $c_2$ & \ours & {$76.08\pm0.28$}  & {$76.37\pm1.39$}  & {$76.01\pm1.83$} & {$93.44\pm0.77$} \\
 & Downstream Task 3 & $c_3$ & \ours & {$75.44\pm0.91$}  & {$74.30\pm0.47$}  & {$75.62\pm1.44$} & {$93.11\pm1.25$} \\
\hline
\hline
\end{tabular}
\end{table*}

\begin{table*}[htpb]
 \centering
 \small
 \caption{\textbf{Performance for attacker and target downstream tasks with and without defense \ours in three attack setups (Pubmed, $\delta=0.25$, $c_i$ represents a community).}
 }
\begin{tabular}{cccccccc} 
 \hline
  & User & Dataset & Defense & GAT & GIN & GraphSAGE & Graph Transformer \\
 \hline
 & N/A & $\mathbf{G}_{test}$ & N/A & {$83.11\pm0.39$} & {$84.51\pm0.43$} & {$85.74\pm0.25$} & {$97.68\pm0.09$} \\
\hline
\hline
\multirow{5}{20mm}{\begin{center}Attack setup A (Probabilities)\end{center}} 
 & Attacker & $\mathbf{G}_{test}$ & NONE & {$83.77\pm0.20$}  & {$84.69\pm0.37$}  & {$85.27\pm0.11$} & {$85.27\pm0.06$} \\
 & Attacker & $\mathbf{G}_{test}$ & \ours & {$\textbf{39.55}\pm0.21$}  & {$\textbf{33.24}\pm0.06$}  & {$\textbf{35.15}\pm0.59$} & {$\textbf{44.63}\pm0.14$} \\
\cline{2-8}
 & Downstream Task 1 & $c_1$ & \ours & {$81.74\pm0.77$}  & {$82.16\pm0.96$}  & {$83.36\pm2.50$} & {$96.45\pm0.43$} \\
 & Downstream Task 2 & $c_2$ & \ours & {$79.27\pm1.52$}  & {$83.07\pm0.84$}  & {$84.52\pm0.49$} & {$96.39\pm1.03$} \\
 & Downstream Task 3 & $c_3$ & \ours & {$80.23\pm0.66$}  & {$81.36\pm1.29$}  & {$85.71\pm0.18$} & {$94.29\pm1.41$} \\
\hline
\hline

\multirow{5}{20mm}{\begin{center}Attack setup B (Embeddings)\end{center}} 
 & Attacker & $\mathbf{G}_{test}$ & NONE & {$83.62\pm0.51$}  & {$85.00\pm0.25$}  & {$85.26\pm0.33$} & {$86.43\pm1.37$} \\
 & Attacker & $\mathbf{G}_{test}$ & \ours & {$\textbf{23.34}\pm0.02$}  & {$\textbf{32.41}\pm0.52$}  & {$\textbf{29.25}\pm0.05$} & {$\textbf{30.51}\pm0.01$} \\
\cline{2-8}
 & Downstream Task 1 & $c_1$ & \ours & {$82.51\pm0.26$}  & {$81.30\pm0.29$}  & {$85.27\pm0.02$} & {$96.69\pm1.02$} \\
 & Downstream Task 2 & $c_2$ & \ours & {$83.27\pm0.41$}  & {$83.56\pm0.47$}  & {$85.07\pm0.19$} & {$97.04\pm0.56$} \\
 & Downstream Task 3 & $c_3$ & \ours & {$83.38\pm1.03$}  & {$83.55\pm0.27$}  & {$84.54\pm0.22$} & {$96.57\pm0.46$} \\
\hline
\hline

\multirow{5}{20mm}{\begin{center}Attack setup C (Projections)\end{center}}  
 & Attacker & $\mathbf{G}_{test}$ & NONE & {$82.99\pm0.70$}  & {$80.34\pm1.99$}  & {$83.32\pm1.60$} & {$86.43\pm1.37$} \\
 & Attacker & $\mathbf{G}_{test}$ & \ours & {$\textbf{41.76}\pm1.32$}  & {$\textbf{41.89}\pm0.16$}  & {$\textbf{41.78}\pm0.36$} & {$\textbf{35.37}\pm0.17$} \\
\cline{2-8}
 & Downstream Task 1 & $c_1$ & \ours & {$80.57\pm0.72$}  & {$82.14\pm1.26$}  & {$82.83\pm1.49$} & {$95.48\pm1.48$} \\
 & Downstream Task 2 & $c_2$ & \ours & {$81.46\pm0.54$}  & {$81.85\pm1.60$}  & {$84.56\pm1.50$} & {$94.98\pm1.48$} \\
 & Downstream Task 3 & $c_3$ & \ours & {$81.65\pm1.92$}  & {$83.87\pm0.90$}  & {$84.56\pm1.40$} & {$94.68\pm2.21$} \\

\hline
\hline
\end{tabular}
\label{table:results_with_without_defenses_pubmed}
\end{table*}

\begin{table*}[htpb]
\small
 \centering
 \caption{\textbf{Performance for attacker and target downstream tasks with and without defense \ours in three attack setups (Citeseer, $\delta=0.25$, $c_i$ represents a community).}
 }
\begin{tabular}{cccccccc} 
 \hline
  & User & Dataset & Defense & GAT & GIN & GraphSAGE & Graph Transformer\\
 \hline
 & N/A & $\mathbf{G}_{test}$ & N/A & {$81.89\pm0.30$} & {$82.49\pm0.85$} & {$83.62\pm1.02$} & {$92.40\pm0.22$} \\
\hline
\hline
\multirow{5}{20mm}{\begin{center}Attack setup A (Probabilities)\end{center}} 
 & Attacker & $\mathbf{G}_{test}$ & NONE & {$80.54\pm0.79$}  & {$75.37\pm2.31$}  & {$79.72\pm1.58$} & {$79.53\pm0.10$} \\
 & Attacker & $\mathbf{G}_{test}$ & \ours & {$\textbf{30.57}\pm0.79$}  & {$\textbf{30.97}\pm1.12$}  & {$\textbf{23.38}\pm0.10$} & {$\textbf{30.85}\pm0.05$} \\
\cline{2-8}
 & Downstream Task 1 & $c_1$ & \ours & {$80.43\pm1.15$}  & {$81.31\pm2.58$}  & {$82.53\pm1.62$} & {$91.86\pm0.55$} \\
 & Downstream Task 2 & $c_2$ & \ours & {$81.57\pm0.64$}  & {$82.47\pm2.30$}  & {$81.50\pm0.48$} & {$91.90\pm0.57$} \\
 & Downstream Task 3 & $c_3$ & \ours & {$80.98\pm0.20$}  & {$79.91\pm1.93$}  & {$80.98\pm0.61$} & {$90.23\pm1.17$} \\

\hline
\hline

\multirow{5}{20mm}{\begin{center}Attack setup B (Embeddings)\end{center}} 
 & Attacker & $\mathbf{G}_{test}$ & NONE & {$82.06\pm0.58$}  & {$75.67\pm1.72$}  & {$78.14\pm1.97$} & {$79.17\pm0.12$} \\
 & Attacker & $\mathbf{G}_{test}$ & \ours & {$\textbf{14.47}\pm0.32$}  & {$\textbf{22.36}\pm0.05$}  & {$\textbf{15.95}\pm0.54$} & {$\textbf{24.67}\pm0.06$} \\
\cline{2-8}
 & Downstream Task 1 & $c_1$ & \ours & {$82.61\pm0.73$}  & {$80.35\pm0.47$}  & {$82.83\pm0.32$} & {$90.98\pm0.58$} \\
 & Downstream Task 2 & $c_2$ & \ours & {$80.80\pm0.20$}  & {$81.56\pm0.07$}  & {$82.08\pm0.48$} & {$91.60\pm0.39$} \\
 & Downstream Task 3 & $c_3$ & \ours & {$80.91\pm0.40$}  & {$82.86\pm1.42$}  & {$81.61\pm0.80$} & {$91.54\pm0.54$} \\

\hline
\hline

\multirow{5}{20mm}{\begin{center}Attack setup C (Projections)\end{center}} 
 & Attacker & $\mathbf{G}_{test}$ & NONE & {$72.39\pm0.08$}  & {$61.98\pm1.85$}  & {$62.98\pm0.66$}  & {$70.01\pm0.19$} \\
 & Attacker & $\mathbf{G}_{test}$ & \ours & {$\textbf{17.25}\pm0.51$}  & {$\textbf{19.13}\pm1.24$}  & {$\textbf{13.66}\pm0.63$} & {$\textbf{16.43}\pm0.21$} \\
\cline{2-8}
 & Downstream Task 1 & $c_1$ & \ours & {$80.72\pm0.59$}  & {$80.84\pm1.63$}  & {$82.55\pm1.25$} & {$90.90\pm0.83$} \\
 & Downstream Task 2 & $c_2$ & \ours & {$80.15\pm1.60$}  & {$81.26\pm1.25$}  & {$80.09\pm2.19$} & {$91.13\pm0.63$} \\
 & Downstream Task 3 & $c_3$ & \ours & {$81.02\pm0.92$}  & {$81.00\pm2.37$}  & {$81.28\pm1.94$} & {$89.94\pm0.55$} \\

\hline
\hline
\end{tabular}
\label{table:results_with_without_defenses_citeseer}
\end{table*}

\begin{table*}[htpb]
\small
 \centering
 \caption{\textbf{Performance for attacker and target downstream tasks with and without defense \ours in three attack setups (Amazon, $\delta=0.25$, $c_i$ represents a community).}
 }
\begin{tabular}{cccccccc} 
 \hline
  & User & Dataset & Defense & GAT & GIN & GraphSAGE & Graph Transformer \\
 \hline
 & N/A & $\mathbf{G}_{test}$ & N/A & {$91.38\pm0.70$} & {$84.97\pm1.54$} & {$91.52\pm0.41$} & {$98.82\pm0.06$} \\
\hline
\hline
\multirow{5}{20mm}{\begin{center}Attack setup A (Probabilites)\end{center}} 
 & Attacker & $\mathbf{G}_{test}$ & NONE & {$91.07\pm0.73$}  & {$86.29\pm1.32$}  & {$90.13\pm1.60$} & {$93.59\pm0.17$} \\
 & Attacker & $\mathbf{G}_{test}$ & \ours & {$\textbf{40.60}\pm0.37$}  & {$\textbf{35.06}\pm0.67$}  & {$\textbf{31.42}\pm3.89$} & {$\textbf{37.18}\pm0.82$} \\
\cline{2-8}
 & Downstream Task 1 & $c_1$ & \ours & {$87.38\pm1.94$}  & {$83.81\pm0.51$}  & {$89.59\pm1.30$} & {$96.23\pm0.57$} \\
 & Downstream Task 2 & $c_2$ & \ours & {$89.62\pm0.96$}  & {$84.25\pm1.45$}  & {$90.48\pm0.33$} & {$97.11\pm0.99$} \\
 & Downstream Task 3 & $c_3$ & \ours & {$88.05\pm0.75$}  & {$84.67\pm1.37$}  & {$88.51\pm2.44$} & {$96.11\pm0.98$} \\
\hline
\hline

\multirow{5}{20mm}{\begin{center}Attack setup B (Embeddings)\end{center}} 
 & Attacker & $\mathbf{G}_{test}$ & NONE & {$80.41\pm8.52$}  & {$79.88\pm1.91$}  & {$90.05\pm0.38$} & {$93.96\pm0.13$} \\
 & Attacker & $\mathbf{G}_{test}$ & \ours & {$\textbf{13.39}\pm1.96$}  & {$\textbf{17.23}\pm0.17$}  & {$\textbf{13.61}\pm5.06$} & {$\textbf{29.45}\pm0.38$} \\
\cline{2-8}
 & Downstream Task 1 & $c_1$ & \ours & {$92.58\pm0.14$}  & {$84.50\pm0.24$}  & {$89.46\pm0.30$} & {$97.46\pm1.23$} \\
 & Downstream Task 2 & $c_2$ & \ours & {$90.75\pm0.13$}  & {$84.40\pm0.28$}  & {$91.27\pm0.05$} & {$97.29\pm0.78$} \\
 & Downstream Task 3 & $c_3$ & \ours & {$90.08\pm0.10$}  & {$81.71\pm0.94$}  & {$91.91\pm0.97$} & {$98.18\pm0.08$} \\
\hline
\hline

\multirow{5}{20mm}{\begin{center}Attack setup C (Projections)\end{center}} 
 & Attacker & $\mathbf{G}_{test}$ & NONE & {$63.56\pm2.07$}  & {$66.51\pm0.69$}  & {$78.54\pm0.23$} & {$94.04\pm0.06$} \\
 & Attacker & $\mathbf{G}_{test}$ & \ours & {$\textbf{15.46}\pm0.10$}  & {$\textbf{17.96}\pm0.04$}  & {$\textbf{18.79}\pm0.09$} & {$\textbf{25.24}\pm0.33$} \\
\cline{2-8}
 & Downstream Task 1 & $c_1$ & \ours & {$90.36\pm2.14$}  & {$84.96\pm0.62$}  & {$90.37\pm1.66$} & {$91.63\pm1.31$} \\
 & Downstream Task 2 & $c_2$ & \ours & {$90.06\pm1.11$}  & {$83.92\pm2.51$}  & {$90.15\pm1.27$} & {$94.88\pm2.26$} \\
 & Downstream Task 3 & $c_3$ & \ours & {$90.80\pm1.70$}  & {$83.13\pm2.64$}  & {$90.63\pm1.67$} & {$93.70\pm1.95$} \\
\hline
\hline
\end{tabular}
\label{table:results_with_without_defenses_amazon}
\end{table*}
\vfill
\begin{table*}[htpb]
\small
 \centering
 \caption{\textbf{Performance for attacker and target downstream tasks with and without defense \ours in three attack setups (Coauthor, $\delta=0.25$, $c_i$ represents a community).}
 }
\begin{tabular}{cccccccc} 
 \hline
  & User & Dataset & Defense & GAT & GIN & GraphSAGE & Graph Transformer \\
 \hline
 & N/A & $\mathbf{G}_{test}$ & N/A & {$94.21\pm0.33$} & {$92.70\pm0.34$} & {$94.27\pm0.12$} & {$99.58\pm0.83$} \\
\hline
\hline
\multirow{5}{20mm}{\begin{center}Attack setup A (Probabilities)\end{center}} 
 & Attacker & $\mathbf{G}_{test}$ & NONE & {$92.82\pm0.23$}  & {$92.88\pm0.19$}  & {$93.82\pm0.48$} & {$93.17\pm0.25$} \\
 & Attacker & $\mathbf{G}_{test}$ & \ours & {$\textbf{55.20}\pm1.31$}  & {$\textbf{52.76}\pm0.19$}  & {$\textbf{51.30}\pm0.26$} & {$\textbf{57.26}\pm0.02$} \\
 \cline{2-8}
 & Downstream Task 1 & $c_1$ & \ours & {$92.73\pm1.04$}  & {$88.51\pm1.67$}  & {$92.99\pm0.73$} & {$97.28\pm0.96$} \\
 & Downstream Task 2 & $c_2$ & \ours & {$91.49\pm2.40$}  & {$92.50\pm1.18$}  & {$92.83\pm1.30$} & {$98.49\pm0.55$} \\
 & Downstream Task 3 & $c_3$ & \ours & {$91.51\pm1.24$}  & {$89.85\pm0.44$}  & {$91.63\pm0.62$} & {$98.46\pm0.38$} \\
\hline
\hline

\multirow{5}{20mm}{\begin{center}Attack setup B (Embeddings)\end{center}} 
 & Attacker & $\mathbf{G}_{test}$ & NONE & {$93.75\pm0.46$}  & {$92.14\pm0.10$}  & {$94.07\pm0.14$} & {$93.32\pm0.16$} \\
 & Attacker & $\mathbf{G}_{test}$ & \ours & {$\textbf{39.36}\pm0.55$}  & {$\textbf{44.49}\pm0.50$}  & {$\textbf{24.01}\pm4.21$} & {$\textbf{51.99}\pm0.01$} \\
\cline{2-8}
 & Downstream Task 1 & $c_1$ & \ours & {$93.22\pm0.10$}  & {$93.58\pm0.28$}  & {$90.71\pm0.08$} & {$98.21\pm1.19$} \\
 & Downstream Task 2 & $c_2$ & \ours & {$92.51\pm0.21$}  & {$90.54\pm0.36$}  & {$93.17\pm0.15$} & {$98.35\pm0.47$} \\
 & Downstream Task 3 & $c_3$ & \ours & {$91.25\pm0.57$}  & {$91.16\pm0.15$}  & {$93.69\pm0.06$} & {$98.55\pm0.90$} \\
\hline
\hline

\multirow{5}{20mm}{\begin{center}Attack setup C (Projections)\end{center}} 
 & Attacker & $\mathbf{G}_{test}$ & NONE & {$87.41\pm0.01$}  & {$87.97\pm0.58$}  & {$89.96\pm0.82$} & {$88.45\pm0.08$} \\
 & Attacker & $\mathbf{G}_{test}$ & \ours & {$\textbf{40.49}\pm0.07$}  & {$\textbf{51.37}\pm0.36$}  & {$\textbf{51.17}\pm0.16$} & {$\textbf{52.49}\pm0.02$} \\
\cline{2-8}
 & Downstream Task 1 & $c_1$ & \ours & {$90.77\pm1.19$}  & {$91.09\pm0.99$}  & {$92.62\pm1.34$} & {$96.18\pm1.69$} \\
 & Downstream Task 2 & $c_2$ & \ours & {$93.60\pm0.26$}  & {$91.16\pm1.33$}  & {$91.85\pm1.82$} & {$96.39\pm1.44$} \\
 & Downstream Task 3 & $c_3$ & \ours & {$93.37\pm1.11$}  & {$87.60\pm3.27$}  & {$91.83\pm1.23$} & {$95.56\pm2.56$} \\
\hline
\hline
\end{tabular}
\label{table:results_with_without_defenses_coauthor}
\end{table*}

\subsection{Downstream Task Performance Covering Multiple Communities}
\label{app:multiple_communities}
The downstream task performance when users cover multiple communities (\eg 1-5 communities) is presented in \Cref{table:multiple_communities}. The results show that the performance for target downstream tasks remains high for users querying several communities. 

\begin{table*}[!h]
 \centering
 \caption{\textbf{Performance for target downstream tasks covering multiple communities with \ours in three attack setups (ACM, $\delta=0.25$, $c_i$ represents community $i$, GT - Graph Transformer).} Overall, with our defense, the performance for downstream tasks remains high for users querying several communities.}
 \label{table:multiple_communities}
\begin{tabular}{ccccccc} 
 \hline
  & Dataset & Defense & GAT & GIN & GraphSAGE & GT \\
 \hline
Baseline & $\mathbf{G}_{test}$ & N/A & $90.04\pm 0.67$ & $88.30\pm0.47$ & $90.75\pm0.92$ & $96.72\pm0.30$ \\
 \hline
 \hline
 \multirow{3}{25mm}{\begin{center}Attack setup A (Probabilities)\end{center}} 
  & $c_1$ & \ours & {$88.69\pm0.67$} & {$84.24\pm2.19$} & {$89.74\pm1.47$} & {$94.42\pm1.24$} \\
  & $c_{1-3}$ & \ours & {$88.10\pm0.11$} & {$83.25\pm0.42$} & {$88.85\pm1.12$}  & {$93.73\pm0.63$} \\
  & $c_{1-5}$ & \ours & {$86.79\pm0.26$} & {$82.86\pm1.27$} & {$87.79\pm0.23$}  & {$93.38\pm1.27$} \\
 \hline
 \hline

 \multirow{3}{25mm}{\begin{center}Attack setup B (Embeddings)\end{center}} 
  & $c_1$ & \ours & {$87.94\pm0.07$} & {$89.61\pm0.20$} & {$88.37\pm0.10$} & {$95.34\pm0.93$} \\
  & $c_{1-3}$ & \ours & {$86.85\pm0.25$} & {$85.79\pm0.07$} & {$87.69\pm0.32$}  & {$94.48\pm0.76$} \\
  & $c_{1-5}$ & \ours & {$85.84\pm0.12$} & {$85.02\pm0.13$} & {$86.37\pm0.79$}  & {$93.17\pm0.38$} \\
 \hline
 \hline

 \multirow{3}{25mm}{\begin{center}Attack setup C (Projections)\end{center}} 
  & $c_1$ & \ours & {$86.59\pm0.39$} & {$86.83\pm1.59$} & {$89.10\pm1.78$} & {$95.35\pm0.93$} \\
  & $c_{1-3}$ & \ours & {$85.12\pm0.22$} & {$85.82\pm0.90$} & {$88.45\pm0.15$}  & {$94.26\pm0.32$} \\
  & $c_{1-5}$ & \ours & {$84.72\pm0.08$} & {$84.62\pm0.19$} & {$87.14\pm0.25$}  & {$93.07\pm0.47$} \\

 \hline
 \hline
\end{tabular}
\end{table*}

\subsection{Ablation Study on Community Detection Algorithm}
The exact results of \ours-greedy are presented in \Cref{table:results_with_without_defenses_greedy_acm}. 
As we can observe, the degradation of the stealing performance applying \ours-greedy is similar to or less than \ours. 
For instance, in the attack setup A on the ACM dataset, the surrogate accuracy of applying \ours is $36.90\%$, $30.45\%$, $31.12\%$, and $32.27\%$ for GAT, GIN, GraphSAGE, and Graph Transformer models, respectively, while that of \ours-greedy is $39.16\%$, $39.04\%$, $39.88\%$ and $37.96\%$ respectively.

\begin{table*}[!h]
\small
 \centering
 \caption{\textbf{Performance for attacker and target downstream task with and without defense \ours-greedy in three attack setups (ACM, $\delta=0.25$)).} Overall, the stealing performance with \ours-greedy is similar with \ours.}
\begin{tabular}{cccccccc} 
 \hline
  & User & Dataset & Defense & GAT & GIN & GraphSAGE & Graph Transformer \\
 \hline
 & N/A & $\mathbf{G}_{test}$ & N/A & {$90.04\pm0.67$} & {$88.30\pm0.47$} & {$90.75\pm0.92$} & $96.72\pm0.30$ \\
\hline
\hline
\multirow{5}{20mm}{\begin{center}Attack setup A (Probabilities)\end{center}} 
 & Attacker & $\mathbf{G}_{test}$ & NONE & {$88.53\pm0.62$}  & {$85.46\pm0.16$}  & {$88.14\pm0.12$} & {$88.30\pm0.49$} \\
 & Attacker & $\mathbf{G}_{test}$ & \ours-greedy & {$\textbf{39.16}\pm0.04$}  & {$\textbf{39.04}\pm0.04$}  & {$\textbf{39.88}\pm0.05$} & {$\textbf{37.96}\pm0.22$} \\
 \cline{2-8}
 & Downstream Task 1 & $c_1$ & \ours-greedy & {$90.29\pm0.72$}  & {$85.78\pm0.56$}  & {$89.46\pm1.86$} & {$95.10\pm0.58$} \\
 & Downstream Task 2 & $c_2$ & \ours-greedy & {$87.72\pm1.44$}  & {$87.73\pm0.35$}  & {$87.92\pm0.87$} & {$94.68\pm0.89$} \\
 & Downstream Task 3 & $c_3$ & \ours-greedy & {$88.34\pm1.20$}  & {$87.01\pm2.81$}  & {$89.87\pm1.57$} & {$95.30\pm0.59$} \\
\hline
\hline

\multirow{5}{20mm}{\begin{center}Attack setup B (Embeddings)\end{center}} 
 & Attacker & $\mathbf{G}_{test}$ & NONE & {$87.26\pm1.09$}  & {$85.00\pm0.34$}  & {$86.67\pm3.16$} & {$78.67\pm0.32$} \\
 & Attacker & $\mathbf{G}_{test}$ & \ours-greedy & {$\textbf{37.15}\pm0.24$}  & {$\textbf{38.75}\pm0.19$}  & {$\textbf{38.65}\pm0.11$} & {$\textbf{35.19}\pm0.28$} \\
\cline{2-8}
 & Downstream Task 1 & $c_1$ & \ours-greedy & {$89.12\pm0.25$}  & {$85.58\pm0.50$}  & {$87.65\pm0.25$} & {$94.97\pm0.20$} \\
 & Downstream Task 2 & $c_2$ & \ours-greedy & {$89.72\pm0.43$}  & {$87.43\pm0.05$}  & {$88.49\pm0.57$} & {$95.85\pm0.27$} \\
 & Downstream Task 3 & $c_3$ & \ours-greedy & {$89.56\pm0.26$}  & {$86.35\pm0.09$}  & {$89.64\pm0.01$} & {$95.50\pm0.65$} \\
\hline
\hline

\multirow{5}{20mm}{\begin{center}Attack setup C (Projections)\end{center}} 
 & Attacker & $\mathbf{G}_{test}$ & NONE & {$87.28\pm0.19$}  & {$84.14\pm2.81$}  & {$83.67\pm0.11$} & {$88.27\pm0.94$} \\
 & Attacker & $\mathbf{G}_{test}$ & \ours-greedy & {$\textbf{43.03}\pm0.22$}  & {$\textbf{25.52}\pm0.91$}  & {$\textbf{33.54}\pm0.03$} & {$\textbf{25.16}\pm0.05$} \\
\cline{2-8}
 & Downstream Task 1 & $c_1$ & \ours-greedy & {$88.69\pm0.67$}  & {$84.24\pm2.19$}  & {$89.74\pm1.47$} & {$94.42\pm1.24$} \\
 & Downstream Task 2 & $c_2$ & \ours-greedy & {$87.73\pm2.03$}  & {$88.10\pm0.72$}  & {$90.16\pm1.53$} & {$95.62\pm0.34$} \\
 & Downstream Task 3 & $c_3$ & \ours-greedy & {$87.34\pm0.48$}  & {$85.68\pm1.07$}  & {$88.56\pm0.79$} & {$95.60\pm0.67$} \\

\hline
\hline
\end{tabular}
\label{table:results_with_without_defenses_greedy_acm}
\end{table*}

\subsection{Adaptive Attacks}
\label{app:subsec_adaptive_attacks}
Figures \ref{fig:average_noise_dblp}, \ref{fig:average_noise_pubmed}, \ref{fig:average_noise_citeseer}, \ref{fig:average_noise_amazon},  and \ref{fig:average_noise_coauthor} 
show the surrogate performance with the adaptive attack of averaging noise on other datasets, on GAT model. 
Similar to the trend on the ACM dataset, our defense can degrade the surrogate performance significantly with REP up to $200$ times. When REP increases to $1000$, the attacker can obtain high surrogate performance, but such performance requires substantial effort, which is impractical for the attacker. 

As for the second adaptive attack, the stealing performance on other datasets, GAT model, is presented in Figures \ref{fig:adaptive_attacker_community_knowledge_dblp},\ref{fig:adaptive_attacker_community_knowledge_pubmed},\ref{fig:adaptive_attacker_community_knowledge_citeseer}, \ref{fig:adaptive_attacker_community_knowledge_amazon}, and \ref{fig:adaptive_attacker_community_knowledge_coauthor}.
It can be seen that even with knowledge about communities in the underlying graph, the adaptive attacker can still not steal a surrogate model of high performance. 

\begin{figure}[htpb]
\centering
    \begin{subfigure}[t]{0.35\textwidth}
         \centering
         \includegraphics[width=\textwidth]{figures/average_noise/legend.pdf}
     \end{subfigure}
     \vfill  
     \begin{subfigure}[t]{0.15\textwidth}
         \centering
         \includegraphics[width=\textwidth]{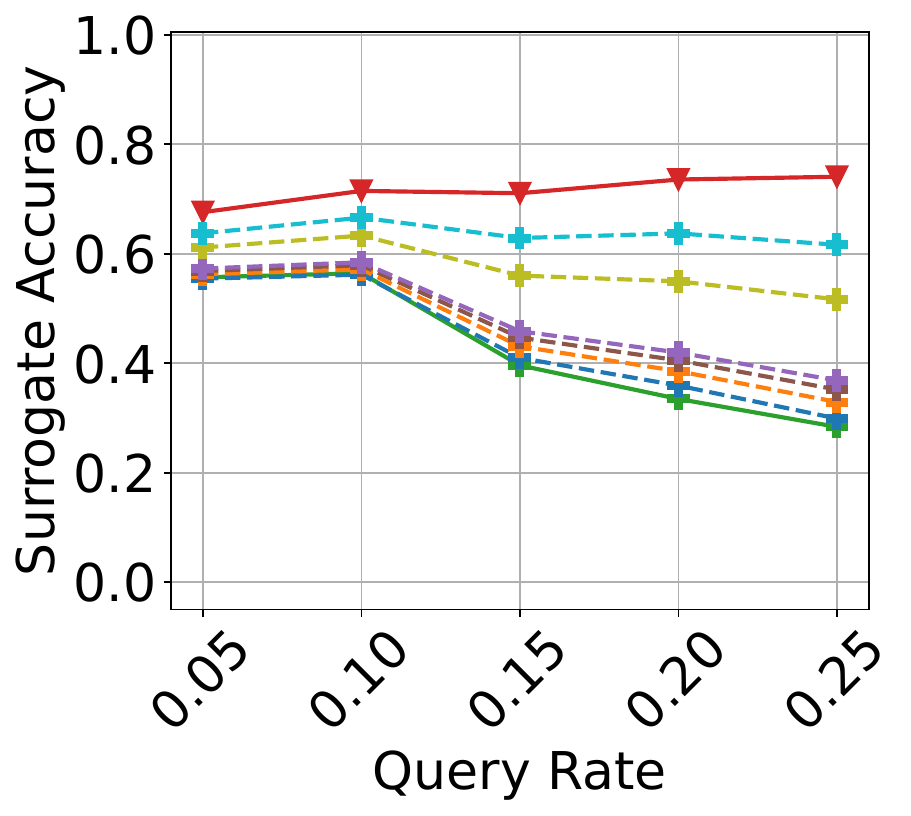}
         \caption{Probabilities.}
     \end{subfigure}
     \begin{subfigure}[t]{0.15\textwidth}
         \centering
         \includegraphics[width=\textwidth]{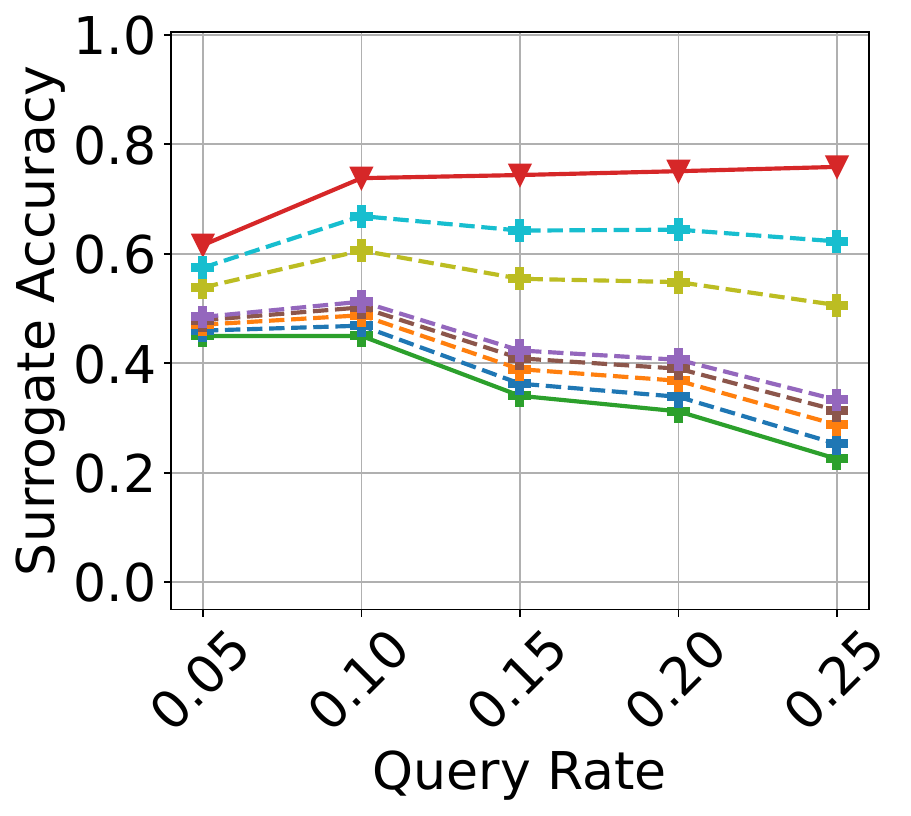}
         \caption{Embeddings.}
     \end{subfigure}
     \begin{subfigure}[t]{0.15\textwidth}
         \centering
         \includegraphics[width=\textwidth]{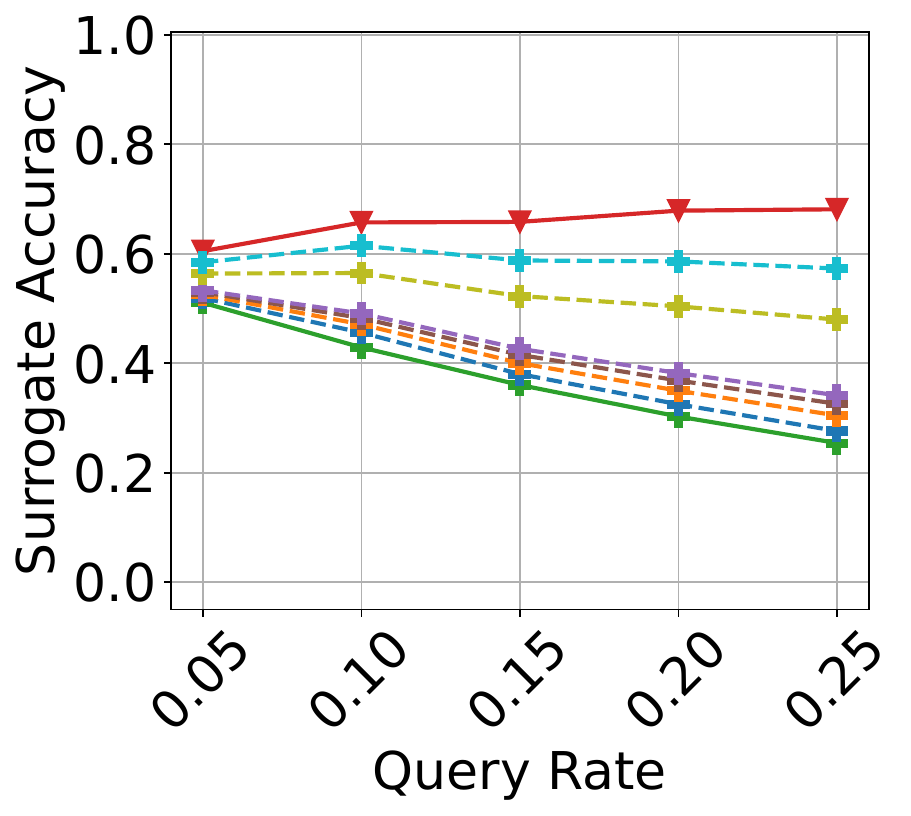}
         \caption{Projections.}
     \end{subfigure}
\caption{\textbf{Performance of the surrogate model with the first adaptive attack (DBLP, GAT).} %
}
\label{fig:average_noise_dblp}
\end{figure}
\vfill
\begin{figure}[htpb]
\centering
    \begin{subfigure}[t]{0.35\textwidth}
         \centering
         \includegraphics[width=\textwidth]{figures/average_noise/legend.pdf}
     \end{subfigure}
     \vfill    
     \begin{subfigure}[t]{0.15\textwidth}
         \centering
         \includegraphics[width=\textwidth]{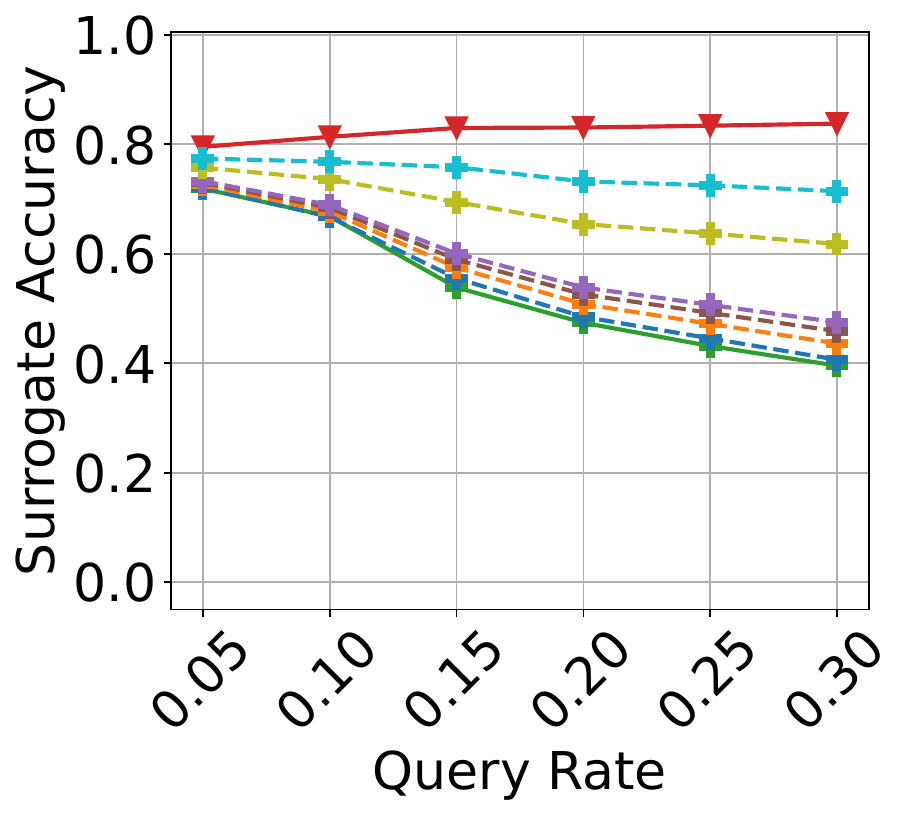}
         \caption{Probabilities.}
     \end{subfigure}
     \begin{subfigure}[t]{0.15\textwidth}
         \centering
         \includegraphics[width=\textwidth]{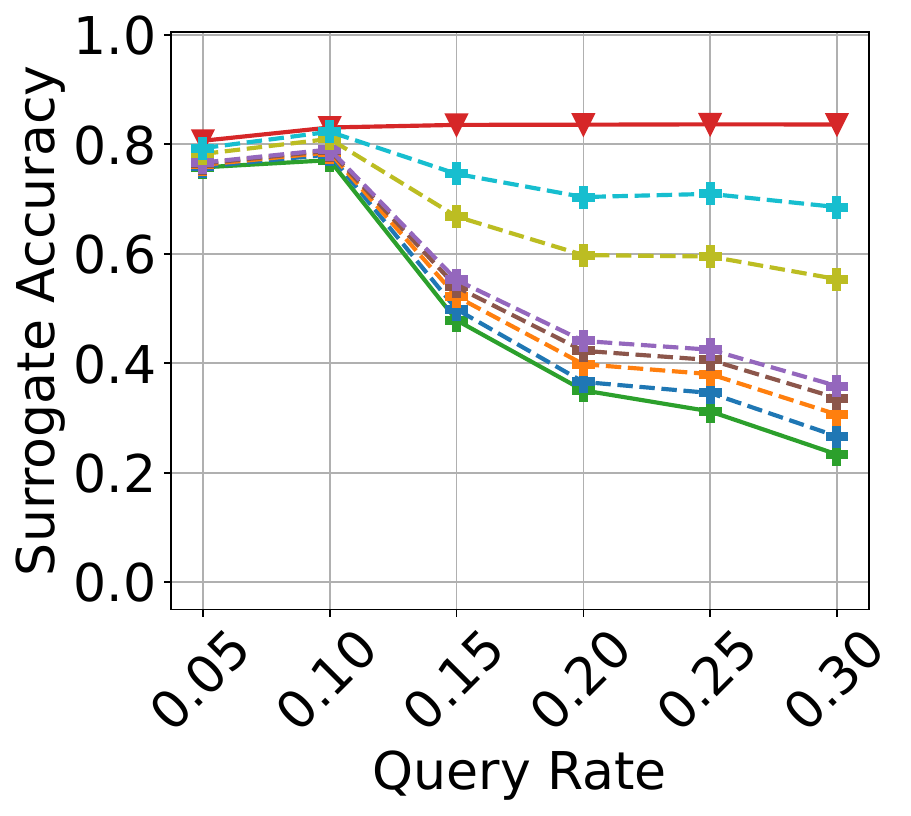}
         \caption{Embeddings.}
     \end{subfigure}
     \begin{subfigure}[t]{0.15\textwidth}
         \centering
         \includegraphics[width=\textwidth]{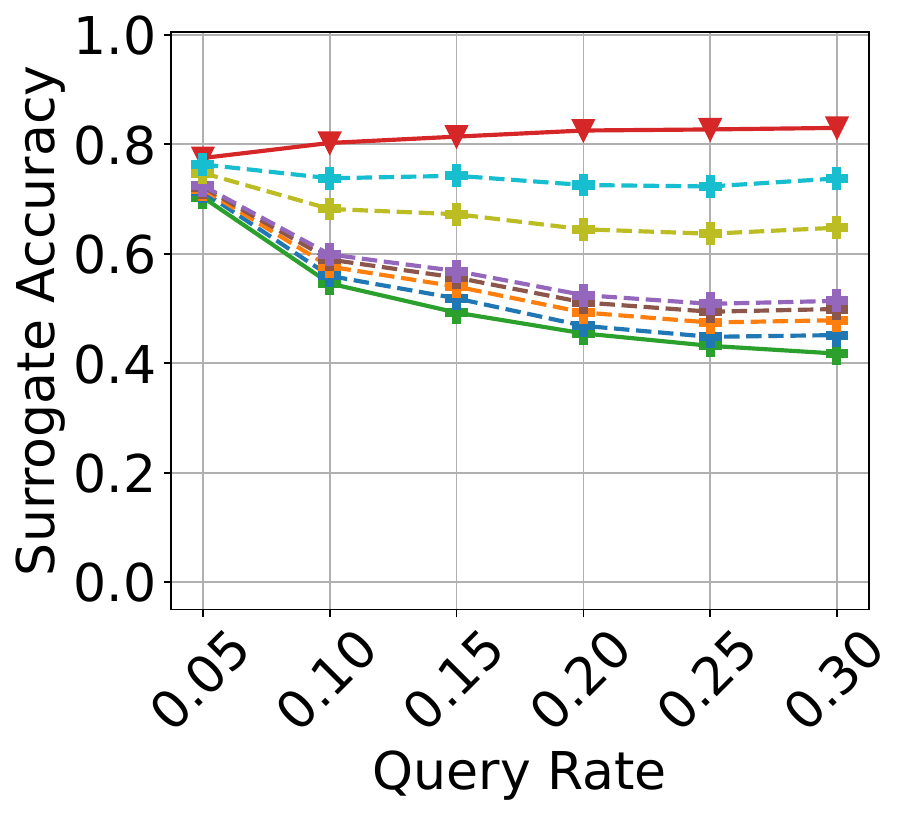}
         \caption{Projections.}
     \end{subfigure}
\caption{\textbf{Performance of the surrogate model with the first adaptive attack (Pubmed, GAT).} %
}
\label{fig:average_noise_pubmed}
\end{figure}
\vfill
\begin{figure}[htpb]
\centering
    \begin{subfigure}[t]{0.35\textwidth}
         \centering
         \includegraphics[width=\textwidth]{figures/average_noise/legend.pdf}
     \end{subfigure}
     \vfill    
     \begin{subfigure}[t]{0.15\textwidth}
         \centering
         \includegraphics[width=\textwidth]{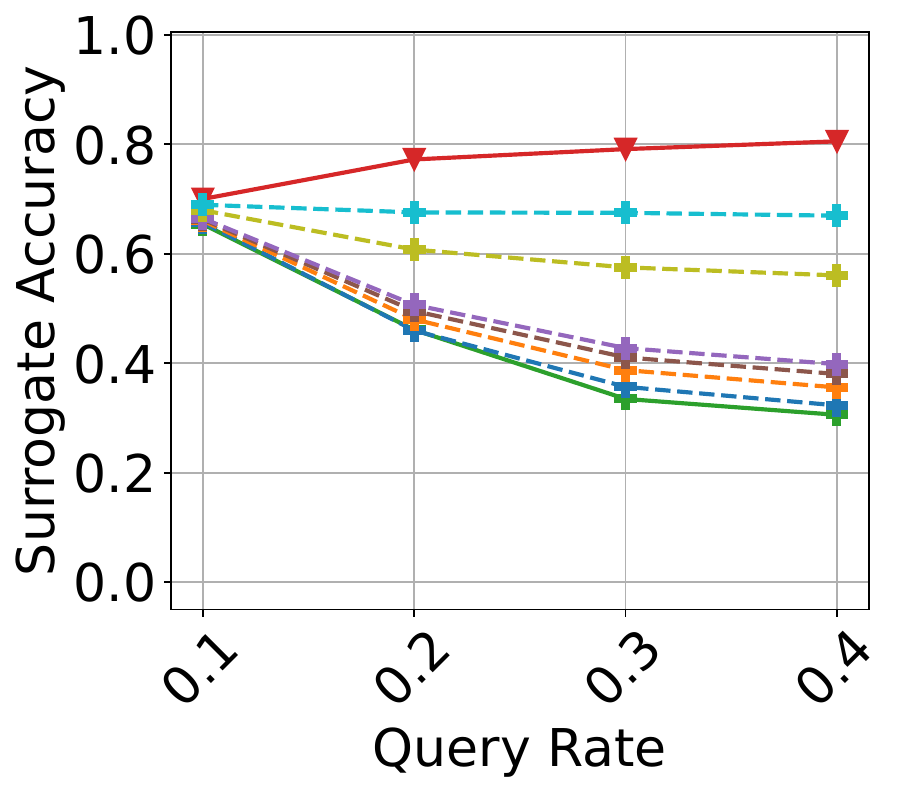}
         \caption{Probabilities.}
     \end{subfigure}
     \begin{subfigure}[t]{0.15\textwidth}
         \centering
         \includegraphics[width=\textwidth]{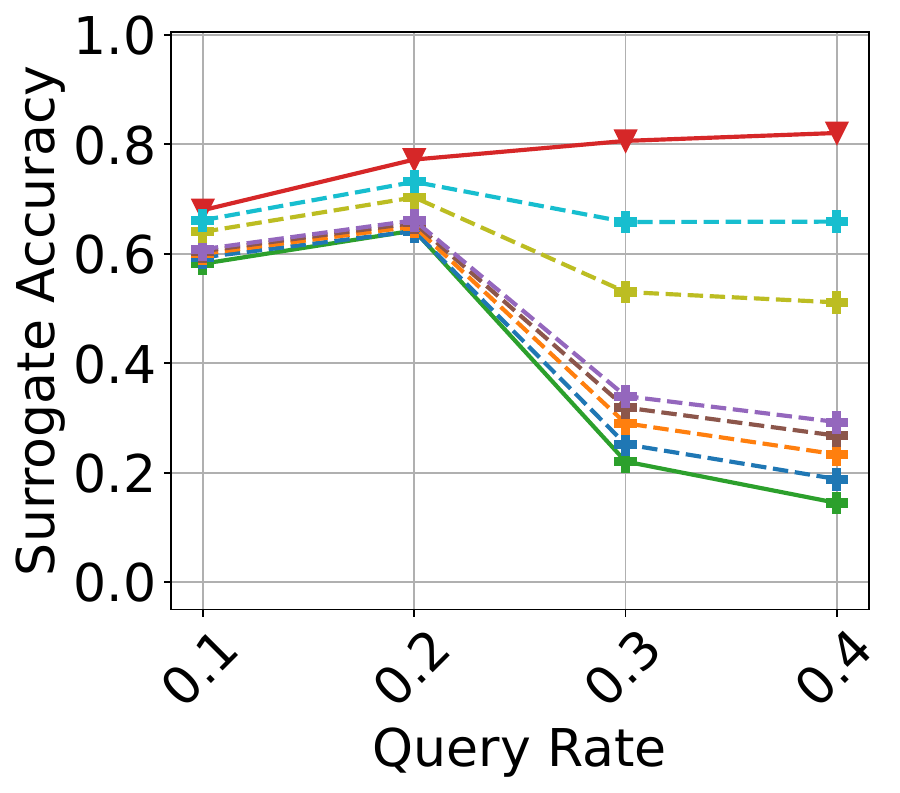}
         \caption{Embeddings.}
     \end{subfigure}
     \begin{subfigure}[t]{0.15\textwidth}
         \centering
         \includegraphics[width=\textwidth]{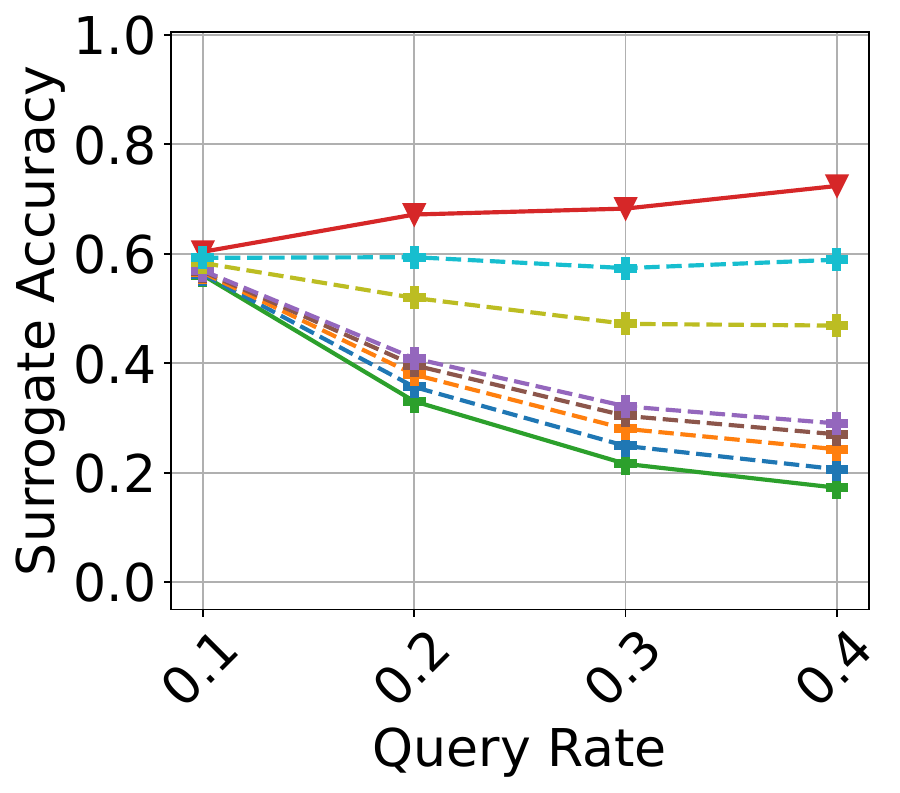}
         \caption{Projections.}
     \end{subfigure}
\caption{\textbf{Performance of the surrogate model with the first adaptive attack (Citeseer, GAT).} %
}
\label{fig:average_noise_citeseer}
\end{figure}

\begin{figure}[htpb]
\centering
    \begin{subfigure}[t]{0.35\textwidth}
         \centering
         \includegraphics[width=\textwidth]{figures/average_noise/legend.pdf}
     \end{subfigure}
     \vfill    
     \begin{subfigure}[t]{0.15\textwidth}
         \centering
         \includegraphics[width=\textwidth]{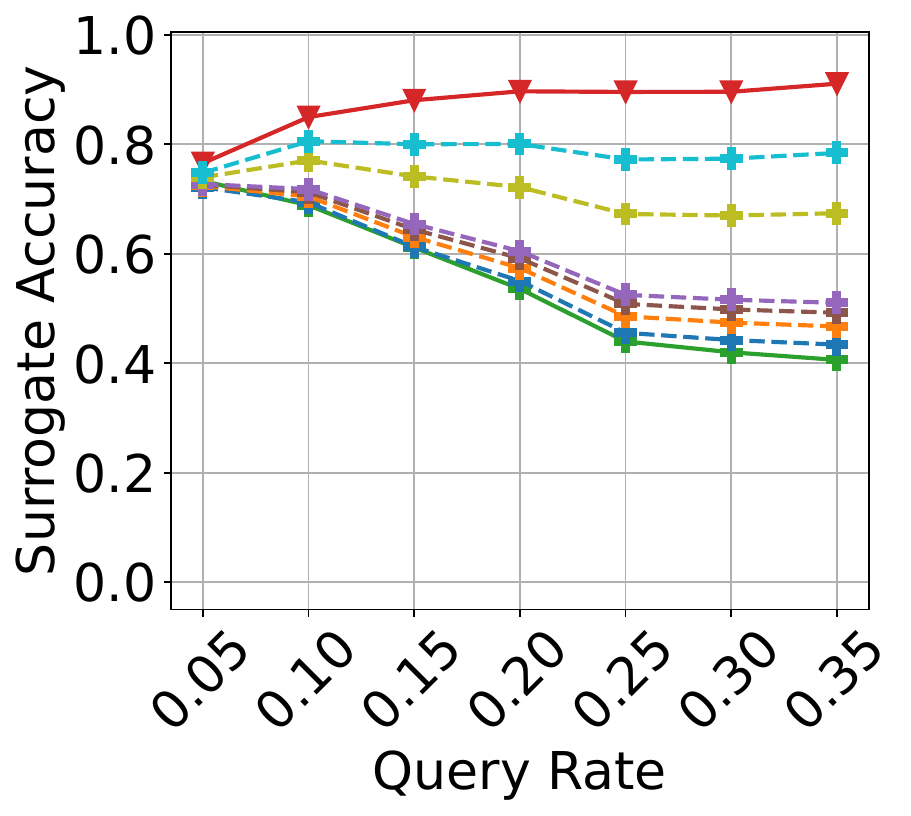}
         \caption{Probabilities.}
     \end{subfigure}
     \begin{subfigure}[t]{0.15\textwidth}
         \centering
         \includegraphics[width=\textwidth]{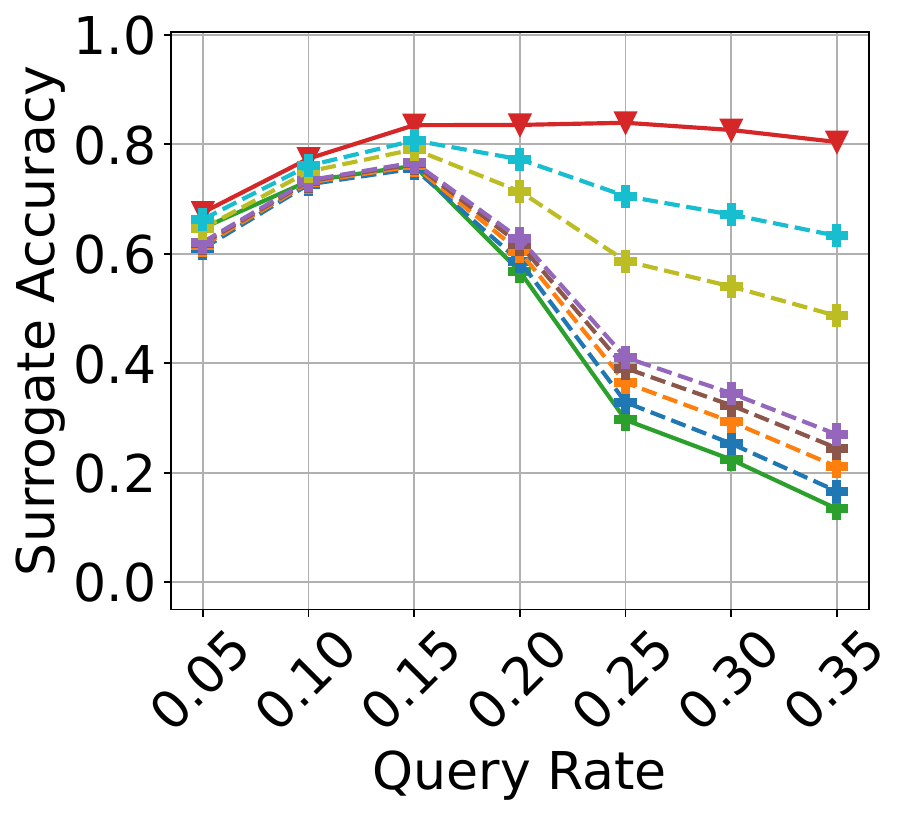}
         \caption{Embeddings.}
     \end{subfigure}
     \begin{subfigure}[t]{0.15\textwidth}
         \centering
         \includegraphics[width=\textwidth]{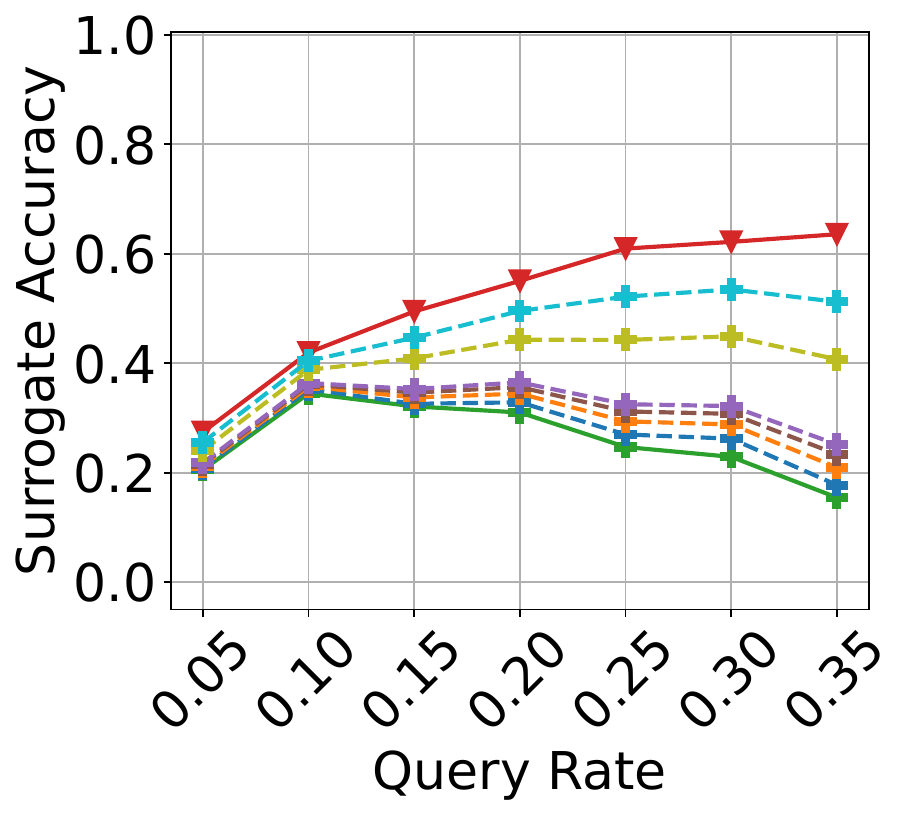}
         \caption{Projections.}
     \end{subfigure}
\caption{\textbf{Performance of the surrogate model with the first adaptive attack (Amazon, GAT).} %
}
\label{fig:average_noise_amazon}
\end{figure}

\begin{figure}[htpb]
\centering
    \begin{subfigure}[t]{0.35\textwidth}
         \centering
         \includegraphics[width=\textwidth]{figures/average_noise/legend.pdf}
     \end{subfigure}
     \vfill    
     \begin{subfigure}[t]{0.15\textwidth}
         \centering
         \includegraphics[width=\textwidth]{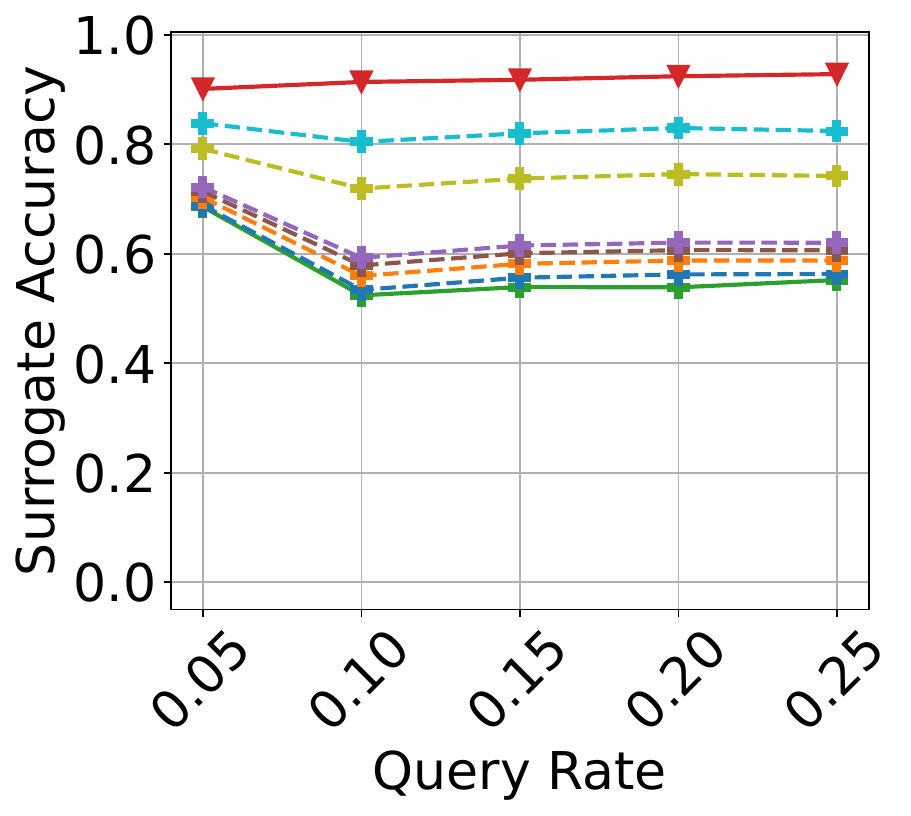}
         \caption{Probabilities.}
     \end{subfigure}
     \begin{subfigure}[t]{0.15\textwidth}
         \centering
         \includegraphics[width=\textwidth]{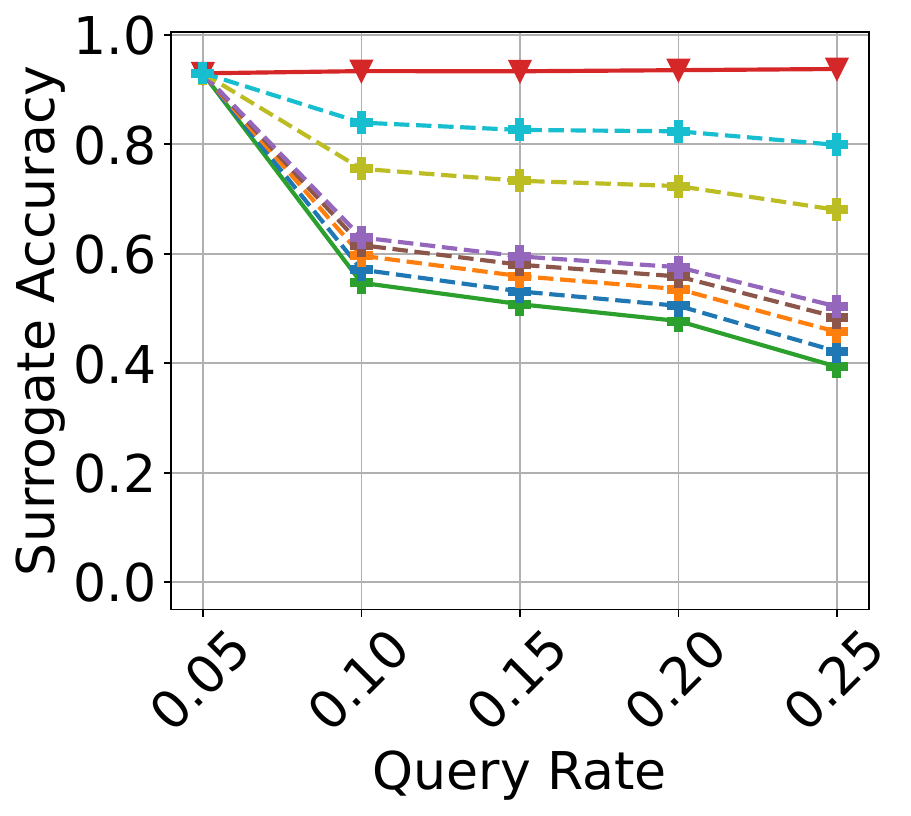}
         \caption{Embeddings.}
     \end{subfigure}
     \begin{subfigure}[t]{0.15\textwidth}
         \centering
         \includegraphics[width=\textwidth]{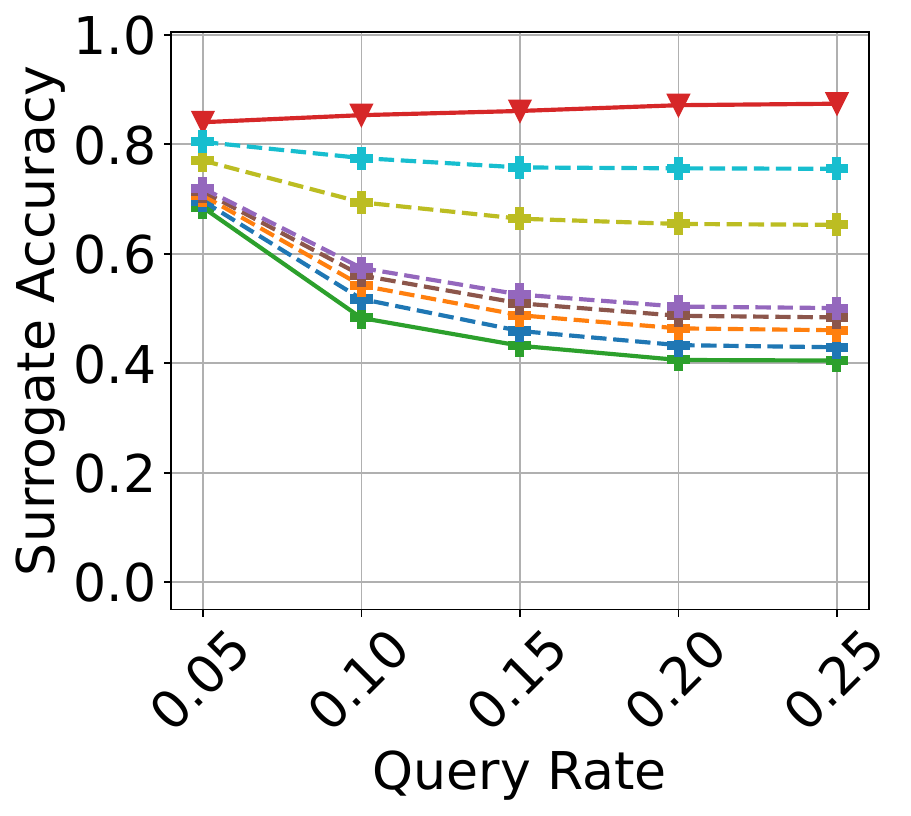}
         \caption{Projections.}
     \end{subfigure}
\caption{\textbf{Performance of the surrogate model with the first adaptive attack (Coauthor, GAT).} %
}
\label{fig:average_noise_coauthor}
\end{figure}

\begin{figure}[htpb]
\centering
    \begin{subfigure}[t]{0.35\textwidth}
         \centering
         \includegraphics[width=\textwidth]{figures/adaptive_attacker_community_knowledge/legend.pdf}
     \end{subfigure}
     \vfill  
     \begin{subfigure}[t]{0.15\textwidth}
         \centering
         \includegraphics[width=\textwidth]{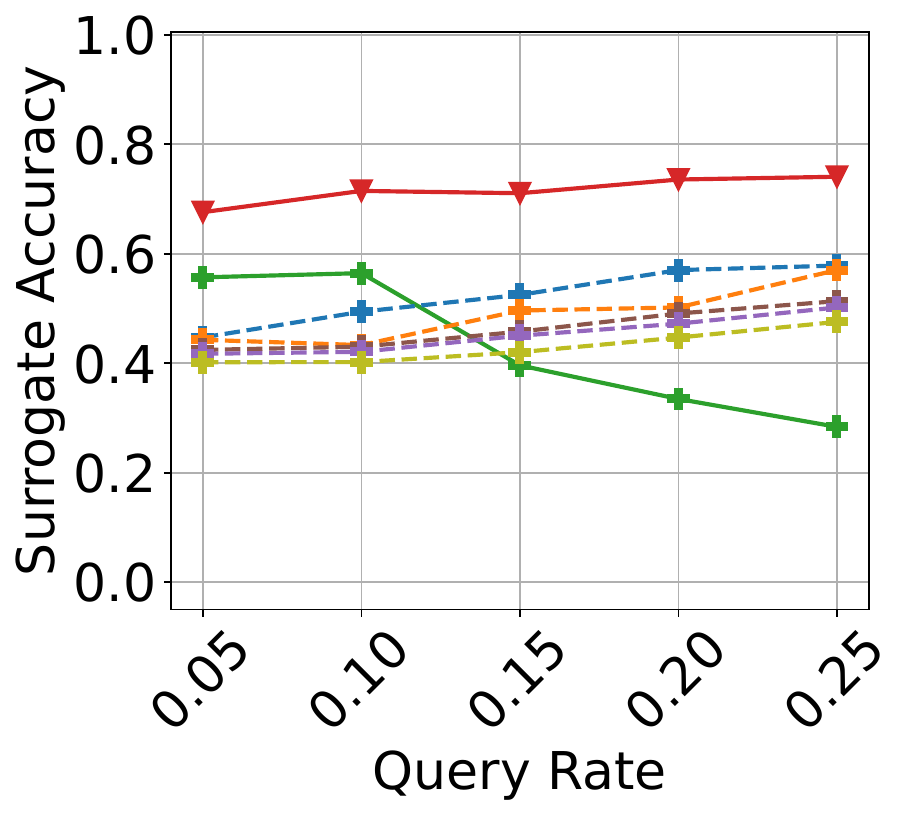}
         \caption{Probabilities.}
     \end{subfigure}
     \begin{subfigure}[t]{0.15\textwidth}
         \centering
         \includegraphics[width=\textwidth]{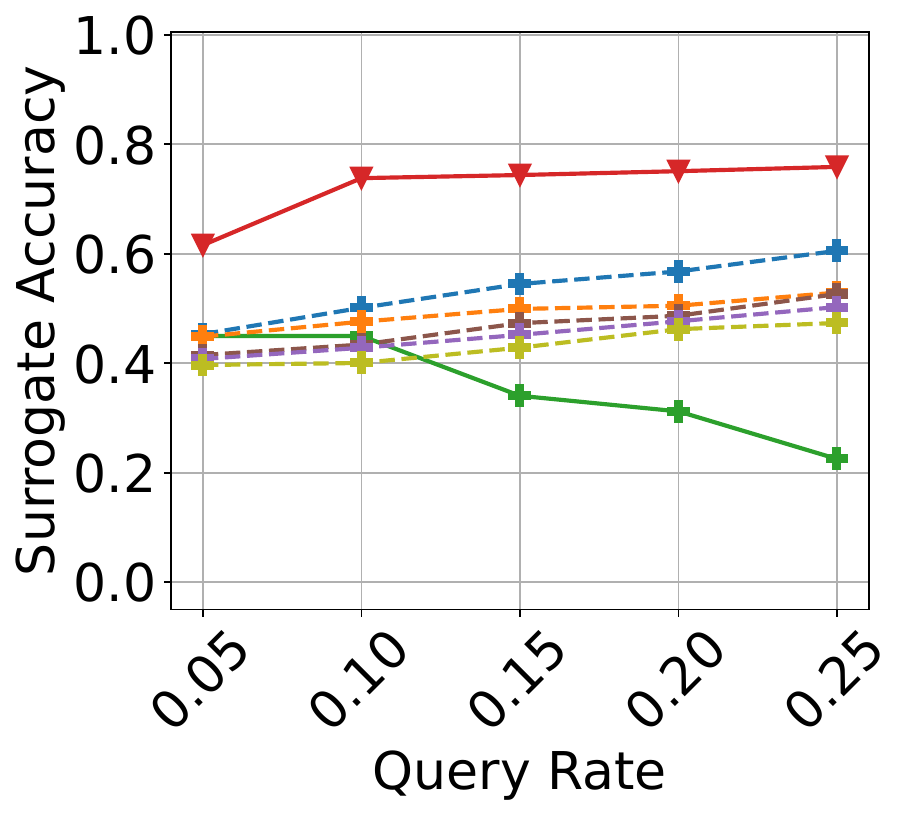}
         \caption{Embeddings.}
     \end{subfigure}
     \begin{subfigure}[t]{0.15\textwidth}
         \centering
         \includegraphics[width=\textwidth]{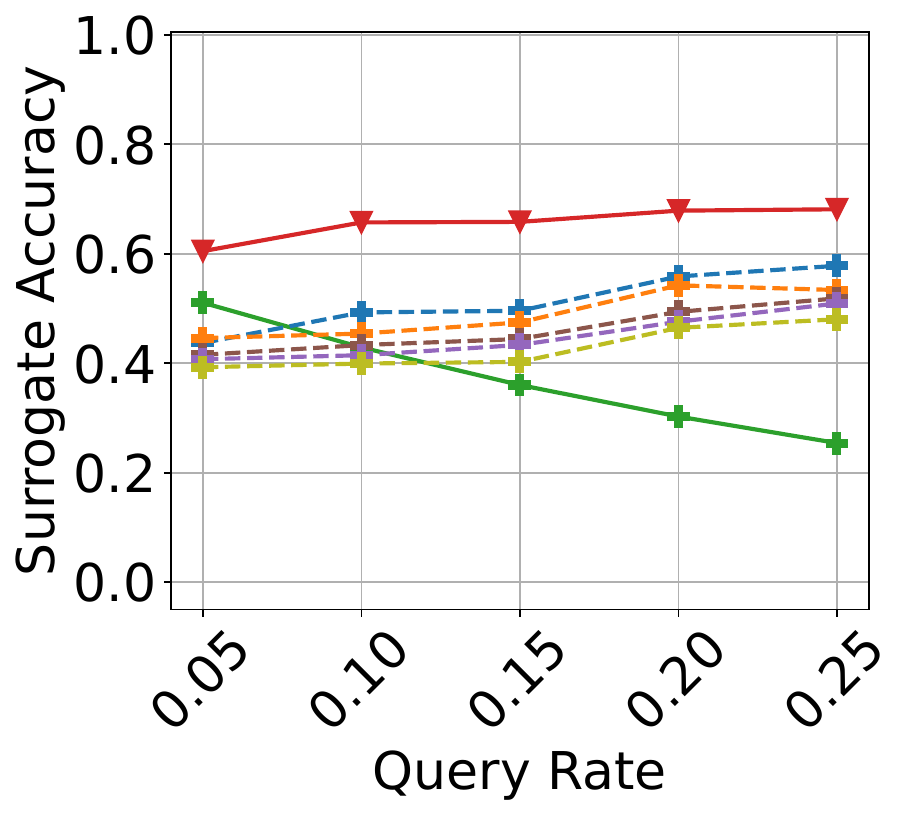}
         \caption{Projections.}
     \end{subfigure}
\caption{\textbf{Performance of the surrogate model with the second adaptive attack (DBLP, GAT).} %
}
\label{fig:adaptive_attacker_community_knowledge_dblp}
\end{figure}

\begin{figure}[htpb]
\centering
    \begin{subfigure}[t]{0.35\textwidth}
         \centering
         \includegraphics[width=\textwidth]{figures/adaptive_attacker_community_knowledge/legend.pdf}
     \end{subfigure}
     \vfill  
     \begin{subfigure}[t]{0.15\textwidth}
         \centering
         \includegraphics[width=\textwidth]{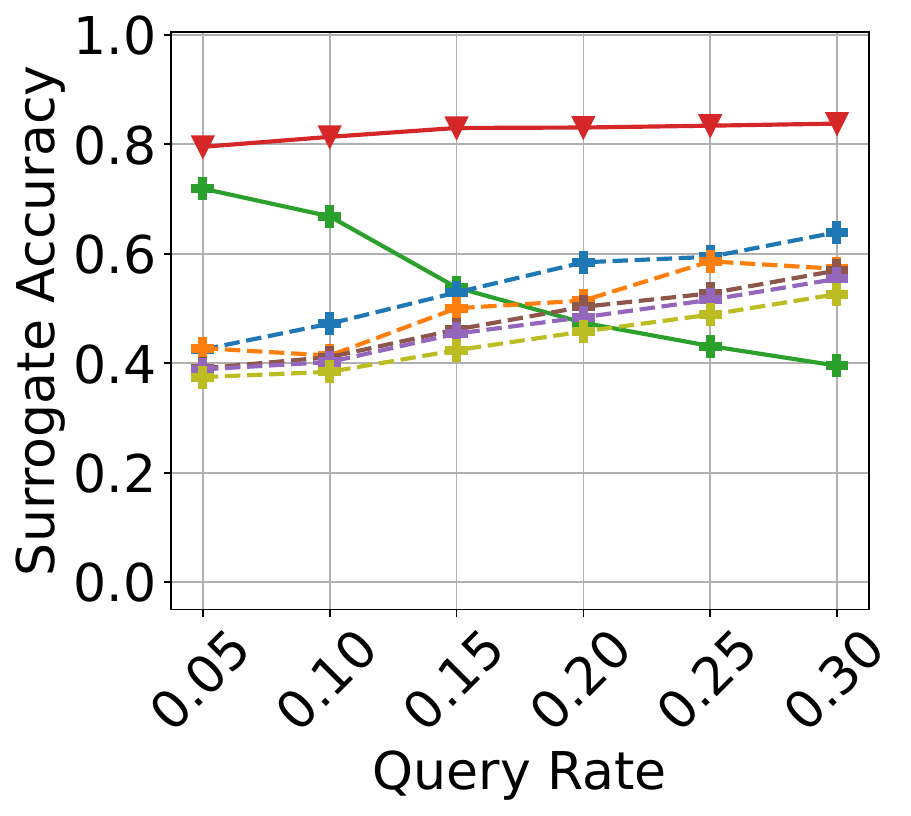}
         \caption{Probabilities.}
     \end{subfigure}
     \begin{subfigure}[t]{0.15\textwidth}
         \centering
         \includegraphics[width=\textwidth]{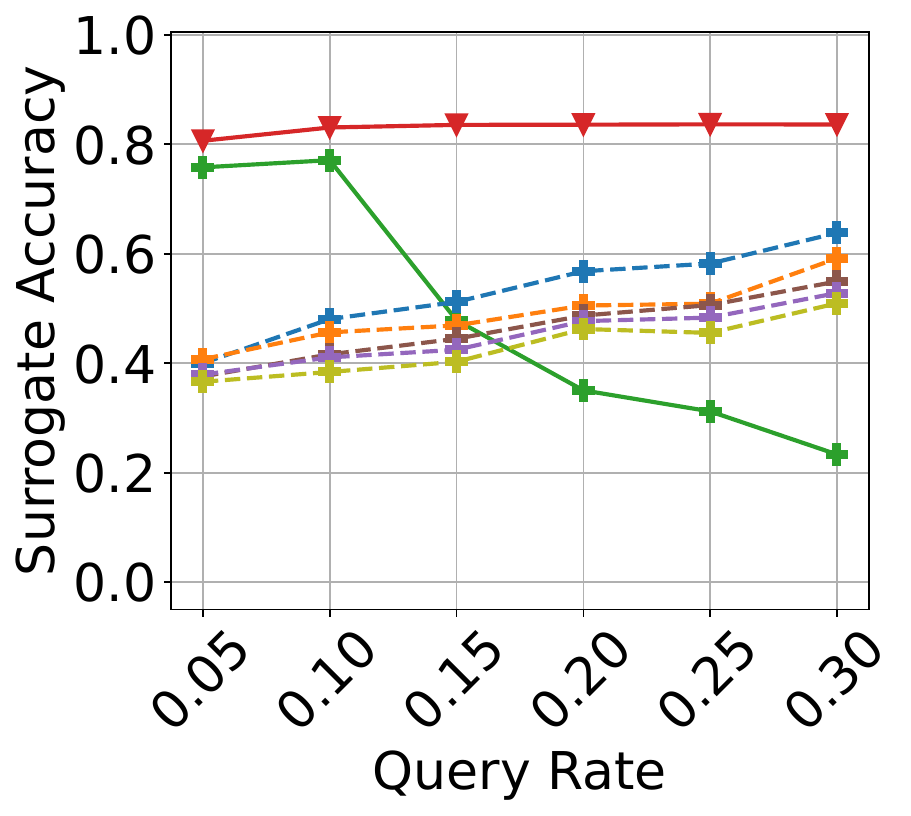}
         \caption{Embeddings.}
     \end{subfigure}
     \begin{subfigure}[t]{0.15\textwidth}
         \centering
         \includegraphics[width=\textwidth]{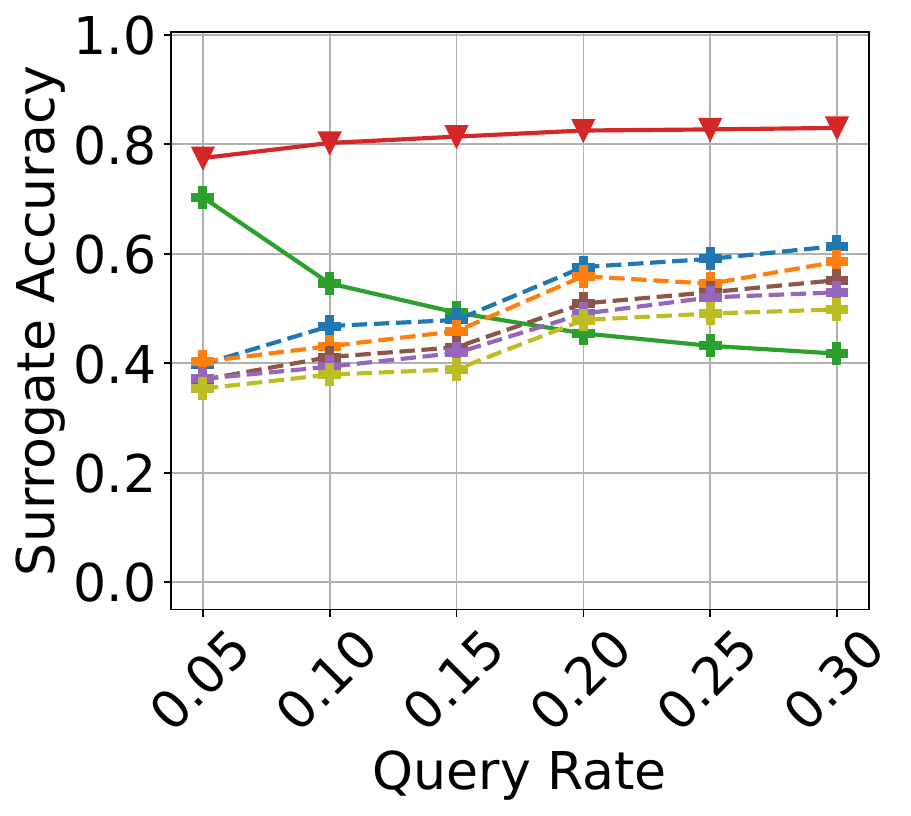}
         \caption{Projections.}
     \end{subfigure}
\caption{\textbf{Performance of the surrogate model with the second adaptive attack (Pubmed, GAT).} %
}
\label{fig:adaptive_attacker_community_knowledge_pubmed}
\end{figure}

\begin{figure}[htpb]
\centering
    \begin{subfigure}[t]{0.35\textwidth}
         \centering
         \includegraphics[width=\textwidth]{figures/adaptive_attacker_community_knowledge/legend.pdf}
     \end{subfigure}
     \vfill  
     \begin{subfigure}[t]{0.15\textwidth}
         \centering
         \includegraphics[width=\textwidth]{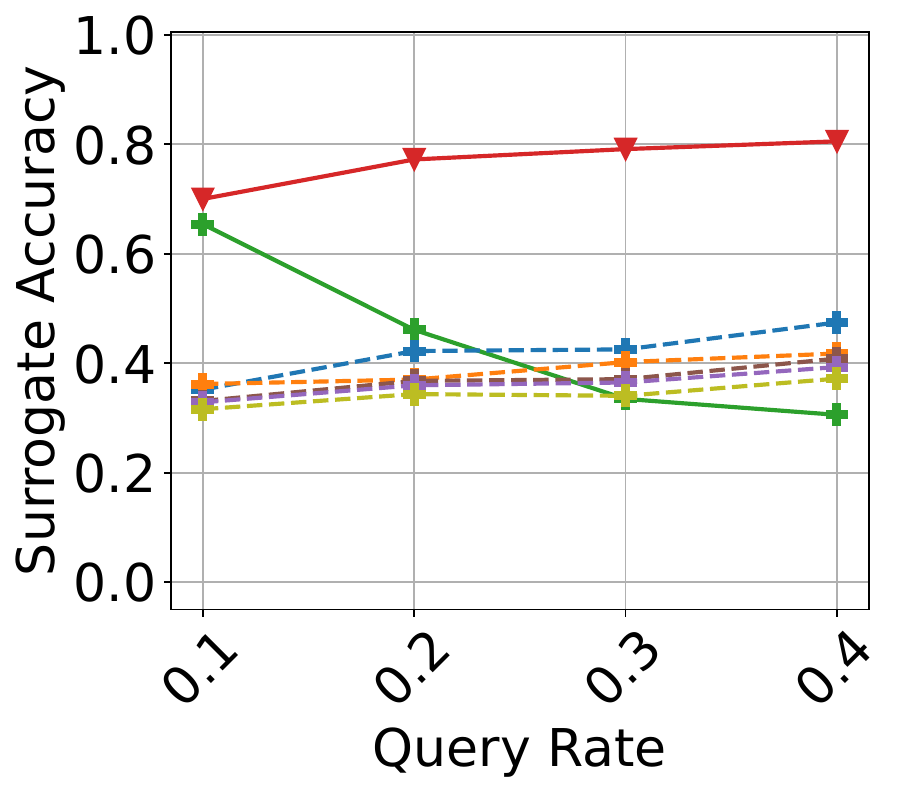}
         \caption{Probabilities.}
     \end{subfigure}
     \begin{subfigure}[t]{0.15\textwidth}
         \centering
         \includegraphics[width=\textwidth]{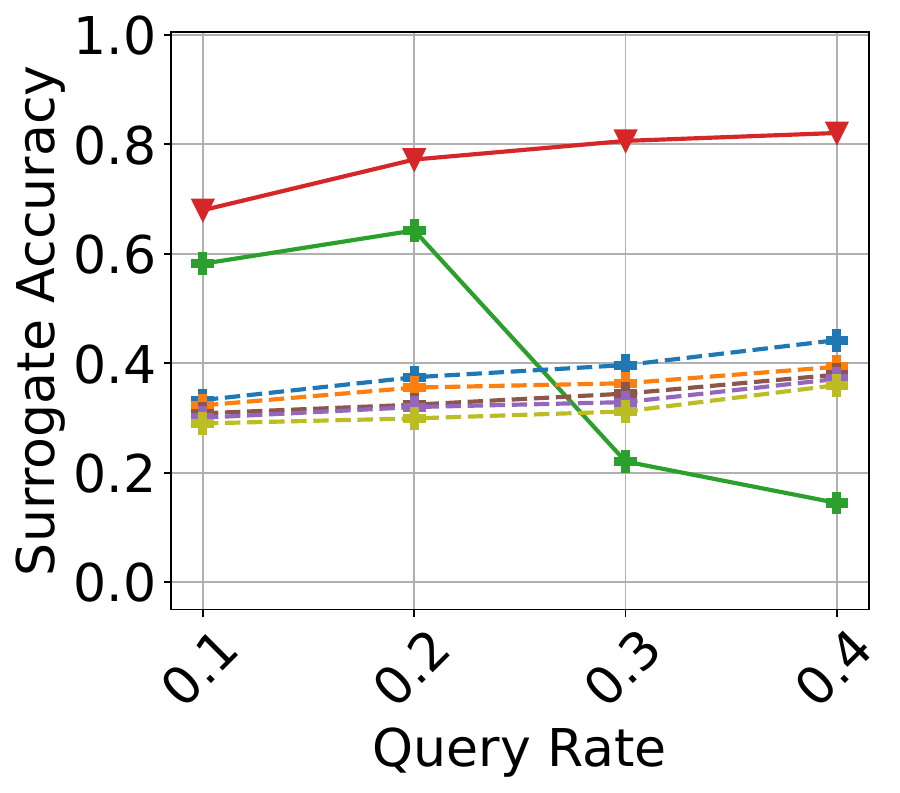}
         \caption{Embeddings.}
     \end{subfigure}
     \begin{subfigure}[t]{0.15\textwidth}
         \centering
         \includegraphics[width=\textwidth]{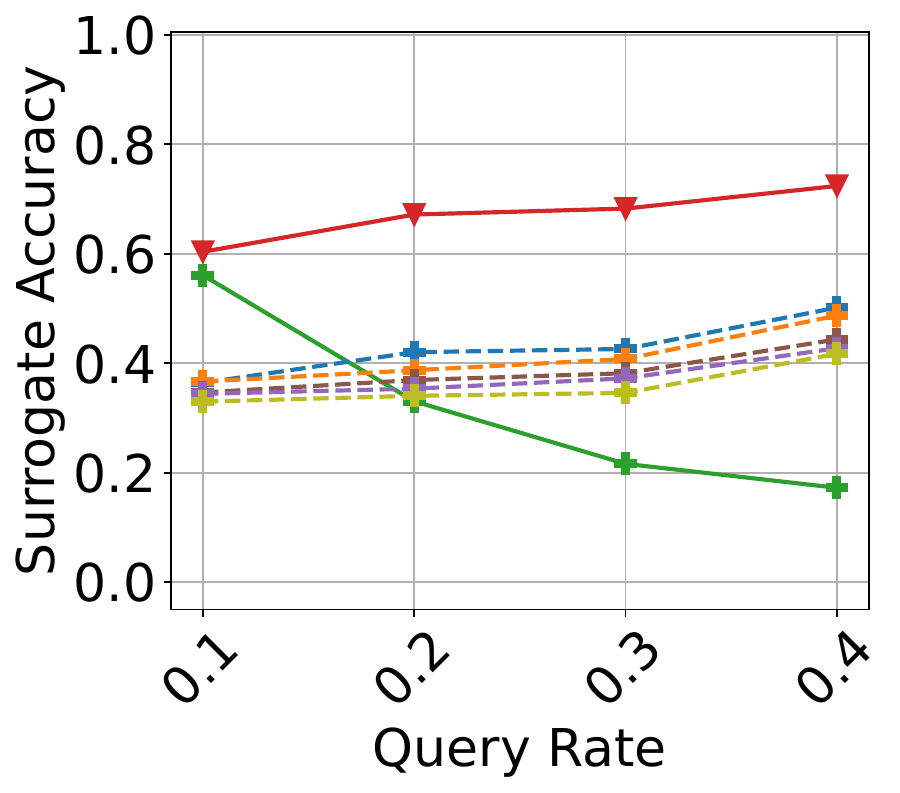}
         \caption{Projections.}
     \end{subfigure}
\caption{\textbf{Performance of the surrogate model with the second adaptive attack (Citeseer, GAT).} %
}
\label{fig:adaptive_attacker_community_knowledge_citeseer}
\end{figure}

\begin{figure}[htpb]
\centering
    \begin{subfigure}[t]{0.35\textwidth}
         \centering
         \includegraphics[width=\textwidth]{figures/adaptive_attacker_community_knowledge/legend.pdf}
     \end{subfigure}
     \vfill  
     \begin{subfigure}[t]{0.15\textwidth}
         \centering
         \includegraphics[width=\textwidth]{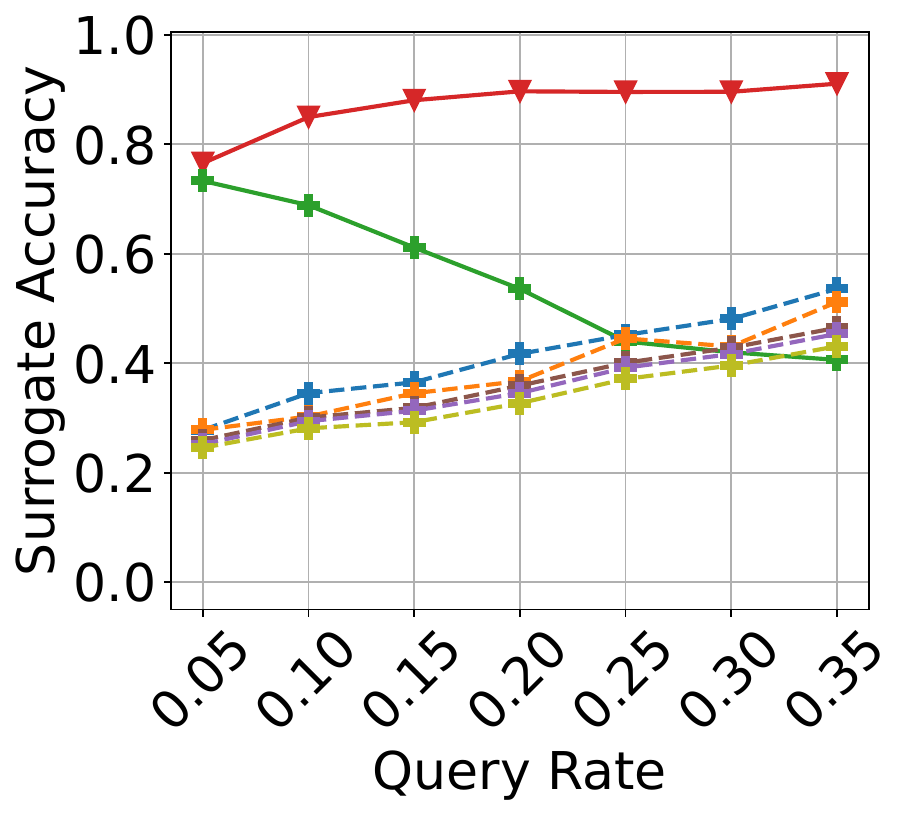}
         \caption{Probabilities.}
     \end{subfigure}
     \begin{subfigure}[t]{0.15\textwidth}
         \centering
         \includegraphics[width=\textwidth]{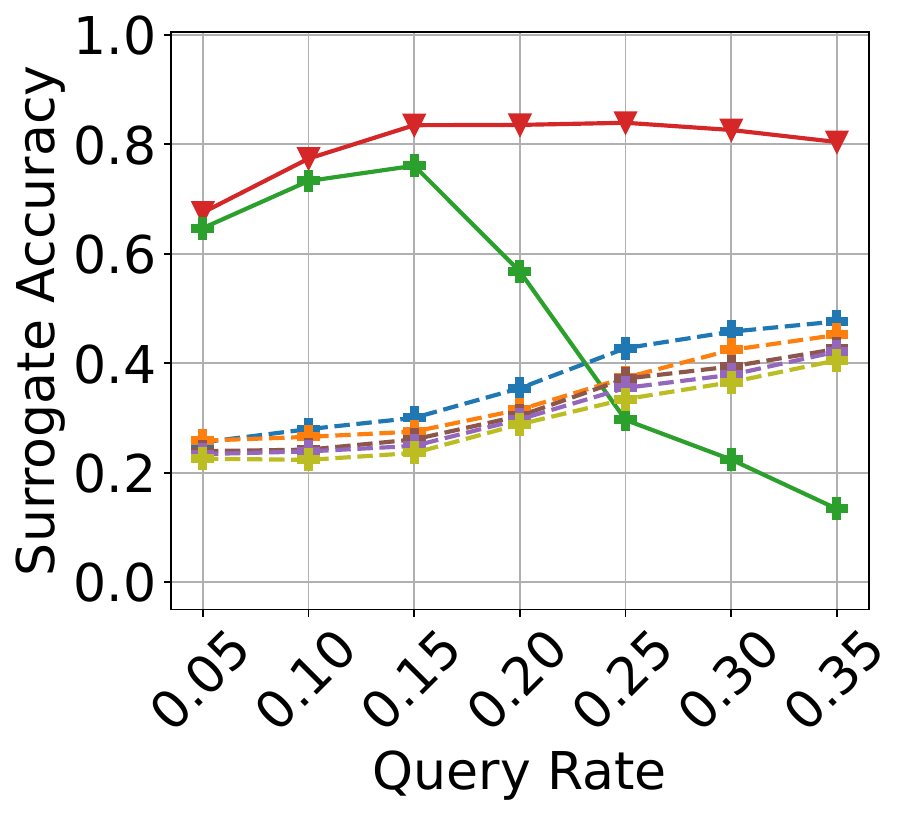}
         \caption{Embeddings.}
     \end{subfigure}
     \begin{subfigure}[t]{0.15\textwidth}
         \centering
         \includegraphics[width=\textwidth]{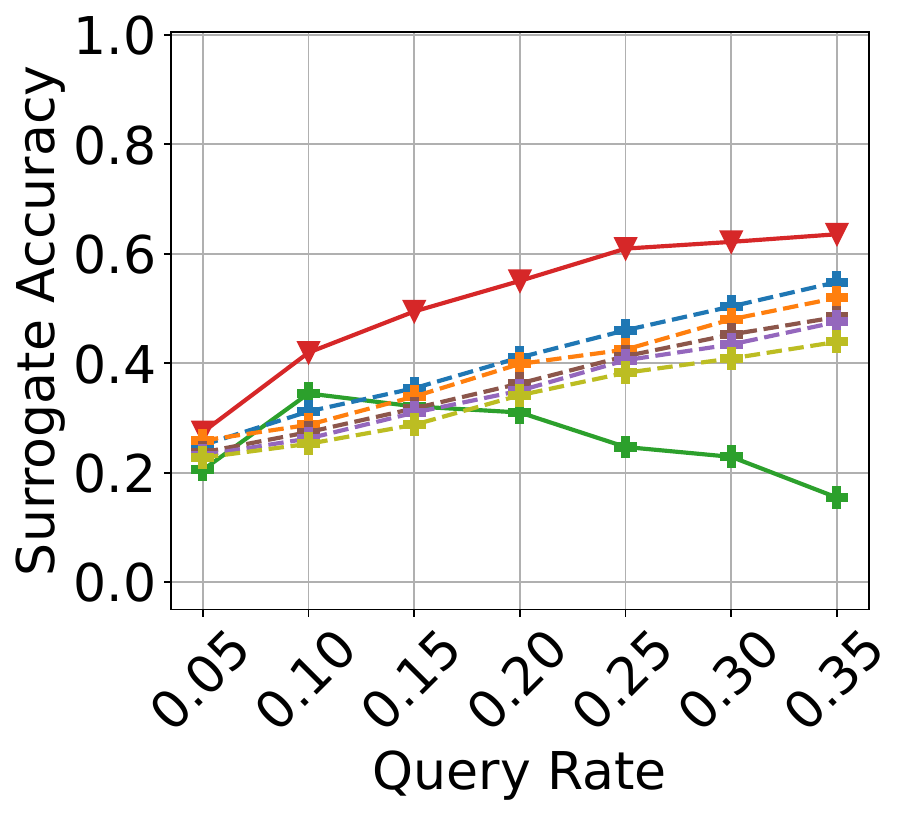}
         \caption{Projections.}
     \end{subfigure}
\caption{\textbf{Performance of the surrogate model with the second adaptive attack (Amazon, GAT).} %
}
\label{fig:adaptive_attacker_community_knowledge_amazon}
\end{figure}

\begin{figure}[htpb]
\centering
    \begin{subfigure}[t]{0.35\textwidth}
         \centering
         \includegraphics[width=\textwidth]{figures/adaptive_attacker_community_knowledge/legend.pdf}
     \end{subfigure}
     \vfill  
     \begin{subfigure}[t]{0.15\textwidth}
         \centering
         \includegraphics[width=\textwidth]{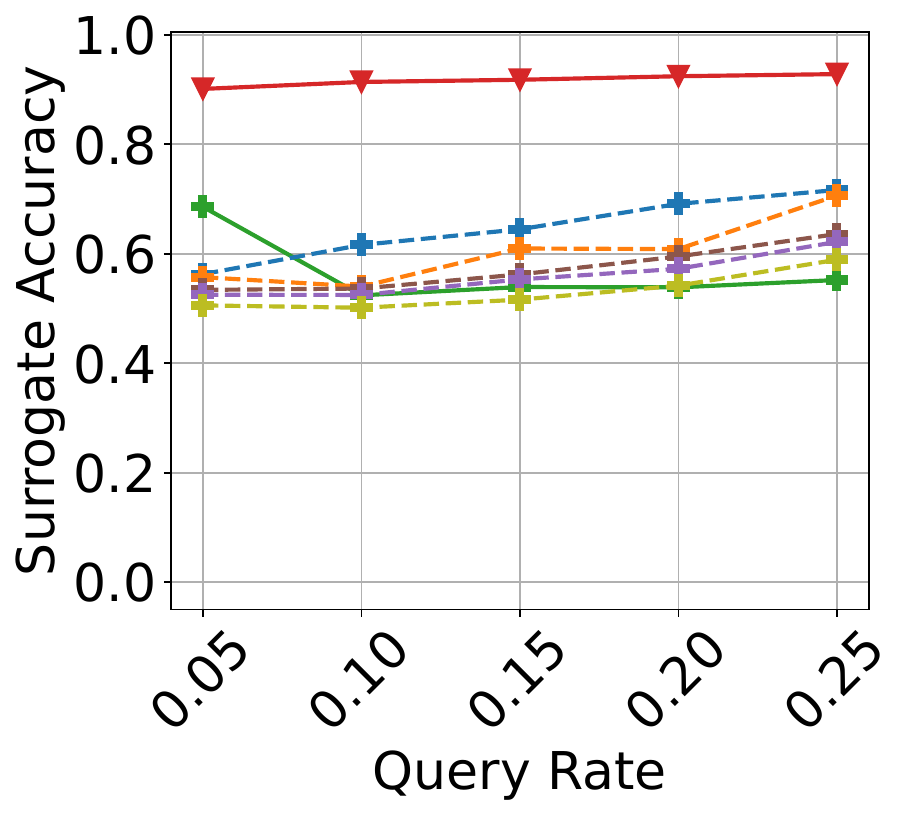}
         \caption{Probabilities.}
     \end{subfigure}
     \begin{subfigure}[t]{0.15\textwidth}
         \centering
         \includegraphics[width=\textwidth]{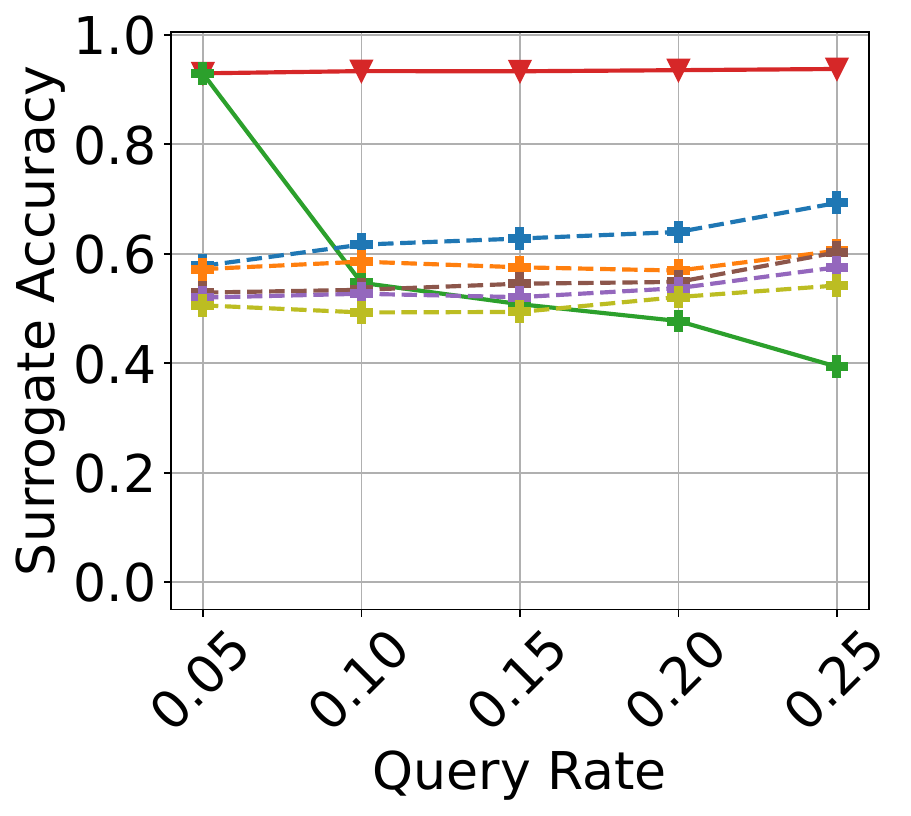}
         \caption{Embeddings.}
     \end{subfigure}
     \begin{subfigure}[t]{0.15\textwidth}
         \centering
         \includegraphics[width=\textwidth]{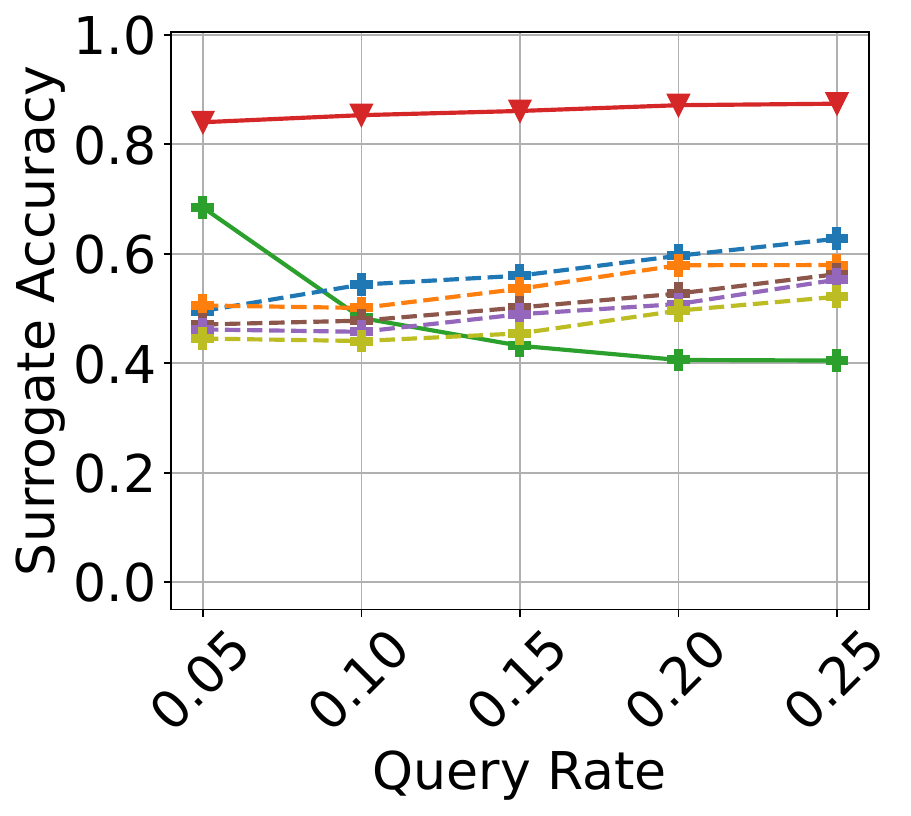}
         \caption{Projections.}
     \end{subfigure}
\caption{\textbf{Performance of the surrogate model with the second adaptive attack (Coauthor, GAT).} %
}
\label{fig:adaptive_attacker_community_knowledge_coauthor}
\end{figure}

To further quantify how many users are required to obtain a successful Sybil attack, we measured the remapping computational cost and attack performance of Sybil attacks under increasing numbers of accounts, ranging from 2 to 20. The results are presented in \Cref{table:remapping_cost}, taking the ACM dataset, attack setup B, GAT, and affine transformation as an example. 
It is noted that the reported remapping computational cost corresponds to a query rate of 0.8, as only when the query rate is more than 0.8 can the attacker remap the embeddings from different accounts into a similar embedding space (\ie with a remapping distance less than 0.2) and then achieve a successful model stealing attack.

\begin{table}[htpb]
\small
 \centering
 \caption{
 \textbf{Remapping computational cost and attack performance of Sybil attacks with various numbers of accounts (ACM, $\delta=0.8$, Embeddings setup, GAT, affine).}}
 \label{table:remapping_cost}
\begin{tabular}{p{4.4cm}cccc} 
 \hline
\textbf{Number of Accounts} & \textbf{2} & \textbf{5} & \textbf{10} & \textbf{20} \\
 \hline
Remapping Computational Cost (hours) & 0.90 & 2.12 & 4.28 & 6.15 \\
 \hline
Attack Performance (\%) & 53.82 & 59.47 & 66.84 & 70.71 \\
\hline
\end{tabular}
\end{table}

These results show a clear trade-off between attack effectiveness and efficiency for Sybil attacks. Specifically, increasing the number of accounts in Sybil attacks improves attack performance marginally but dramatically increases computational cost, \ie more than 6 hours for 20 accounts. 
Under the Sybil threat, it costs the attacker approximately 162 times the compute to steal the model over retraining the model from scratch when using 2 accounts, and about 1,107 times when using 20 accounts, disincentivizing stealing.

While remapping cost between representations from one account to another is costly for attackers, the transformations themselves are extremely efficient, as shown in \Cref{table:per_user_transformation_cost}. The results demonstrate that the per-user transformations we use to defend against Sybil attacks are very efficient, \ie less than $1 \times 10^{-4}$ seconds for per embedding or projection, making it practical for real-time use.

\begin{table*}[htpb]
 \centering
 \caption{\textbf{Computational cost ($\times 10^{-5} $seconds) of per-user transformation in B) and C) setups (ACM).} Overall, the per-user transformations used to defend against the Sybil attacks are very efficient.}
 \label{table:per_user_transformation_cost}
\begin{tabular}{ccccccc} 
 \hline
Attack setup & Transformation & GAT & GIN & GraphSAGE & Graph Transformer \\
 \hline
 \hline

\multirow{3}{*}{Embeddings} & Affine & {$ 2.96 \pm 0.49 $} & {$ 2.58 \pm 0.01 $} & {$ 2.53 \pm 0.00 $}& {$ 2.60 \pm 0.03 $}\\
 & Shuffle & {$ 0.70 \pm 0.00 $} & {$ 0.68 \pm 0.00 $} & {$ 0.68 \pm 0.00 $}& {$ 0.68 \pm 0.01 $}\\
 & Affine+Shuffle & {$ 3.00 \pm 0.05 $} & {$ 2.98 \pm 0.00 $}& {$ 2.97 \pm 0.02 $} & {$ 2.96 \pm 0.02 $}\\
 \hline
 \hline

\multirow{3}{*}{Projections} & Affine & {$ 2.23 \pm 0.39 $} & {$ 1.96 \pm 0.01 $} & {$ 1.96 \pm 0.02 $} & {$ 2.00 \pm 0.04 $}\\
 & Shuffle & {$ 0.28 \pm 0.00 $} & {$ 0.28 \pm 0.00 $} & {$ 0.26 \pm 0.00 $} & {$ 0.26 \pm 0.00 $}\\
 & Affine+Shuffle & {$ 2.24 \pm 0.01 $} & {$ 2.29 \pm 0.02 $} & {$ 2.27 \pm 0.01 $} & {$ 2.23 \pm 0.01 $}\\
 \hline
 \hline
\end{tabular}
\end{table*}

\section{Additional Insights into \ours}
\label{app:method_extended}
\subsection{Details on Designing Calibration Function in \Cref{equ:cost_label_stealing}}
\label{appendix:cost_label_flip}

Here, we present our motivation for designing the calibration function for label flipping probability, \ie \Cref{equ:cost_label_stealing}. 
First, we need to guarantee that the output of $h(\tau)$ is between $0$ and $1$ (since it is a probability).
Additionally, we want to yield low penalties for small fractions of occupied communities and high penalties for large fractions. This behavior can be best modeled with an exponential function that has a long flat area, and then a very steep increase. 
Therefore, we instantiate an exponential calibration function that maps the estimated information leakage to a label flipping probability $\rho$ as 

\begin{equation}
    \rho(\tau) = h_{a, b}(\tau) = \frac{1}{1+\exp^{a\tau+b}} \text{.}
    \label{equ:label_flipping_a_b}
\end{equation}
where $a, b$ are two hyperparameters.

Here, \Cref{equ:label_flipping_a_b} has two constraints: (1) when $\tau=0$ which means no community is occupied, the label flipping probability is $0$, (2) when $\tau=1$ which means that all communities are occupied, the label flipping probability should be $1$. Specifically, these two constraints are as follows:
\begin{equation}
\begin{split}
    \rho(\tau=0) = h_{a, b}(\tau=0) &= \frac{1}{1+\exp^{b}} = 0 \Rightarrow \exp^b = \infty\\
    \rho(\tau=1) = h_{a, b}(\tau=1) &= \frac{1}{1+\exp^{a+b}} = 1 \Rightarrow \exp^{a+b} = 0 \text{.}
\end{split}
\label{equ:restricts}
\end{equation}

If we define $\exp^{-b} = \lim_{\varepsilon \to 0}\varepsilon$ where $\varepsilon \in \mathbb{R}$, then
\begin{equation}
\begin{split}
    \exp^b &= \infty \\
    \exp^{a+b} &= \exp^{-b} \Rightarrow \exp^a = \exp^{-2b} \Rightarrow a=-2b \text{.}
\end{split}
\label{equ:a_2b}
\end{equation}

Thus, based on \Cref{equ:label_flipping_a_b}, \Cref{equ:restricts}, \Cref{equ:a_2b}, we can get calibration function as

\begin{equation}
    h_\eta(\tau) = \frac{1}{1+\exp^{\eta \times (1-2\times\tau)}} \text{.}
\end{equation}

\subsection{Query Diversity Metric}
\label{app:query_diversity_metric}
Our current query diversity metric is defined as $\tau=\frac{|I|}{K}$, which is intentionally simple and efficient to compute in real-time. To better capture both the fraction of nodes queried within each community and the community size and density, we additionally explore and discuss a weighted diversity metric.

Let $q_i$ be the number of queries assigned to community $C_i$, and $|C_i|$ is the size of the community. Instead of treating each community as binary (\ie occupied vs. not occupied), we first define a per-community occupancy score:
\begin{equation}
    u_i = 1-\exp\Big(-\gamma\frac{q_i}{|C_i|}\Big)
\end{equation}
where $\gamma>0$ is a scaling parameter controlling how fast the occupancy happens. When only one or a few queries are assigned to $C_i$, the ratio $q_i/|C_i|$ is small, leading to a small $u_i$, while when more queries are assigned to $C_i$, $u_i$ approaches 1, but each additional query increases $u_i$ less and less.

Also, instead of treating each community equally, the occupied score of each community is weighted according to its community size and density:
\begin{equation}
    w_i \;=\; \frac{\big(\tfrac{\lvert C_i\rvert}{\lvert V\rvert}\big)^{\eta}\, \delta_i^{\zeta}}{\sum_{j} \big(\tfrac{\lvert C_{j}\rvert}{\lvert V\rvert}\big)^{\eta}\, \delta_{j}^{\zeta}}
\end{equation}
where $|V|$ is the size of the whole graph, $|C_i|/|V|$ is the relative size of community $C_i$ compared to the whole graph. $\delta_i$ denotes a density measure for community $C_i$, \eg internal edge density. The exponents $\eta, \zeta>0$ control how strongly the community size and density influence the weight, respectively. For instance, if $\eta=\zeta=0$, all communities are treated equally and uniform weights are used (as in our current diversity metric), while larger values emphasize larger and/or denser communities. The denominator normalizes the weights so that $\sum_iw_i=1$. Thus, $w_i$ acts as an important weight that can prioritize structurally significant communities in the diversity metric.

The resulting weighted diversity is $\tau_w = \sum_{i=1}^Kw_iu_i$. This weighted query diversity metric considers not only the fraction of nodes queried within each community, but also the community size and density. 

\subsection{Per-query Complexity of Comparing Embeddings to All Community Centroids}
\label{app:complexity_comparing_emgeddings}
To further evaluate the computational complexity and scalability of our \ours , we also report the per-query complexity of comparing embeddings to all community centroids for each dataset (with a different number of communities), as presented in \Cref{tab:per_query_complexity}. We can observe that the per-query complexity of comparing embeddings to all community centroids is very low, \ie less than $1\times 10^{-5}$ seconds. Even for much larger real-world graphs, \ie Amazon Co-purchasing, which contains around 50K communities, the per-query computational cost is only $5.5775 \times 10^{-5}$ seconds, confirming the scalability of ADAGE. Furthermore, we can always set a proper value for $K$ to balance the defense strength and the per-query complexity.

\begin{table}[ht]
\scriptsize
\centering
\caption{\textbf{Per-query computational cost ($\times 10^{-6}$ seconds) of comparing embeddings to all community centroids for each dataset (GT - Graph Transformer).} We can observe that the complexity is very low.}
\begin{tabular}{ccccccc}
\hline
Model & ACM & DBLP & Pubmed & Citeseer & Amazon & Coauthor \\ \hline \hline
GAT & {$ 7.2598 $} & {$ 4.2510 $} & {$ 6.7782 $} & {$ 6.2704 $} & {$ 4.5919 $} & {$ 6.9714 $} \\ \hline
GIN & {$ 2.0576 $} & {$ 1.4305 $} & {$ 2.1672 $} & {$ 1.8001 $} & {$ 1.4305 $} & {$ 3.9816 $}\\ \hline
GraphSAGE & {$ 3.3307 $} & {$ 2.0099 $} & {$ 2.5392 $} & {$ 2.8300 $} & {$ 2.0719 $} & {$ 2.6488 $}\\ \hline
GT & {$ 2.3770 $} & {$ 1.8692 $} & {$ 2.0218 $} & {$ 1.8597 $} & {$ 1.5187 $} & {$ 2.1410 $}\\ \hline
\end{tabular}
\label{tab:per_query_complexity}
\end{table}

\begin{table*}[!h]
\small
 \centering
 \caption{\textbf{Performance for attacker and target downstream task with and without \ours for link prediction task, in three attack setups (Cora, $\delta=0.25$, $c_i$ represents a community).} Overall, with our defense, the performance for target downstream tasks remains high while the performance of the surrogate model is significantly degraded.
 }
 \label{table:results_with_without_defenses_link_cora}
\begin{tabular}{ccccccc} 
 \hline
  & User & Dataset & Defense & GAT & GIN & GCN \\
 \hline
Baseline & N/A & $\mathbf{G}_{test}$ & N/A & $68.82\pm 4.27$ & $76.95\pm2.05$ & $65.10\pm3.27$ \\
 \hline
 \hline
 \multirow{5}{22mm}{\begin{center}Attack setup A (Probabilities)\end{center}} 
   & Attacker & $\mathbf{G}_{test}$ & NONE & {$53.00\pm0.76$} & {$76.05\pm0.38$} & {$61.80\pm4.69$} \\
  & Attacker & $\mathbf{G}_{test}$ & \ours & {$\textbf{48.89}\pm0.98$} & {$\textbf{32.94}\pm2.03$} & {$\textbf{45.34}\pm1.21$} \\
\cline{2-7}
& Downstream Task 1 & $c_1$ & \ours & {$67.33\pm0.38$}  & {$76.54\pm0.60$}  & {$63.16\pm0.29$} \\
 & Downstream Task 2 & $c_2$ & \ours & {$67.19\pm0.21$}  & {$75.56\pm1.96$}  & {$63.70\pm1.33$} \\
 & Downstream Task 3 & $c_3$ & \ours & {$66.10\pm0.72$}  & {$72.71\pm0.46$}  & {$60.70\pm1.06$} \\
 \hline
 \hline

\multirow{5}{22mm}{\begin{center}Attack setup B (Embeddings)\end{center}} 
 & Attacker & $\mathbf{G}_{test}$ & NONE & {$61.31\pm2.56$} & {$76.71\pm0.79$} & {$63.38\pm0.66$} \\
 & Attacker & $\mathbf{G}_{test}$ & \ours & {$\textbf{50.03}\pm0.05$} & {$\textbf{50.52}\pm0.47$} & {$\textbf{49.70}\pm1.52$} \\
\cline{2-7}
 & Downstream Task 1 & $c_1$ & \ours & {$64.25\pm0.60$}  & {$76.25\pm1.85$}  & {$64.40\pm0.01$} \\
 & Downstream Task 2 & $c_2$ & \ours & {$68.09\pm0.20$}  & {$71.21\pm0.74$}  & {$59.93\pm0.17$} \\
 & Downstream Task 3 & $c_3$ & \ours & {$68.14\pm0.34$}  & {$76.75\pm1.76$}  & {$60.83\pm0.32$} \\
 \hline
 \hline

\multirow{5}{22mm}{\begin{center}Attack setup C (Projections)\end{center}} 
 & Attacker & $\mathbf{G}_{test}$ & NONE & {$64.55\pm2.04$} & {$76.60\pm1.03$} & {$70.66\pm2.33$} \\
 & Attacker & $\mathbf{G}_{test}$ & \ours & {$\textbf{51.93}\pm3.08$} & {$\textbf{54.20}\pm1.95$} & {$\textbf{51.61}\pm2.90$} \\
\cline{2-7}
 & Downstream Task 1 & $c_1$ & \ours & {$64.08\pm1.06$}  & {$74.65\pm0.82$}  & {$63.69\pm1.22$} \\
 & Downstream Task 2 & $c_2$ & \ours & {$67.13\pm0.12$}  & {$73.91\pm0.30$}  & {$64.16\pm1.48$} \\
 & Downstream Task 3 & $c_3$ & \ours & {$67.19\pm0.39$}  & {$76.62\pm0.68$}  & {$61.65\pm0.61$} \\

 \hline
 \hline
\end{tabular}
\end{table*}

\begin{table*}[!h]
\small
 \centering
 \caption{\textbf{Performance for attacker and target downstream tasks with and without \ours for link prediction task, in three attack setups (CiteSeer, $\delta=0.25$, $c_i$ represents a community).} Overall, with our defense, the performance for the target downstream tasks remains high while the performance of the surrogate model is significantly degraded.
 }
 \label{table:results_with_without_defenses_link_citeseer}
\begin{tabular}{ccccccc} 
 \hline
  & User & Dataset & Defense & GAT & GIN & GCN \\
 \hline
Baseline & N/A & $\mathbf{G}_{test}$ & N/A & $73.50\pm 0.75$ & $82.03\pm1.87$ & $65.88\pm3.78$ \\
 \hline
 \hline
\multirow{5}{22mm}{\begin{center}Attack setup A (Probabilities)\end{center}}  
  & Attacker & $\mathbf{G}_{test}$ & NONE & {$61.09\pm0.50$} & {$77.11\pm0.79$} & {$69.01\pm2.81$} \\
  & Attacker & $\mathbf{G}_{test}$ & \ours & {$\textbf{47.38}\pm0.58$} & {$\textbf{31.15}\pm1.86$} & {$\textbf{45.57}\pm1.52$} \\
\cline{2-7}
 & Downstream Task 1 & $c_1$ & \ours & {$71.42\pm0.10$}  & {$76.95\pm3.26$}  & {$64.92\pm0.73$} \\
 & Downstream Task 2 & $c_2$ & \ours & {$67.64\pm2.30$}  & {$80.35\pm0.01$}  & {$60.01\pm1.39$} \\
 & Downstream Task 3 & $c_3$ & \ours & {$70.11\pm0.13$}  & {$75.04\pm1.01$}  & {$62.65\pm4.55$} \\

 \hline
 \hline

\multirow{5}{22mm}{\begin{center}Attack setup B (Embeddings)\end{center}} 
 & Attacker & $\mathbf{G}_{test}$ & NONE & {$70.20\pm2.44$} & {$81.30\pm1.52$} & {$67.40\pm0.59$} \\
 & Attacker & $\mathbf{G}_{test}$ & \ours & {$\textbf{48.11}\pm3.27$} & {$\textbf{52.40}\pm2.86$} & {$\textbf{52.58}\pm3.23$} \\
 \cline{2-7}
 & Downstream Task 1 & $c_1$ & \ours & {$70.28\pm0.05$}  & {$81.96\pm1.65$}  & {$64.30\pm3.11$} \\
 & Downstream Task 2 & $c_2$ & \ours & {$72.63\pm0.57$}  & {$80.00\pm1.28$}  & {$61.99\pm2.23$} \\
 & Downstream Task 3 & $c_3$ & \ours & {$68.54\pm1.34$}  & {$79.79\pm3.50$}  & {$63.71\pm2.35$} \\
 \hline
 \hline

\multirow{5}{22mm}{\begin{center}Attack setup C (Projections)\end{center}} 
 & Attacker & $\mathbf{G}_{test}$ & NONE & {$64.47\pm1.08$} & {$80.73\pm0.19$} & {$64.03\pm0.41$} \\
 & Attacker & $\mathbf{G}_{test}$ & \ours & {$\textbf{51.54}\pm2.14$} & {$\textbf{52.38}\pm1.23$} & {$\textbf{50.43}\pm0.89$} \\
\cline{2-7}
 & Downstream Task 1 & $c_1$ & \ours & {$68.38\pm0.80$}  & {$79.39\pm1.22$}  & {$64.06\pm0.13$} \\
 & Downstream Task 2 & $c_2$ & \ours & {$72.07\pm0.63$}  & {$78.88\pm3.53$}  & {$64.95\pm1.87$} \\
 & Downstream Task 3 & $c_3$ & \ours & {$71.56\pm3.46$}  & {$76.75\pm1.68$}  & {$60.10\pm0.18$} \\

 \hline
 \hline
\end{tabular}
\end{table*}

\section{Extension to Other Graph Tasks}
\label{app:other_tasks}
In addition to the node classification task, we also evaluate \ours on the link prediction task. The general idea of \ours remains the same, where we still use the query diversity to design the penalty function. However, the community detection in the link prediction task is slightly different from that in the node classification task. In the node classification task, we detect communities based on the graph structure of the training graph $G_{train}$, while in the link prediction task, we detect communities based on the representations of links in the training graph. Specifically, we first obtain the representations of each link in the training graph, and then we apply the community detection algorithm, \ie k-means, on these link representations. Similar to the node classification task, once the communities in $G_{train}$ are determined, we calculate the centroids of these communities and use them to calculate the query diversity. Then, when one query comes, based on the internal representation of the link, we can determine which community it belongs to according to the distance to the centroids of the communities. The penalty function is then designed based on the percentage of occupied communities, which is similar to the node classification task. 

We evaluate \ours on the link prediction task using the Cora and CiteSeer datasets, and we use GAT, GIN, and GCN for the target and surrogate models. 
The experimental results on the link prediction task are presented in \Cref{table:results_with_without_defenses_link_cora} and \Cref{table:results_with_without_defenses_link_citeseer} for Cora and CiteSeer datasets, respectively.
The results show that \ours can effectively degrade the stealing performance of the surrogate model while maintaining the performance on downstream tasks, similar to the node classification task.

\end{document}